\overfullrule=0pt
\overfullrule=0pt
 
\newcount\mgnf  
\mgnf=0
 
\ifnum\mgnf=0
\def\openone{\leavevmode\hbox{\ninerm 1\kern-3.3pt\tenrm1}}%
\def\*{\vglue0.3truecm}\fi
\ifnum\mgnf=1
\def\openone{\leavevmode\hbox{\ninerm 1\kern-3.63pt\tenrm1}}%
\def\*{\vglue0.5truecm}\fi
\ifnum\mgnf=0
\magnification=\magstep1
\hsize=15truecm\vsize=23truecm
\parindent=4.pt\baselineskip=0.45cm
\font\titolo=cmbx12
\font\titolone=cmbx10 scaled\magstep 2
\font\cs=cmcsc10
\font\ottorm=cmr8

\font\msytw=msbm10

\font\indbf=cmbx10 scaled\magstep1

\fi
\ifnum\mgnf=2
   \magnification=\magstep1\hoffset=0.truecm
   \hsize=15truecm\vsize=24.truecm
   \baselineskip=18truept plus0.1pt minus0.1pt \parindent=0.9truecm
   \lineskip=0.5truecm\lineskiplimit=0.1pt      \parskip=0.1pt plus1pt
\font\titolo=cmbx12 scaled\magstep 1
\font\titolone=cmbx10 scaled\magstep 3
\font\cs=cmcsc10 scaled\magstep 1
\font\ottorm=cmr8 scaled\magstep 1

\font\msytw=msbm10 scaled\magstep1

\font\indbf=cmbx10 scaled\magstep2
\fi
 
\global\newcount\numsec\global\newcount\numapp
\global\newcount\numfor\global\newcount\numfig\global\newcount\numsub
\numsec=0\numapp=0\numfig=1
\def\veroparagrafo{\number\numsec}\def\veraformula{\number\numfor}
\def\veraappendice{\number\numapp}\def\verasub{\number\numsub}
\def\verafigura{\number\numfig}
 
\def\section(#1,#2){\advance\numsec by 1\numfor=1\numsub=1%
\SIA p,#1,{\veroparagrafo} %
\write15{\string\Fp (#1){\secc(#1)}}%
\write16{ sec. #1 ==> \secc(#1)  }%
\hbox to \hsize{\titolo\hfill \number\numsec. #2\hfill%
\expandafter{\alato(sec. #1)}}\*}
 
\def\appendix(#1,#2){\advance\numapp by 1\numfor=1\numsub=1%
\SIA p,#1,{A\veraappendice} %
\write15{\string\Fp (#1){\secc(#1)}}%
\write16{ app. #1 ==> \secc(#1)  }%
\hbox to \hsize{\titolo\hfill Appendix A\number\numapp. #2\hfill%
\expandafter{\alato(app. #1)}}\*}
 
\def\senondefinito#1{\expandafter\ifx\csname#1\endcsname\relax}
 
\def\SIA #1,#2,#3 {\senondefinito{#1#2}%
\expandafter\xdef\csname #1#2\endcsname{#3}\else
\write16{???? ma #1#2 e' gia' stato definito !!!!} \fi}
 
\def \Fe(#1)#2{\SIA fe,#1,#2 }
\def \Fp(#1)#2{\SIA fp,#1,#2 }
\def \Fg(#1)#2{\SIA fg,#1,#2 }
 
\def\etichetta(#1){(\veroparagrafo.\veraformula)%
\SIA e,#1,(\veroparagrafo.\veraformula) %
\global\advance\numfor by 1%
\write15{\string\Fe (#1){\equ(#1)}}%
\write16{ EQ #1 ==> \equ(#1)  }}
 
\def\etichettaa(#1){(A\veraappendice.\veraformula)%
\SIA e,#1,(A\veraappendice.\veraformula) %
\global\advance\numfor by 1%
\write15{\string\Fe (#1){\equ(#1)}}%
\write16{ EQ #1 ==> \equ(#1) }}
 
\def\getichetta(#1){Fig. \verafigura%
\SIA g,#1,{\verafigura} %
\global\advance\numfig by 1%
\write15{\string\Fg (#1){\graf(#1)}}%
\write16{ Fig. #1 ==> \graf(#1) }}
 
\def\etichettap(#1){\veroparagrafo.\verasub%
\SIA p,#1,{\veroparagrafo.\verasub} %
\global\advance\numsub by 1%
\write15{\string\Fp (#1){\secc(#1)}}%
\write16{ par #1 ==> \secc(#1)  }}
 
\def\etichettapa(#1){A\veraappendice.\verasub%
\SIA p,#1,{A\veraappendice.\verasub} %
\global\advance\numsub by 1%
\write15{\string\Fp (#1){\secc(#1)}}%
\write16{ par #1 ==> \secc(#1)  }}
 
\def\Eq(#1){\eqno{\etichetta(#1)\alato(#1)}}
\def\eq(#1){\etichetta(#1)\alato(#1)}
\def\Eqa(#1){\eqno{\etichettaa(#1)\alato(#1)}}
\def\eqa(#1){\etichettaa(#1)\alato(#1)}
\def\eqg(#1){\getichetta(#1)\alato(fig. #1)}
\def\sub(#1){\0\palato(p. #1){\bf \etichettap(#1)\hskip.3truecm}}
\def\asub(#1){\0\palato(p. #1){\bf \etichettapa(#1)\hskip.3truecm}}
 
\def\equv(#1){\senondefinito{fe#1}$\clubsuit$#1%
\write16{eq. #1 non e' (ancora) definita}%
\else\csname fe#1\endcsname\fi}
\def\grafv(#1){\senondefinito{fg#1}$\clubsuit$#1%
\write16{fig. #1 non e' (ancora) definito}%
\else\csname fg#1\endcsname\fi}
\def\secv(#1){\senondefinito{fp#1}$\clubsuit$#1%
\write16{par. #1 non e' (ancora) definito}%
\else\csname fp#1\endcsname\fi}
 
\def\equ(#1){\senondefinito{e#1}\equv(#1)\else\csname e#1\endcsname\fi}
\def\graf(#1){\senondefinito{g#1}\grafv(#1)\else\csname g#1\endcsname\fi}
\def\secc(#1){\senondefinito{p#1}\secv(#1)\else\csname p#1\endcsname\fi}
\def\sec(#1){{\S\secc(#1)}}
 
\def\BOZZA{
\def\alato(##1){\rlap{\kern-\hsize\kern-1.2truecm{$\scriptstyle##1$}}}
\def\palato(##1){\rlap{\kern-1.2truecm{$\scriptstyle##1$}}}
}
 
\def\alato(#1){}
\def\galato(#1){}
\def\palato(#1){}

{\count255=\time\divide\count255 by 60 \xdef\hourmin{\number\count255}
        \multiply\count255 by-60\advance\count255 by\time
   \xdef\hourmin{\hourmin:\ifnum\count255<10 0\fi\the\count255}}
 
\def\oramin{\hourmin }
 
\def\data{\number\day/\ifcase\month\or gennaio \or febbraio \or marzo \or
aprile \or maggio \or giugno \or luglio \or gosto \or settembre
\or ottobre \or novembre \or dicembre \fi/\number\year;\ \oramin}
\setbox200\hbox{$\scriptscriptstyle \data $}
\footline={\rlap{\hbox{\copy200}}\tenrm\hss \number\pageno\hss}

\let\a=\alpha \let\b=\beta  \let\g=\gamma     \let\d=\delta  \let\e=\varepsilon
\let\z=\zeta  \let\h=\eta    \let\k=\kappa   \let\l=\lambda
    \let\n=\nu    \let\x=\xi        \let\p=\pi      
\let\s=\sigma \let\t=\tau        \let\c=\chi
   \let\o=\omega 
 \let\D=\Delta     \let\L=\Lambda  
           
\let\O=\Omega 
 
\def\\{\hfill\break} \let\==\equiv

\let\io=\infty 

\let\0=\noindent \def\pagina{{\vfill\eject}}

\def\ie{\hbox{\it i.e.\ }}
\let\dpr=\partial 
\let\bs=\backslash
 
\def\tende#1{\,\vtop{\ialign{##\crcr\rightarrowfill\crcr
 \noalign{\kern-1pt\nointerlineskip}
 \hskip3.pt${\scriptstyle #1}$\hskip3.pt\crcr}}\,}
\def\otto{\,{\kern-1.truept\leftarrow\kern-5.truept\to\kern-1.truept}\,}
\def\fra#1#2{{#1\over#2}}
 
\def\PP{{\cal P}}\def\EE{{\cal E}}\def\VV{{\cal V}}
\def\FF{{\cal F}}\def\HH{{\cal H}}
\def\TT{{\cal T}}\def\NN{{\cal N}}\def\BB{{\cal B}}
\def\RR{{\cal R}}\def\LL{{\cal L}}
\def\DD{{\cal D}}\def\SS{{\cal S}}
 
\def\T#1{{#1_{\kern-3pt\lower7pt\hbox{$\widetilde{}$}}\kern3pt}}
\def\VVV#1{{\underline #1}_{\kern-3pt
\lower7pt\hbox{$\widetilde{}$}}\kern3pt\,}
\def\W#1{#1_{\kern-3pt\lower7.5pt\hbox{$\widetilde{}$}}\kern2pt\,}

\def\indica{\leaders \hbox to 0.5cm{\hss.\hss}\hfill}
\def\guida{\leaders\hbox to 1em{\hss.\hss}\hfill}
\mathchardef\oo= "0521

\def\pp{{\bf p}}\def\xx{{\bf x}}
\def\yy{{\bf y}}\def\kk{{\bf k}}
\def\dd{{\bf d}}\def\zz{{\bf z}}\def\uu{{\bf u}}
 \def\bP{{\bf P}}
\def\tt{{\bf t}}
\def\ss{{\underline \sigma}}\def\oo{{\underline \omega}}

\def\qed{\raise1pt\hbox{\vrule height5pt width5pt depth0pt}}

 \def\bh{{\bar h}}  

\def\indic{\hbox{\raise-2pt \hbox{\indbf 1}}}
\def\bk#1#2{\bar\kk_{#1#2}}

\def\RRR{\hbox{\msytw R}}

%
%
%
\def\ins#1#2#3{\vbox to0pt{\kern-#2 \hbox{\kern#1 #3}\vss}\nointerlineskip}
%
%
%
\newdimen\xshift \newdimen\xwidth \newdimen\yshift
 
\def\insertplot#1#2#3#4#5{\par%
\xwidth=#1 \xshift=\hsize \advance\xshift by-\xwidth \divide\xshift by 2%
\yshift=#2 \divide\yshift by 2%
\line{\hskip\xshift \vbox to #2{\vfil%
#3 \includegraphics{#4.ps}}\hfill \raise\yshift\hbox{#5}}}
 
\def\initfig#1{%
\catcode`\%=12\catcode`\{=12\catcode`\}=12
\catcode`\<=1\catcode`\>=2
\openout13=#1.ps}
 
\def\endfig{%
\closeout13
\catcode`\%=14\catcode`\{=1
\catcode`\}=2\catcode`\<=12\catcode`\>=12}
 
 
\initfig{fig51}
\write13<
\write13<
\write13<gsave .5 setlinewidth 40 20 260 {dup 0 moveto 140 lineto} for stroke
grestore>
\write13</punto { gsave  
\write13<2 0 360 newpath arc fill stroke grestore} def>
\write13<40 75 punto>
\write13<60 75 punto>
\write13<80 75 punto>
\write13<100 75 punto 120 68 punto 140 61 punto 160 54 punto 180 47 punto 200
40 punto>
\write13<220 33 punto 240 26 punto 260 19 punto>
\write13<120 82.5 punto>
\write13<140 90 punto>
\write13<160 80 punto>
\write13<160 100 punto>
\write13<180 110 punto>
\write13<180 70 punto>
\write13<200 60 punto>
\write13<200 120 punto>
\write13<220 110 punto>
\write13<220 50 punto>
\write13<240 100 punto>
\write13<240 60 punto>
\write13<120 50 punto>
\write13<260 20 punto>
\write13<240 40 punto>
\write13<240 50 punto>
\write13<260 70 punto>
\write13<200 80 punto>
\write13<260 90 punto>
\write13<260 110 punto>
\write13<220 130 punto>
\write13<40 75 moveto 100 75 lineto 140 90 lineto 200 120 lineto 220 130
lineto>
\write13<200 120 moveto 240 100 lineto 260 110 lineto>
\write13<240 100 moveto 260 90 lineto>
\write13<140 90 moveto 180 70 lineto 200 80 lineto>
\write13<180 70 moveto 220 50 lineto 260 70 lineto>
\write13<220 50 moveto 240 40 lineto>
\write13<220 50 moveto 240 50 lineto>
\write13<100 75 moveto 260 20 lineto>
\write13<100 75 moveto 120 50 lineto stroke>
\write13<grestore>
\endfig

\openin14=\jobname.aux \ifeof14 \relax \else
\input \jobname.aux \closein14 \fi
\openout15=\jobname.aux
{\baselineskip=12pt
\centerline{\titolone Non-universality in Ising models}
\centerline{\titolone with four spin interaction.}
\vskip1.truecm
\centerline{{\titolo V. Mastropietro}}%
\centerline{Dipartimento di Matematica,}
\centerline{Universit\`a di Roma ``Tor Vergata''}
\centerline{Via della Ricerca Scientifica, I-00133, Roma}
\vskip1.truecm
\line{\vtop{
\line{\hskip1.5truecm\vbox{\advance \hsize by -3.1 truecm
\vskip.5cm
\0{\cs Abstract.}
{\it 
We consider two bidimensional
classical Ising models,
coupled by a weak interaction bilinear in the energy
densities of the two systems; the model contains,
as limiting cases, the Ashkin-Teller and the Eight-vertex
models for certain values of their parameters. 
We write the energy-energy correlations and the specific
heat as Grassman integrals
formally describing Dirac 
1+1 dimensional interacting massive fermions
on a lattice, and an expansion based on Renormalization Group
is written for them,
convergent up to temperatures
very close to the critical temperature for small
coupling.
The asymptotic behaviour is determined
by critical indices which
are continuous functions of the coupling.
}}
\hfill} }}
}
\line{\vtop{
\line{\hskip1.5truecm\vbox{\advance \hsize by -3.1 truecm

\vskip.5cm
\0{\cs Key words}
{\it Coupled Ising models, non-universality, 
Renormalization Group, 
fermions, critical indices.}
\hfill} }}
}

\pagina
\vskip1.2truecm
\section(1,Introduction and main results)
\sub(1.1) {\it Motivations.} 
It is well known that the
partition function and many correlations of the $d=2$
classical {\it Ising model} can be written
as Grassman integrals formally describing
a system of $d=1+1$ {\it free} fermions, 
see [H],[SML],[K],[MW],[S],[ID].
This mapping of the Ising model in terms of a fermionic theory
is quite useful; 
as the corresponding 
action is quadratic in the Grassman variables,
it can be diagonalized by a Bogoliubov
transformation so that
the partition function and many observables can be 
exactly computed. 
However, even a slight perturbation of the Ising model
makes the corresponding fermionic action {\it non-quadratic}.
The relationship between
spin Ising-like models and interacting fermionic
models is deeply investigated in the literature;
it is for instance claimed that
models of  
coupled Ising models with an interaction quartic in the spins,
like the {\it Eight-vertex} model,
are in the same {\it class of universality}
of models of {\it interacting} $d=1+1$
fermions in the continuum, like the {\it massive Thirring}
model or the {\it massive Luttinger} model, see
for instance [N], [LP], [PB]. This relationship
is however found under many approximations.
In this paper we consider 
two Ising models coupled by a general interaction
bilinear in the spin densities; the model contains
as particular cases the Ashkin-Teller and Eight-vertex model
for certain values of their parameters. 
We clarify the relation between
this general spin model
and systems of interacting fermions by writing
the partition function and some correlations
as Grassman integrals with a formal action
resembling but not coinciding 
with the action of the massive Thirring or Luttinger model;
the main difference (in addition to the fact that 
the formal action describes fermions in a lattice,
not in the continuum) is that
it does not verify
some special symmetries 
(like invariance under local Gauge transformations)
which are generally verified 
in models of interacting fermions
and which play an important role in
their analysis. The main interest in this representation
in terms of Grassman integrals is that we can apply the
Renormalization Group techniques developed for 
interacting fermions
(see for instance [BG] or [GM])
for writing a convergent perturbative expansion
for the partition function and some correlations
of the spin model we are considering.
A naive power series expansion 
in the coupling
is not expected to be 
convergent
close to the critical temperature (the free and interacting model
are not analytically close)
and one needs a {\it resummed}
or {\it renormalized} expansion
which is provided by Renormalization Group.
\*
\sub(1.2) {\it Spin systems with quartic interactions.}
We consider two Ising models 
coupled via a four spin interaction bilinear in the 
energy densities of the two sublattices. 
Given $\L_M\in Z^2$ a square lattice with
side $M$ and periodic boundary condition, 
we call $\xx=(x,x_0)$ a site of $\L_M$. 
If $\s^{(1)}_\xx=\pm 1$ and
$\s^{(2)}_\xx=\pm 1$, we write the following Hamiltonian
$$H_\L(\s^{(1)},\s^{(2)})=
H_I(\s^{(1)})+H_I(\s^{(2)})+V(\s^{(1)},\s^{(2)})\Eq(ff)$$
where, if $\a=1,2$ 
$$H_I(\s^{(\a)})=-
\sum_{x,x_0=1}^M [J_1^{(\a)}\s^{(\a)}_{x,x_0}\s^{(\a)}_{x+1,x_0}+
J_2^{(\a)}\s^{(\a)}_{x,x_0}\s^{(\a)}_{x,x_0+1}]\Eq(I)$$
$$V(\s^{(1)},\s^{(2)})=-\sum_{x,x_0=1}^M\{\l a [
\s^{(1)}_{x,x_0}\s^{(1)}_{x+1,x_0}
\s^{(2)}_{x,x_0}\s^{(2)}_{x+1,x_0}+
\s^{(1)}_{x,x_0}\s^{(1)}_{x,x_0+1}
\s^{(2)}_{x,x_0}\s^{(2)}_{x,x_0+1}]$$
$$+\l b[\s^{(1)}_{x,x_0}\s^{(1)}_{x+1,x_0}
\s^{(2)}_{x,x_0}\s^{(2)}_{x,x_0+1}+
\s^{(1)}_{x,x_0}\s^{(1)}_{x,x_0+1}
\s^{(2)}_{x-1,x_0+1}\s^{(2)}_{x,x_0+1}]
\Eq(int)$$
$$+\l c\sum_{\a=1,2}[\s^{(\a)}_{x,x_0}\s^{(\a)}_{x+1,x_0}
\s^{(\a)}_{x,x_0}\s^{(\a)}_{x,x_0+1}+
\s^{(\a)}_{x,x_0}\s^{(\a)}_{x+1,x_0}
\s^{(\a)}_{x+1,x_0-1}\s^{(\a)}_{x+1,x_0}]\}$$

The above Hamiltonian can be a model
for a number of physical
problems; for instance two magnetic spin planes
coupled by an interplanar interaction 
quartic in the spins (in this interpretation
$\s^{(1)}$ and $\s^{(2)}$ are the spin of the 
first or second plane). 
Moreover it 
contains, as limiting cases, two well known models,
the {\it Ashkin-Teller} and the {\it Eight-vertex} model. 
\*
\sub(1.3a) {\it The Ashkin-Teller model.}
Ashkin and Teller introduced their model 
as a generalization of the Ising model to a four state system.
Each site $\xx$ is occupied by one of the four kinds of atom,
said A,B,C,D, and two neighboring atoms interact with an energy $\e_0$
for $AA$, $BB$, $CC$, $DD$; $\e_1$ for $AB,CD$,
$\e_2$ for $AC,BD$, $\e_3$ for $AD, BC$. 
It is quite easy to express this model in terms of Ising spins, see [F].
One associates to each site of the 
lattice two spins $\s^{(1)}_\xx$ and $\s^{(2)}_\xx$, see Fig.1; 
then
$\s^{(1)}_\xx,\s^{(2)}_\xx=(+,+)$ if there is an $A$ 
atom at site $\xx$, $(+,-)$ if there is an atom $B$, $(-,+)$
if there is an atom $C$ and $(-,-)$ if there is an atom $D$.
Then the hamiltonian is given by
$$H_\L(\s^{(1)},\s^{(2)})=H_I(\s^{(1)})+H_I(\s^{(2)})$$
$$-J''\sum_{x,x_0=1}^M[\s^{(2)}_{x,x_0}\s^{(2)}_{x,x_0+1}
\s^{(1)}_{x,x_0}\s^{(1)}_{x,x_0+1}+
\s^{(2)}_{x,x_0}\s^{(2)}_{x+1,x_0}
\s^{(1)}_{x,x_0}\s^{(1)}_{x+1,x_0}]-J_0\Eq(3a)$$
where $H_I(\s^{(\a)})$ is given by \equ(I)
with $J_1^{(\a)}=J_2^{(\a)}=J^{(\a)}$ and
$$-J^{(1)}=(\e_0+\e_1-\e_2-\e_3)/4\quad-J^{(2)}=(\e_0+\e_2-\e_3-\e_1)/4\Eq(cc1)$$
$$-J''=(\e_0+\e_3-\e_1-\e_2)/4\quad -J_0=(\e_0+\e_1+\e_2+\e_3)/4$$
The model is not soluble, except in some particular case,
and we refer to [Ba] chapt.12 and references therein
for a review of the main results about it; it is of the
form \equ(ff) with $b=c=0$.
\*
\sub(1.3b) {\it The Eight-vertex model.}
In such model one associates to each site of a square lattice
a vertex with four arrows with different orientations, see [Ba]
chapt. 10.
Among the sixteen possibilities, only eight
kind of vertices are allowed, and an energy
$\e_i$, $i=1,..,8$, is associated to each of them.
It was solved in [Ba] if $\e_1=\e_2$, $\e_3=\e_4$,
$\e_5=\e_6$, $\e_7=\e_8$
and
for suitable values of the $\e_i$ it reduces to
the {\it ice-model}, solved in [Li]. Such models
are called
{\it vertex models} and 
were introduced to describe crystal with hydrogen bonding.
As explained in [Ba]
or [W] they can be written
in terms of the following Hamiltonian, see Fig. 2
$$H_\L(\s^{(1)},\s^{(2)})=-\sum_{x,x_0}
[J\s_{x,x_0+1}\s_{x+1,x_0}+J'\s_{x,x_0}\s_{x+1,x_0+1}$$
$$+J''\s_{x,x_0}\s_{x+1,x_0+1}\s_{x,x_0+1}\s_{x+1,x_0}]\Eq(1)$$
with
$$\e_1=\e_2=-J-J'-J''\quad\e_3=\e_4=J+J'-J''\Eq(cc2)$$
$$\e_5=\e_6=J'-J+J''\quad\e_7=\e_8=-J'+J+J''$$
If $J''=0$ the above hamiltonian
decouples in the hamiltonian of two Ising models,
involving spins located in two sublattices, one with $x+x_0$
equal to an even number and the other to an odd number.
Each site of the sublattices is in the center of the
unit square of the other one.
Calling $\s^{(1)}$ and $\s^{(2)}$ the spin located
in the first or the second sublattice, relabeling the spins
and performing a rotation of ${\pi\over 4}$
the hamiltonian can be written as, see Fig.3
$$H_\L(\s^{(1)},\s^{(2)})=H_I(\s^{(1)})+H_I(\s^{(2)})$$
$$-J''\sum_{x,x_0=1}^M[\s^{(2)}_{x,x_0}\s^{(2)}_{x,x_0+1}
\s^{(1)}_{x,x_0}\s^{(1)}_{x+1,x_0}+
\s^{(1)}_{x,x_0}\s^{(1)}_{x,x_0+1}
\s^{(2)}_{x-1,x_0+1}\s^{(2)}_{x,x_0+1}]\Eq(3)$$
where
$$H_I(\s^{(\a)})=-
\sum_{x,x_0=1}^MJ[\s^{(\a)}_{x,x_0}\s^{(\a)}_{x+1,x_0}+
\s^{(\a)}_{x,x_0}\s^{(\a)}_{x,x_0+1}]$$
which is again of the form \equ(ff) with $a=c=0$ (of course this identification
is exact if boundary conditions in the Eight-vertex model
are chosen properly).
\*
\sub(1.4) {\it Main results.} 
We will consider in this paper the case
$$J^{(2)}_1=J^{(1)}_2=J^{(2)}_1=J^{(2)}_2=J\Eq(malw)$$
If $\l=0$ the problem reduces to the Ising model
whose critical temperature is 
given by
$\tanh\b_c J_0=\sqrt{2}-1\equiv t_c$
where $J\equiv\b J_0$ and $\b^{-1}$ is the temperature.
The interaction changes in general the value
of the critical temperature
by terms $O(\l)$, and 
it is instead technically convenient 
to fix its value still in correspondence of 
$t_c=\sqrt{2}-1$, by choosing properly
the molecular energy parameter
$J$ as a function of $\l$; in this way the critical
temperature of the system with $\l=0$ and $\l\not=0$
is the same.
We consider then
the model \equ(ff) with
$J_{r}$ replacing $J$,
and we will choose $J_r=J+O(\l)$ so that the critical temperature
is in correspondence $t_c=\sqrt{2}-1$.

We define
$$<O(\xx)O(\yy)>_\L={1\over Z_\L}
\sum_{\s^{(1)}_\xx,\s^{(2)}_\xx=\pm 1\atop\xx\in\L_M} O(\xx) O(\yy)
e^{- H_\L(\s^{(1)},
\s^{(2)})}\Eq(corr)$$
where
$Z_\L= \sum_{\s^{(1)}_\xx,\s^{(2)}_\xx=\pm 1\atop\xx\in\L_M}
e^{-H_\L(\s^{(1)},
\s^{(2)})}$
is the {\it partition function}, 

$f^\l(J)=\lim_{|\L|\to\io}
{1\over|\L|}\log Z_\L$ is the {\it free energy},
${\partial^2\over\partial J^2} f^\l(J)$ is the
specific heat, 
and the correlation function of the observable $O(\xx)$ is
$$<O(\xx)O(\yy)>_{\L,T}=<O(\xx)O(\yy)>_\L-<O(\xx)>_\L<O(\yy)>_\L.\Eq(xzaqr)$$
In the Ising model close to the critical temperature
${\partial^2\over\partial J^2} f^0(J)\simeq C\log|t-t_c|^{-1}$
and for large distances $|<O(\xx)O(\yy)>_{\L,T}|\le 
{Ce^{-A|t-t_c||\xx-\yy|}\over |\xx-\yy|^{2}}$, with $A,C$ 
suitable constants.

We will prove the following theorem.
\*
{\bf Theorem 1.} {\it Consider the hamiltonian \equ(ff), 
with $J_r$ replacing $J$, 
and assume that $|a+b|>0$
and \equ(malw). 
There exist $\e>0, a_3>0, \bar a>0$ constants
and a function $\nu(\l)=\tanh J_r-\tanh J$ 
such that, 
for $|\l|\le\e$
and $\bar a\ge |t-t_c|\ge e^{-{1\over a_3\l^2}}$,
the energy-energy correlation 
verifies
$$\lim_{|\L|\to\io}<\s^{(\a)}_{\xx}\s_{\xx'}^{(\a)}
\s^{(\a)}_{\yy}\s_{\yy'}^{(\a)}>_{\L,T}=
\O^{(\a),a}(\xx,\yy)+\O^{(\a),b}(\xx,\yy)+
\O^{(\a),c}(\xx,\yy)\Eq(fonfo)$$
where $\xx,\xx'$ and $\yy,\yy'$ 
are nearest neighbor pairs respectively and,
for any integer positive $N$  
$$|\O^{(\a)a}(\xx,\yy)|\le {1\over |\xx-\yy|^{2+\h_1}}{C_N\over 1+(\D|\xx-\yy|)^N}$$
$$|\O^{(\a),b}(\xx,\yy)|\le {1\over |\xx-\yy|^{2+\h_3}}
{C_N\over 1+(\D|\xx-\yy|)^N}\Eq(jkazz)$$
$$|\O^{(\a),c}(\xx,\yy)|\le {1\over |\xx-\yy|^{2+\t}}{C_N\over 1+(\D|\xx-\yy|)^N}$$
where $C_N,\t$ are positive constants,  
$\D=|t-t_c|^{1+\h_2}$, and $\nu(\l)=O(\l)$,
$$\h_1(\l)=-a_1\l+O(\l^2)\quad
\h_2(\l)= a_2\l+O(\l^2)\quad
\h_3(\l)= a_1\l+O(\l^2)\Eq(camp)$$
with $a_1>0,a_2>0$ constants.
If $1\le|\xx|\le\D^{-1}$
$$\O^{(\a)a}(\xx,\yy)={1+f^{(\a),a}_\l(\xx,\yy)\over 
\tilde Z_0^2(x_0^2+x^2)^{1+\h_1}}\quad
\O^{(\a)b}(\xx,\yy)=
{1+f^{(\a),b}_\l(\xx,\yy)\over \tilde Z_0^2(x_0^2+x^2)^{1+\h_3}}\Eq(bvqii)$$
and $|\O^{(\a),c}(\xx,\yy)|\le {C\over |\xx-\yy|^{2+\t}}$,
where $\tilde Z_0>0$ is a constant and $f_\l(\xx,\yy)$ are smooth 
bounded $O(\l)$ functions.
Finally the specific heat verifies
$$C_1|{1\over\h_1}[1-|\D|^{2\h_1}]+
{1\over\h_3}[1-|\D|^{2\h_3}]|\le 
|{\partial^2\over\partial J^2} f^\l(J)|\le 
C_2|{1\over\h_1}[1-|\D|^{2\h_1}]+
{1\over\h_3}[1-|\D|^{2\h_3}]|
\Eq(cv)$$
where $C_1,C_2$ are positive constants.} 
\vskip.5cm
\sub(1.1xcz) {\it Remarks.} The above theorem 
describes the behaviour of 
the energy-energy correlation and the specific heat
near the critical temperature 
of Ising models in two dimensions
weakly coupled by a four spin interaction.  
One can distinguish two
different regimes in the asymptotic behaviour of the energy-
energy correlation function, discriminated by
an intrinsic correlation length 
$\xi$ of order $|t-t_c|^{-1-\h_2}$ with $\h_2=O(\l)$.
If $1<<|\xx-\yy|<<\xi$, the bounds 
for the correlation function is power-like (with a $\l$
dependent exponent)
while if $\xi<<|\xx|$, there is a faster than
any power decay with rate of order $\xi^{-1}$. In the first region we can
obtain the exact large distance asymptotic behaviour of 
the energy-energy correlation function, see (1.15), while
in the second region only an upper bound is
obtained. The logarithmic behaviour of the specific
heat in the Ising model getting closer and closer
to the critical temperature 
is changed by the four spin interaction
in a power law, at least up to temperatures
very close to $t_c$ for small
four spin coupling. 
A bound like
$C_1\log(|t-t_c|^{-1})\le |{\partial^2\over\partial J^2} f^\l(J)| \le 
C_2\log(|t-t_c|^{-1})$, for some positive $\l$-independent
constants $C_1$ and $C_2$, which would be true
if the model would be in the same {\it universality class} 
as the Ising model, cannot be true by 
\equ(cv) at least for $|t-t_c|\ge e^{-{1\over a_3\l^2}}$.

In the particular case $a=c=0$ our model reduces to the 
Eight-vertex model,
which is exactly soluble [Ba]; our results agree 
with the informations obtained by the exact solution,
in which non universal critical behaviour is found and it is
believed that the specific heat diverges with a power law [W]. 
In the case $a=b=0$ our model reduces to a model
of two {\it non interacting}
Ising models
with nearest neighbor 
and four spin interaction within each copy,
which was studied in 
[PS] (which is indeed the first paper in which fermionic RG methods were 
applied
to classical Ising-like models, and it is a major
source of inspiration for the present work). 
In this case $C_1\log(|t-t_c|^{-1})\le |{\partial^2\over\partial J^2} f^\l(J)|\le 
C_2\log(|t-t_c|^{-1})$ (up to $t=t_c$)
so that {\it universality} indeed holds
\*
\sub(1.1xczz) {\it Sketch of the proof.}
The proof of the theorem is based on  
Grassmann variables combined with renormalization group techniques.
In \S 2 we briefly recall the well known representation
of the bidimensional Ising model in terms of Grassmann
variables, mainly due to [K], [H], [MW], [S] and recently
rederived in a more cohesive way in [PS]. The partition
function and the correlation can be written in terms of 
Grassmann integrals with a quadratic action which can be
explicitly computed in terms of Pfaffians. The exact
solvability of the Ising model is hence related to the
fact that it can be expressed in terms of {\it free} fermions,
as it was first noted in [SML]. In the Grassmann representation
there are four independent Grassmann variables associated to each point
of the lattice, two with a large $O(1)$ mass
and the other two with a continuously vanishing mass at the 
critical temperature. In \S 2 we show that
two Ising models
weakly coupled by four spin interactions, with hamiltonian
(1.1), can be also written in terms of Grassmannn integrals,
with the difference that there are now {\it eight}
independent Grassmannn variables (four for each Ising model)
and that the formal action contains now terms which are 
{\it quartic} in the fields, corresponding to
a short range interaction among fermions. By performing
a suitable linear transformation in the Grassmann variables
one obtains that in the new variables
the quadratic part of the action strongly
resembles the action of two massive Dirac fermions in $d=1+1$
dimensions
(it would coincide with that if the continuum limit 
would be taken); again one Dirac fermion 
has a large $O(1)$ mass and the other is vanishing at $t_c$, 
and we call them heavy and light fermions.
We are essentially
exploiting in this transformation the relation ${\rm Pf}^2 A=\det A$.

In  \S 3 we integrate out the heavy fields 
in the Grassmann integral for the partition
function, so obtaining 
a Grassmann integral whose 
formal action contains monomials of every degree in
the light Dirac fermions.
In order to do this we use the representation in [Le] of fermionic
truncated expectations and Gram-Hadamard inequality.

In \S 4 we apply renormalization group methods 
to integrate the light fermions,
which in a sense are the critical modes.
We will use a suitable modification
of the multiscale 
expansion used in [BM] (see also [BG] or [GM]
for a general introduction to the formalism) 
to study the correlation
functions of the {\it Heisemberg-Ising} $XYZ$ chain;
the close relationship
between Ising models with quartic coupling and the $XYZ$ times
has been pointed out many times in the literature, see
for instance [Ba]. A power counting analysis says 
that the terms {\it bilinear} in the Grassmann variables
are {\it relevant} in a RG sense, while the quartic
terms, or the bilinear with an extra derivative are {\it marginal}.
One can understand here
why universality is still present if one considers
decoupled Ising models
with nearest neighbor 
and four spin interaction within each copy, and why on the 
contrary it is lost if 
the Ising models are coupled.
In the first case (which is the one treated
in [SP]) one can easily check that the local
part of the quartic terms is vanishing, so the effective
interaction is indeed {\it irrelevant} in the RG sense.
On the other hand in the second case the quartic interactions
is truly {\it marginal} and this produces
a line of fixed points for the RG transformation
(instead of the gaussian fixed point
as in the previous case) continuously
depending on the coupling $\l$.
We decompose the Grassmann integration $P(d\psi)$
as a product
of independent Grassmann integrations $P(d\psi^{(h)})$, with covariance
with non vanishing support only for momenta with modulus
between $\g^{h-1}$ and $\g^{h+1}$, with $\g>1$
and $h=0,-1,-2,-3,...$. We integrate each $P(d\psi^{(h)})$
iteratively starting from $P(d\psi^{(1)})$
obtaining a sequence of effective
potentials $\VV^{(h)}$
describing the theory at momentum scale $\g^h$;
at each step new contributions to the mass and the wave
function renormalization are obtained which are included
in the fermionic integration; hence $P(d\psi^{(h)})$
has a covariance with mass $m_h$ and wave function renormalization
$Z_h$ with a non trivial dependence on $h$, \ie 
$m_h\simeq |t-t_c|\g^{\h_2 h}$ and
$Z_h\simeq\g^{\h_4 h}$, with $\h_4=O(\l^2)$. 
The iteration stops as soon as the mass $m_h$
becomes of order $\g^h$; we will call $h^*$ the last scale
to be integrated (of course $h^*\to-\io$ 
at the critical temperature 
$t=t_c$). The iterative procedure allows to write 
the effective potential $\VV^{(h)}$ as sum of monomials
in the Grassmann variables, with coefficients which are
(convergent) perturbative expansions in terms of a few
{\it running coupling constant}, $\l_h$, the effective
coupling of the interaction between fermions, $\nu_h$,
which takes into account the renormalization of the value of
the critical
temperature, $\d_h$ related to the renormalization of the fermionic velocity
and other couplings $\z_{i,\o,h}, \tilde \z_{i,\o,h}$ which
are the coefficients of quadratic terms in the Grassmann fields with 
a derivated field. Despite there are many similarities
between our Grassmann integrals and the one describing
relativistic fermions, there is a crucial difference;
the interaction term in the action
is {\it not} invariant under gauge transformation,
hence terms which were absent in the free action can be generated
in the RG iterations. One can check that indeed peculiar symmetries
of the model (1.2) ensure that there is only one relevant term
quadratic in the fermions (if there were more one gets in troubles,
as there is only one free parameter in the hamiltonian).
On the other hand we cannot exclude the generation
of marginal quadratic terms which were absent in the free action;
they are the terms $\z_{i,\o,h}, \tilde \z_{i,\o,h}$.
At the end the result of this iterative integration is an 
expansion
for the partition function in terms of the running coupling
constants which is proved in \S 4 to be convergent {\it provided that}
the running coupling constants are small for any $h$.

In \S 5 we prove indeed that it is possible to choose
the counterterm $\nu$ as a function of $\l$ 
so that the running coupling constants are
indeed small; the condition on the temperature 
$|t-t_c|\ge e^{-{1\over a_3\l^2}}$ is used 
to control the flow of the running coupling constants, and
some cancellations 
at the lower order in our expansion are also used
to be as close as possible 
to the critical temperature.

Finally in \S 6 we define an expansion for the correlation
functions and the specific heat; 
it is similar to the one for the partition function, 
with the main difference that
one has to introduce new  
fields associated to the external fields.
There are additional
marginal terms in RG expansion to which
other {\it renormalization constants}, with a non trivial behaviour in
$h$, are associated, \ie
$Z_h^{(1)}\simeq \g^{\h_1 h}$ and $Z_h^{(2)}\simeq \g^{\h_3 h}$.
By such expansion the statements in the theorem are derived.

Among interesting open problems
there is the reaching of
the critical temperature; surely 
one has to exploit 
suitable cancellations 
at every order
of the expansion for $\l_h,\d_h$, as in the theory of $d=1$
interacting Fermi systems, see for instance
[BGPS] and [BM], but at the moment the main
difficulty is in the flow of the couplings $\z_{i,\o,h}$
and $\tilde \z_{i,\o,h}$. Another interesting problem
would be the analysis of the spin-spin correlation function
whose
expression in terms of Grassmann variables is unfortunately quite
complicated and not easy to manage. A similar problem
appears in considering two Ising models
coupled by a weak interaction {\it bilinear} in the spins; in such
a case the interaction in terms of Grassmann variables has
a quite complex expression and it seems difficult to study. 
On the other hand, for {\it large} coupling,
such bilinear interaction should be {\it irrelevant}
and universality should hold.
Finally it should be interesting to study the case
of two coupled Ising models {\it at different temperatures}
or the case of {\it four} coupled Ising models; 
in this last case interacting {\it spinning} $d=1$
fermions appear in the fermionic description, 
which are known to have a behaviour
quite different from the spinless one (like in the
$d=1$ {\it Hubbard} model).

\vskip.5cm
\section(2,Fermionic representation)
\vskip.5cm
\sub(1.1) {\it Grassmann integrals.}
If $\l=0$ the hamiltonian \equ(ff) is given by the sum of two 
independent Ising model hamiltonians, and the partition function
is given by $Z_I^{(1)} Z_I^{(2)}$ where
$$Z_{I}^{(\a)}=\sum_{\s_\xx^{(\a)}=
\pm 1\atop\xx\in\L_M } e^{-H_I(\s^{(\a)})}.\Eq(c1)$$
It is well known that such partition function can be written 
in terms of Grassmann integrals, so we recall first their 
definition. Grassmann variables $\h_\a$, $\a=1,2,..,2n$,
,$n$ even, are anticommuting variables satisfying
$$\{\h_\a,\h_{\a'}\}=0\qquad\h_\a^2=0\Eq(x1)$$
The Grassmann integration $\int d\h_\a$
is a linear operation defined as 
$$\int d\h_\a=0\qquad\int d\h_\a \h_\a=1\Eq(x2)$$
It holds that
$$\int \prod_\a d\h_\a 
e^{{1\over 2}\sum_{\a,\b}\h_\a A_{\a,\b}\h_\b}={\rm Pf}A\Eq(c2)$$
where $A$ is an even antisymmetric $2n$-matrix
and ${\rm Pf}A$ denotes the {\it Pfaffian}. 
It holds
$$\int d\h_1...d\h_{2n}\exp
{1\over 2}\sum_{\a,\b}\h_\a A_{\a,\b}\h_\b=\Eq(c5)$$
$$\int d\h_{2n}...d\h_1\prod_{\a<\b}(1+A_{\a,\b}\h_\a\h_\b)=
{1\over 2^n n!}\sum_p (-1)^p
A_{p_1,p_2}A_{p_3,p_4}...A_{p_{2n-1},p_{2n}}\equiv {\rm Pf}A$$
where the sum is over all the permutations.
We can consider another set of Grassmann variables 
$\h^+_\a$, $\a=1,..,2n$, and 
$$\int \prod_\a d\h_\a \prod_\a d\h^+_\a 
\exp[\sum_{\a,\b}\h_\a B_{\a,\b}\h^+_\b]={\rm Det}B\Eq(c4)$$
The well known relation $({\rm Pf}A)^2={\rm det} A$ 
can be quite 
easily deduced by the above Grassmann integrals;
it can be written as 
$$\int \prod_\a d\h_\a d\h^+_{\a}e^{\h_\a A_{\a,\b}\h^+_\b}=
\int \prod_\a d\h_\a^{(1)}e^{{1\over 2}\h^{(1)}_\a A_{\a,\b}\h^{(1)}_\b}
\int \prod_\a d\h_\a^{(2)}e^{{1\over 2}\h^{(2)}_\a A_{\a,\b}\h^{(2)}_\b}
\Eq(fon)$$
which can be proved by
the change of variables
$$\h^+_\a={1\over\sqrt{2}}(\h^{(1)}_\a+i\h^{(2)}_\a)\quad
\h_\a={1\over\sqrt{2}}(\h^{(1)}_\a-i\h^{(2)}_\a)\Eq(c7)$$
in
$\int d\h_\a d\h^+_{\a}e^{\h_\a A_{\a,\b}\h^+_\b}$.
Then $d\h_\a d\h^+_\a=i d\h^{(1)}_\a d\h^{(2)}_\a$
and
$$\h_\a A_{\a,\b}\h^+_\b={1\over 2}\h^{(1)}_\a A_{\a,\b}
\h^{(1)}_\b+{1\over 2}\h^{(2)}_\a A_{\a,\b}
\h^{(2)}_\b\Eq(c9)$$
as 
$$\h^{(1)}_\a A_{\a,\b}\h^{(2)}_\b-
\h^{(2)}_\a A_{\a,\b}\h^{(1)}_\b=
\h^{(1)}_\a A_{\a,\b}\h^{(2)}_\b-
\h^{(1)}_\b A_{\b,\a}\h^{(2)}_\a=0
\Eq(c10)$$
\*
\sub(1.2) {\it The fermionic representation of the Ising model.}
It is well known that the partition function can be written
as a Grassmann integral. It is proved in [K],[H],[MW],[S] that
$$Z_I^{(\a)}=(\cosh J_r)^B 2^S {1\over
2}\int \prod_{\xx\in\L_M} dH^{(\a)}_\xx 
d\bar H^{(\a)}_\xx d V^{(\a)}_\xx d\bar V^{(\a)}_\xx
[-e^{S_{+,+}}+e^{S_{+,-}}+e^{S_{-,+}}+e^{S_{-,-}}]\Eq(9a)$$
where $B$ is the total number of bonds and
$S$ is the total number of sites, 
$$S_{\e,\e'}^{(\a)}=\sum_{\xx\in\L_M} \tanh J_r [\bar H^{(\a)}_{x,x_0} H^{(\a)}_{x+1,x_0}+
\bar V^{(\a)}_{x,x_0} V^{(\a)}_{x,x_0+1}]+$$
$$\sum_{\xx\in\L_M}[\bar H^{(\a)}_{x,x_0} H^{(\a)}_{x,x_0}+
\bar V^{(\a)}_{x,x_0} 
V^{(\a)}_{x,x_0}+\bar V^{(\a)}_{x,x_0} \bar H^{(\a)}_{x,x_0}+ 
V^{(\a)}_{x,x_0} \bar H^{(\a)}_{x,x_0}+
H^{(\a)}_{x,x_0} \bar V^{(\a)}_{x,x_0}+
V^{(\a)}_{x,x_0} H^{(\a)}_{x,x_0}]\Eq(c20)$$
and $H^{(\a)}_\xx,\bar H^{(\a)}_\xx,V^{(\a)}_\xx,
\bar V^{(\a)}_\xx$ are {\it Grassmann variables}
such that
$$\bar H^{(\a)}_{x,x_0+M}=\e\bar H^{(\a)}_{x,x_0}
\qquad\bar H^{(\a)}_{x+M,x_0}=\e'\bar H^{(\a)}_{x,x_0}$$
$$
H^{(\a)}_{x,x_0+M}=\e H^{(\a)}_{x,x_0}\quad
H^{(\a)}_{x+M,x_0}=\e' H^{(\a)}_{x,x_0}\Eq(c31)$$
and identical relations hold for the variables $V^{(\a)},\bar V^{(\a)}$.
The Grassmann integration 

$\int \prod_x dH^{(\a)}_x d \bar H^{(\a)}_x$
is defined as 
the linear functional on the Grassmannian algebra, such
that, given a monomial $Q( H^{(\a)}, \bar H^{(\a)})$ 
in the variables $H^{(\a)}_\xx, \bar H^{(\a)}_\xx$,
$\xx\in\L_M$, its value is $0$, except in the case $Q( H^{(\a)},
\bar H^{(\a)})=
\prod_\xx H^{(\a)}_\xx \bar H^{(\a)}_\xx$, 
up to a permutation of the variables.
In this case the value of the functional is determined, by using the
anticommuting properties of the variables, by the condition
$$\int \;\left[\prod_{\xx\in\L_M}d \bar H^{(\a)}_\xx dH^{(\a)}_\xx\right]\;
\prod_{\xx\in\L_M} H^{(\a)}_\xx \bar H^{(\a)}_\xx = 1\;.\Eq(2.11)$$
In a similar way is defined the Grassmann integration for
$V^{(\a)},\bar V^{(\a)}$, just exchanging $H,\bar H$ with $V,\bar V$.
In order to prove \equ(9a)
the starting point is the high temperature
expansion
$$Z^{(\a)}_I=(\cosh J_r)^B 2^S\sum (\tanh J_r)^l\Eq(ht)$$
where the sum is over all the closed polygons
which may have points but not sides in common, see
Fig.4. 
One then replaces the Ising lattice
with another lattice, called {\it second lattice}, 
in which each site is replaced by four
surroundings sites or {\it terminals}, see Fig.5. 
If we associate $\bar V_\xx$ with $N_\xx$, $V_\xx$ with $S_\xx$,
$\bar H_\xx$ with $E_\xx$ and $H_\xx$ with $W_\xx$, see Fig.5, 
it is easy to verify, by the rules of Grassmann integrations,
that $\int e^{S_{\e,\e'}}$ can be represented
as a sum over polygon configurations over the second lattice.
Such polygon configurations are defined such that each
terminal must be coupled to another terminal, $E_{\xx}$
may be coupled to $W_{x+1,x_0}$; $N_{x_0,x}$ may
be coupled $S_{x,x_0+1}$,;
$E_\xx$ may be coupled to $N_\xx,W_\xx,S_\xx$;
$N_\xx$ may be coupled to $W_\xx$ or $S_\xx$;
$W_\xx$ may be coupled to $S_\xx$, and no other types
of couplings are allowed. An example
of polygon configuration in the second lattice is in Fig.5.
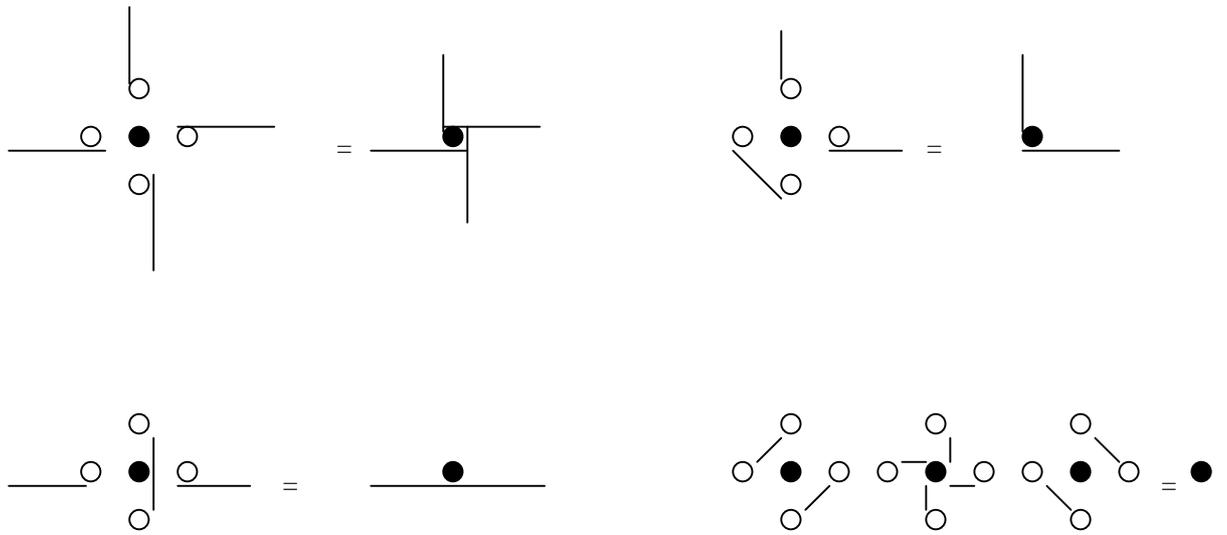
By a local identification of Ising configurations
and configurations over the second lattice (see Fig.6),
there is a correspondence between polygon configurations
on the Ising lattice and polygon configurations
on the second lattice. The correspondence
is not one to one, as the configuration with zero
lines in the Ising model corresponds to three configurations
in the second lattice. However the signs add
so that the sum over polygon configurations in the Ising lattice in \equ(ht)
is exactly equal to the sum over polygon configurations in
\equ(9a); the proof is based on a subtle combinatorial
analysis first done in [K]. In order to ensure periodic 
boundary conditions, one has to sum over four terms as
in \equ(9a), as explained in [MW].
We write $e^{S^{(\a)}_{\e,\e'}}=e^{S^{(\a),0}_{\e,\e'}}
e^{S^{(\a),\nu}_{\e,\e'}}$
where $S^{(\a),0}_{\e,\e'}$ is given by \equ(c20) with $J$
replacing $J_r$ and
$$S^{(\a),\nu}_{\e,\e'}=\nu
\sum_{\xx\in\L_M} [\bar H^{(\a)}_{x,x_0} H^{(\a)}_{x+1,x_0}+
\bar V^{(\a)}_{x,x_0} V^{(\a)}_{x,x_0+1}]\Eq(nun)$$

If $J$ is not constant but it depends 
on the bounds one obtains a similar formula
in which $S^{(\a),0}_{\e,\e'}$ is given by
$$S^{(\a),0}_{\e,\e'}=\sum_\xx \tanh J^{(\a)}_{1;x,x_0;x+1,x_0} 
\bar H^{(\a)}_{x,x_0} H^{(\a)}_{x+1,x_0}+
\tanh J^{(\a)}_{2;x,x_0;x,x_0+1}
\bar V^{(\a)}_{x,x_0} V^{(\a)}_{x,x_0+1}]+\Eq(c21)$$
$$\sum_\xx[\bar H^{(\a)}_{x,x_0} H^{(\a)}_{x,x_0}+\bar V^{(\a)}_{x,x_0} 
V^{(\a)}_{x,x_0}+\bar V^{(\a)}_{x,x_0} \bar H^{(\a)}_{x,x_0}+ 
V^{(\a)}_{x,x_0} \bar H^{(\a)}_{x,x_0}+
H^{(\a)}_{x,x_0} \bar V^{(\a)}_{x,x_0}+
V^{(\a)}_{x,x_0} H^{(\a)}_{x,x_0}]$$
and the factor $(\cosh J)^B$ is replaced by $\prod_b \cosh J_b$,
where the product is over all the possible nearest neighbor bounds.
We will call $Z^{(\a)}_I(J_{\xx,\xx'})$ the Ising model partition function
with non constant $J$.
\*
\sub(1.3) {\it The fermionic representation of coupled Ising models.}
The partition function of \equ(ff) is
$$Z_{2I}=\sum_{\s^{(1)}_\xx=\pm 1\atop\xx\in\L_M }\sum_{\s^{(2)}_\xx=\pm 1\atop\xx=\L_M} 
e^{-H_I(\s^{(1)})}e^{-H_I(\s^{(2)})}e^{-V(\s^{(1)},\s^{(2)})}\Eq(int)$$
Let us consider first the essentially equivalent 
expression
$$\hat Z_{2I}=\sum_{\s^{(1)}=\pm 1\atop\xx\in\L_M} 
\sum_{\s^{(2)}=\pm 1\atop\xx\in\L_M} 
e^{-H_I(\s^{(1)})}e^{-H_I(\s^{(2)})}\times$$
$$\prod_{\xx}
[1+\l a 
\s^{(1)}_{x,x_0}\s^{(1)}_{x+1,x_0}
\s^{(2)}_{x,x_0}\s^{(2)}_{x+1,x_0}]\prod_\xx[1+\l a
\s^{(1)}_{x,x_0}\s^{(1)}_{x,x_0+1}
\s^{(2)}_{x,x_0}\s^{(2)}_{x,x_0+1}]$$
$$\prod_\xx[1+\l
b \s^{(1)}_{x,x_0}\s^{(1)}_{x+1,x_0}
\s^{(2)}_{x,x_0}\s^{(2)}_{x,x_0+1}]
\prod_\xx[1+\l b \s^{(1)}_{x,x_0}\s^{(1)}_{x,x_0+1}
\s^{(2)}_{x-1,x_0+1}\s^{(2)}_{x,x_0+1}]\Eq(equiv)$$
$$\prod_\a\prod_\xx[1+\l c
\s^{(\a)}_{x,x_0}\s^{(\a)}_{x+1,x_0}
\s^{(\a)}_{x,x_0}\s^{(\a)}_{x,x_0+1}]\prod_\a\prod_\xx[1+\l c
\s^{(\a)}_{x,x_0}\s^{(\a)}_{x+1,x_0}
\s^{(\a)}_{x+1,x_0-1}\s^{(\a)}_{x+1,x_0})]$$

Noting that
$$\s^{(\a)}_{x,x_0}
\s^{(\a)}_{x+1,x_0} e^{-H_I(\s^{(\a)})}=
{\partial\over \partial J^{(\a)}_{1;x,x_0;x+1,x_0}} 
Z_{I}^{(\a)}(J^{(\a)}_{\xx,\xx'})|_{
\{J^{(\a)}_{\xx,\xx'}\}=\{J^{(\a)}\}}\Eq(c41)$$
and from \equ(c21)
this derivative gives an extra 
factor $\tanh J^{(\a)}+{\rm sech}^2 J^{(\a)} \bar H^{(\a)}_{x,x_0} 
H^{(\a)}_{x+1,x_0}$ in \equ(9a).
In the same way
$$\s^{(\a)}_{x,x_0}
\s^{(\a)}_{x,x_0+1} e^{-H_I(\s^{(\a)})}=
{\partial\over \partial J^{(\a)}_{2;x,x_0;x,x_0+1}} 
Z_{I}^{(\a)}(J^{(\a)}_{\xx,\xx'})|_{
\{J^{(\a)}_{\xx,\xx'}\}=\{J^{(\a)}\}}\Eq(c42)$$
and this derivative gives a factor
$\tanh J^{(\a)}+{\rm sech}^2 J^{(\a)} \bar V^{(\a)}_{x,x_0} V^{(\a)}_{x,x_0+1}$.
We can write than, if
 $\d_{+,+}=1$ and $\d_{+,-}=\d_{-,+}=\d_{-,-}=2$
$$\hat Z_{2I}=\sum_{\e^{(1)},\e^{'(1)}}
(-1)^{\d_{\e^{(1)},\e^{'(1)}}}
\sum_{\e^{(2)},\e^{'(2)}}(-1)^{\d_{\e^{(2)},\e^{'(2)}}} 
\hat Z_{2I}^{\e^{(1)},\e^{'(1)},\e^{(2)},\e^{'(2)}}\Eq(nm)$$
where
$$\hat Z_{2I}^{\e^{(1)},\e^{'(1)},\e^{(2)},\e^{'(2)}}=
(\cosh J)^{2 B} 2^{2S}{1\over4}\Eq(nm1)$$
$$\int \prod_{\a=1}^2[\prod_{\xx} dH^{(\a)}_\xx d\bar H^{(\a)}_\xx d 
V^{(\a)}_\xx d\bar V^{(\a)}_\xx]
e^{S^{(1)}_{\e^{(1)},\e^{'(1)}}} e^{S^{(2)}_{\e^{(2)},\e^{'(2)}}}$$
$$\prod_\xx [1+\l a
(\tanh J+{\rm sech}^2 J \bar H^{(1)}_{x,x_0} H^{(1)}_{x+1,x_0})
(\tanh J+{\rm sech}^2 J \bar H^{(2)}_{x,x_0} H^{(2)}_{x+1,x_0})]$$
$$\prod_\xx[1+\l a (\tanh J+{\rm sech}^2 J \bar V^{(1)}_{x,x_0} 
V^{(1)}_{x,x_0+1})
(\tanh J+{\rm sech}^2 J \bar V^{(2)}_{x,x_0} V^{(2)}_{x,x_0+1})]$$
$$\prod_\xx[1+\l b (\tanh J+{\rm sech}^2 J \bar H^{(1)}_{x,x_0} H^{(1)}_{x+1,x_0})
(\tanh J+{\rm sech}^2 J \bar V^{(2)}_{x,x_0} 
V^{(2)}_{x,x_0+1})]$$
$$\prod_\xx[1+\l b (\tanh J+{\rm sech}^2 J \bar V^{(1)}_{x,x_0} 
V^{(1)}_{x,x_0+1})
(\tanh J+{\rm sech}^2 J \bar H^{(2)}_{x-1,x_0+1} 
H^{(2)}_{x,x_0+1})]$$
$$\prod_\a\prod_\xx[1+ \l c
(\tanh J+{\rm sech}^2 J \bar H^{(\a)}_{x,x_0} H^{(\a)}_{x+1,x_0})
(\tanh J+{\rm sech}^2 J \bar V^{(\a)}_{x,x_0} 
V^{(\a)}_{x,x_0+1})]$$
$$\prod_\a\prod_\xx[1+\l c (\tanh J+{\rm sech}^2 J \bar H^{(\a)}_{x,x_0} 
H^{(\a)}_{x+1,x_0})
(\tanh J+{\rm sech}^2 J \bar V^{(\a)}_{x+1,x_0-1} 
V^{(\a)}_{x+1,x_0})]$$

The above expression can be rewritten as
$$\hat Z_{2I}^{\e^{(1)},\e^{'(1)},\e^{(2)},\e^{'(2)}}=
(\cosh J)^{2 B} 2^{2S}{1\over 4} 
\int [\prod_{\a=1}^2 \prod_{\xx} dH^{(\a)}_\xx d
\bar H^{(\a)}_\xx d V^{(\a)}_\xx 
d\bar V^{(\a)}_\xx
e^{S^{(\a)}_{\e^{(\a)},\e^{'(\a)}}}] 
e^{\VV}\Eq(c60)$$
with
$$\VV=\VV_a+\VV_b+\VV_c\Eq(c61)$$
and, if $f_{i}=log(1+\l [i]\tanh^2 J)$
and $[i]=a,b,c$
$$\VV_a=\sum_\xx [f_a
+\tilde\l_a
[\bar H^{(1)}_{x,x_0} H^{(1)}_{x+1,x_0}+
\bar H^{(2)}_{x,x_0} 
H^{(2)}_{x+1,x_0}]+\l_a \bar 
H^{(1)}_{x,x_0} H^{(1)}_{x+1,x_0}
\bar H^{(2)}_{\xx} H^{(2)}_{x+1,x_0}]$$
$$+\sum_\xx [f_a
+\tilde\l_a
[\bar V^{(1)}_{x,x_0} V^{(1)}_{x,x_0+1}+ 
\bar V^{(2)}_{x,x_0} 
V^{(2)}_{x,x_0+1}]+\l_a \bar V^{(1)}_\xx V^{(1)}_{x,x_0+1}
\bar V^{(2)}_{x,x_0} V^{(2)}_{x,x_0+1}]\Eq(c62)$$
$$\VV_b=\sum_\xx [f_b
+\tilde\l_b
[\bar H^{(1)}_{x,x_0} H^{(1)}_{x+1,x_0}+ \bar V^{(2)}_{x,x_0} 
V^{(2)}_{x,x_0+1}]+\l_b
\bar H^{(1)}_{x,x_0} H^{(1)}_{x+1,x_0}
\bar V^{(2)}_{\xx} V^{(2)}_{x,x_0+1}]$$
$$+\sum_\xx[f_b
+\tilde\l_b
[\bar V^{(1)}_{x,x_0} V^{(1)}_{x,x_0+1}+
\bar H^{(2)}_{x-1,x_0+1} 
H^{(2)}_{x,x_0+1}]
+\l_b \bar V^{(1)}_{x,x_0} V^{(1)}_{x,x_0+1}
\bar H^{(2)}_{x-1,x_0+1} 
H^{(2)}_{x,x_0+1}]$$
$$\VV_c=\sum_\xx \sum_\a [f_c
+\tilde\l_c
[\bar H^{(\a)}_{x,x_0} H^{(\a)}_{x+1,x_0}+ \bar V^{(\a)}_{x,x_0} 
V^{(\a)}_{x,x_0+1}]+\l_c
\bar H^{(\a)}_{x,x_0} H^{(\a)}_{x+1,x_0}
\bar V^{(\a)}_{\xx} V^{(\a)}_{x,x_0+1}]$$
$$+\sum_\xx\sum_\a [f_c
+\tilde\l_c
[\bar V^{(\a)}_{x+1,x_0-1} V^{(\a)}_{x+1,x_0}+
\bar H^{(\a)}_{x,x_0} 
H^{(\a)}_{x+1,x_0}]
+\l_c\bar V^{(\a)}_{x+1,x_0-1} V^{(\a)}_{x+1,x_0}
\bar H^{(\a)}_{\xx} H^{(\a)}_{x+1,x_0}]$$
It is  easy in fact to verify that

$$e^{f_i+
\tilde\l_i [\bar H^{(\a)}_\xx H^{(\a)}_{x+1,x_0}+ 
\bar V^{(\a)}_\xx V^{(\a)}_{x,x_0+1}]+
\l_i\bar H^{(\a)}_\xx H^{(\a)}_{x+1,x_0}
\bar V^{(\b)}_\xx V^{(\b)}_{x,x_0+1}}=\{
(1+\l [i] \tanh^2 J)\Eq(l12)$$
$$[1+\tilde\l_i
[\bar H^{(\a)}_\xx H^{(\a)}_{x+1,x_0}+\bar V^{(\b)}_\xx V^{(\b)}_{x,x_0+1}]
+(\l_i+(\tilde\l_i)^2) \bar H^{(\a)}_\xx H^{(\a)}_{x+1,x_0}
\bar V^{(\b)}_\xx V^{(\b)}_{x,x_0+1}]\}$$
hence the equality between \equ(nm1) and \equ(c60)
holds with the identification
$$\tilde\l_i(1+\l[i])\tanh^2 J)=\l [i] 
{\rm sech}^2 J \tanh J$$
$$(1+\l [i] \tanh^2 J)
(\l_i+(\tilde\l_i)^2)
=\l [i] {\rm sech}^4 J\Eq(l1)$$
An expression identical 
to \equ(nm),\equ(nm1) holds for $Z_{2I}$, the only 
difference being that the relation with respect
$\tilde\l_i, \l_i$ is slightly more complicated
than \equ(l1), but for small $\l$ again 
$$\tilde\l_i=[i] \l(\tanh J {\rm sech}^2 J+O(\l))\quad
\l_i=[i] \l({\rm sech}^4 J+O(\l))\Eq(l4)$$
\*
\sub(1.2) {\it Massive and massless fermions.}
We define
$$P^{(\a)}_{\e^{(\a)},\e^{'(\a)}}
(dH^{\a},dV^{(\a)})=
\prod_{\xx} dH^{(\a)}_\xx d\bar H^{(\a)}_\xx d V^{(\a)}_\xx d\bar V^{(\a)}_\xx
e^{S^{(\a),0}_{\e^{(\a)},\e^{'(\a)}}}\Eq(l9)$$
It is convenient to perform the following change of variables [ID], [SP]
$$\bar H_\xx^{(\a)}+i H_\xx^{(\a)}=e^{i{\pi\over 4}}\psi_\xx^{(\a)}-
e^{i{\pi\over 4}}\chi_\xx^{(\a)}\qquad 
\bar H_\xx^{(\a)}-i H_\xx^{(\a)}=e^{-i{\pi\over 4}}\bar\psi_\xx^{(\a)}-
e^{-i{\pi\over 4}}\bar\chi_\xx^{(\a)}$$
$$\bar V_\xx^{(\a)}+i V_\xx^{(\a)}=\psi_\xx^{(\a)}+\chi_\xx^{(\a)}\qquad
\bar V_\xx^{(\a)}-i V_\xx^{(\a)}=\bar\psi_\xx^{(\a)}+\bar\chi_\xx^{(\a)}\Eq(c63)$$
It holds that
$$P^{(\a)}_{\e^{(\a)},\e^{'(\a)}}(dH^{\a},dV^{(\a)})
=
P^{(\a)}_{\e^{(\a)},\e^{'(\a)}}(d\psi^{\a})
P^{(\a)}_{\e^{(\a)},\e^{'(\a)}}(d\chi^{\a})e^{Q(\chi^{(\a)},\psi^{(\a)})}\Eq(19aa)$$

where, if $t=\tanh J$

$$P^{\a}_{\e,\e'}(d\psi^{(\a)})=[\prod_{\xx\in\L_M} d\psi^{(\a)}_\xx 
d\bar\psi^{(\a)}_\xx]
\exp\{{t\over 4}\sum_{\xx\in\L_M}
\psi^{(\a)}_\xx(\partial_1-i\partial_0)\psi^{(\a)}_\xx+
\bar\psi_\xx^{(\a)}(\partial_1+i\partial_0)\bar\psi^{(\a)}_\xx)$$
$$+{t\over 4}
\sum_{\xx\in\L_M} [-i\bar\psi^{(\a)}_\xx(\partial_1\psi^{(\a)}_x+
\partial_0\psi^{(\a)}_\xx)+i\psi_\xx^{(\a)}
(\partial_1\bar\psi_\xx^{(\a)}+\partial_0\bar\psi_\xx^{(\a)})]$$
$$+\sum_{\xx\in
\L_M}
i(\sqrt{2}-1-t)\bar\psi_\xx^{(\a)}\psi_\xx^{(\a)}\}\Eq(20)$$
where
$$\partial_1\psi_\xx^{(\a)}=\psi_{x+1,x_0}^{(\a)}-\psi_\xx^{(\a)}\qquad
\partial_0\psi_x^{(\a)}=\psi_{x,x_0+1}^{(\a)}-\psi_\xx^{(\a)}\Eq(20a)$$
Moreover 
$$P_{\e,\e'}^{(\a)}(d\chi^{(\a)})=[\prod_{\xx\in\L_M} d\chi^{(\a)}_\xx d\bar\chi^{(\a)}_\xx]
\exp\{{t\over 4}\sum_{\xx\in\L_M}
\chi^{(\a)}_\xx(\partial_1-i\partial_0)\chi^{(\a)}_\xx+
\bar\chi_\xx^{(\a)}(\partial_1+i\partial_0)\bar\chi^{(\a)}_\xx+$$
$${t\over 4}
\sum_{\xx\in\L_M} [-i\bar\chi^{(\a)}_\xx(\partial_1\chi^{(\a)}_x+
\partial_0\chi^{(\a)}_\xx)+
i\chi_\xx^{(\a)}
(\partial_1\bar\chi_\xx^{(\a)}+\partial_0\bar\chi_\xx^{(\a)})]
-\sum_{\xx\in
\L_M}
i(\sqrt{2}+1+t)\bar\chi_\xx^{(\a)}\chi_\xx^{(\a)}\}\Eq(21)$$
and finally
$$Q^{(\a)}(\psi^{(\a)},\chi^{(\a)})=\sum_\xx {t\over 4}
\{-\psi^{(\a)}_\xx(\partial_1\chi^{(\a)}_\xx+
i\partial_0\chi^{(\a)}_\xx)
-\bar\psi^{(\a)}_\xx(\partial_1\bar\chi^{(\a)}_\xx-i\partial_0\bar\chi^{(\a)}_\xx)-$$
$$\chi^{(\a)}_\xx(\partial_1\psi^{(\a)}_\xx+i\partial_0\psi^{(\a)}_\xx)-
\bar\chi^{(\a)}_\xx(\partial_1\bar\psi^{(\a)}_\xx-i\partial_0
\bar\psi^{(\a)}_\xx)
+i\bar\psi^{(\a)}_\xx(\partial_1\chi^{(\a)}_\xx-
\partial_0\chi^{(\a)}_\xx)\Eq(c69)$$
$$+\bar\psi^{(\a)}_\xx(-\partial_1\chi^{(\a)}_\xx-\partial_0\chi^{(\a)}_\xx)
+\bar\chi^{(\a)}_\xx(-\partial_1\psi^{(\a)}_\xx-\partial_0\psi^{(\a)}_\xx)
+\bar\chi^{(\a)}_\xx(-\partial_1\psi^{(\a)}_\xx+\partial_0\psi^{(\a)}_\xx)\}$$

We find convenient to rewrite the Grassmann variables in momentum space.
We call $D_{\e,\e'}$ the set of $\kk$ such that
$$k={2\pi n_1\over M}+{(\e-1)\pi\over 2M}\quad
k_0={2\pi n_0\over M}+{(\e'-1)\pi\over 2M}
\Eq(p1)$$
and $-[M/2]\le n_0\le [(M-1)/2]$, 
$-[M/2]\le n_1\le [(M-1)/2]$, $n_0,n_1\in Z$.
We can write if $\kk=(k_0,k)$ 
$$\psi_{\xx}^{(\a)}={1\over M^2}\sum_{\kk\in D_{\e,\e'}}
\psi^{(\a)}_{\kk} e^{-i\kk\xx}
\quad\bar\psi^{(\a)}_\xx=
{1\over M^2}\sum_{\kk\in D_{\e,\e'}}\bar\psi_{\kk}^{(\a)} 
e^{-i\kk\xx}\Eq(p2)$$
Hence
$$P_{\e,\e'}^{(\a)}(d\psi)=
[\prod_{\kk\in D_{\e,\e'}} 
d\bar\psi^{(\a)}_\kk d\psi^{(\a)}_\kk]\exp
[{t\over 4M^2}\sum_{\kk\in D_{\e,\e'}}
[\psi_\kk^{(\a)}\psi_{-\kk}^{(\a)}(i\sin k+
\sin k_0)$$
$$+\bar\psi_{\kk}^{(\a)}\bar\psi_{-\kk}^{(\a)}(i\sin k-\sin k_0)
+i 2 m_\psi(\kk)\bar\psi_\kk^{(\a)}\psi_{-\kk}^{(\a)}]]\Eq(fm)$$
where $m_{\psi}(\kk)={2\over t}
(\sqrt{2}-1-t)+(2\cos k_0+2\cos k-4)=2{(t-t_c)\over t}+O(\kk^2)$.
In deriving the above formula we have used that
$$\sum_{\xx}\psi^{(\a)}_\xx\partial_0\psi^{(\a)}_\xx=
\sum_{\kk\in D_{\e,\e'}} (e^{ik_0}-1)\psi_\kk^{(\a)}\psi_{-\kk}^{(\a)} 
=\sum_{\kk\in D_{\e,\e'}} i\sin k_0 \psi_\kk^{(\a)}\psi_{-\kk}^{(\a)}\Eq(71)$$
as
$$\sum_{\kk\in D_{\e,\e'}} (\cos k_0-1) \psi_\kk^{(\a)}\psi_{-\kk}^{(\a)}=-
\sum_{\kk\in D_{\e,\e'}} (\cos k_0-1) \psi_{-\kk}^{(\a)}\psi_{\kk}^{(\a)}$$
$$=-\sum_{\kk\in D_{\e,\e'}} (\cos k_0-1) \psi_\kk^{(\a)}
\psi_{-\kk}^{(\a)}=0\Eq(72)$$
\*
\sub(1.4a) {\it Majorana and Dirac fermions.} We write
$\psi^{(\a)}_\xx$ in \equ(p2) as
$$\psi_{\xx}^{(\a)}=e^{-i\pp_{\e,\e'}\xx}
{1\over M^2}\sum_{\kk'\in D_{-,-}}
\psi^{(\a)}_{\kk'+\pp_{\e,\e'}} e^{-i\kk'\xx}
=e^{-i\pp_{\e,\e'}\xx}\psi_{\xx}^{'(\a)}\Eq(mnzwww)$$
where $\pp_{\e}=({\pi(\e+1)\over 2 M}, {\pi(\e'+1)\over 2 M})$,
and $\psi^{'(\a)}_{\kk'}=\psi^{(\a)}_{\kk'+\pp_{\e,\e'}}$;
moreover
$$P_{\e,\e'}^{(\a)}(d\psi)=
P^{(\a)}(d\psi')e^{\tilde Q_{\psi,\e,\e'}^{(\a)}}\Eq(clal)$$
where
$$P^{(\a)}(d\psi')=[\prod_{\kk'\in D_{-,-}} 
d\bar\psi^{'(\a)}_{\kk'} d\psi^{'(\a)}_{\kk'}]
\exp[ {t\over 4 M^2}\sum_{\kk'\in D_{-,-}}
[\psi_{\kk'}^{'(\a)}\psi_{-\kk'}^{'(\a)}(i\sin k'+
\sin k'_0)\Eq(fm00)$$
$$+\bar\psi_{\kk'}^{'(\a)}\bar\psi_{-\kk'}^{'(\a)}
(i\sin k'-\sin k'_0)
+i 2 m_\psi(\kk')\bar\psi_\kk'^{'(\a)}\psi_{-\kk'}^{'(\a)}]]$$
and $\tilde Q_{\psi,\e,\e'}^{\a}$ is defined by \equ(clal)
and is formally vanishing in the limit $M\to\io$.
Proceeding in the same way for $P(d\chi)$ and $Q$ we find 
$$\int [\prod_{\a=1}^2 P^{(\a)}_{\e^{(\a)},\e^{'(\a)}}(d\psi^{\a})
P^{(\a)}_{\e^{(\a)},\e^{'(\a)}}(d\chi^{\a})]
e^{Q(\chi,\psi)}e^{\VV(\psi,\chi)}=\Eq(boc)$$
$$\int [\prod_{\a=1}^2 P^{(\a)}(d\psi') P^{(\a)}(d\chi')]
e^{\tilde Q_{\e}(\psi',\chi')}
e^{Q(\psi',\chi')}
e^{\VV(\psi',\chi')}$$
where 
$$\tilde Q_{\e}(\psi',\chi')=\sum_{\a=1}^2[\tilde Q^{(\a)}_{\psi,\e^{(\a)},\e^{'(\a)}}+
\tilde Q^{(\a)}_{\chi,\e^{(\a)},\e^{'(\a)}}+
\tilde Q^{(\a)}_{\psi\chi,\e^{(\a)},\e^{'(\a)}}]
\Eq(2.36xx);$$
of course $\tilde Q_{-,-,-,-}(\psi',\chi')=0$.
If 
$$D_{-,-}^+=\{\kk'\in D_{+,-}:k'>0\}\Eq(set)$$
we can write $P^{(\a)}(d\psi')$ in the following way
$$P^{(\a)}(d\psi^{'(\a)})=
\prod_{\kk'\in D_{-,-}^+} d\psi^{'(\a)}_{\kk'}
d\psi^{'(\a)}_{-\kk'}d\bar\psi_{\kk'}^{'(\a)}
d\bar\psi_{-\kk'}^{('\a)}\exp[{t\over 4 M^2}\sum_{\kk'\in D_{-,-}^+}
{\bf\x^{(\a)}}^T_{\kk'} A(\kk'){\bf\x^{(\a)}}_{\kk'}]\Eq(bn5)$$
where
$${\bf\x^{(\a)}}^T_{\kk'}=(\psi^{'(\a)}_{\kk'},
\psi^{'(\a)}_{-\kk'},
\bar\psi^{'(\a)}_{\kk'},\bar\psi_{-\kk'}^{'(\a)})\Eq(bn6)$$
and $A(\kk')$ is
$$
\left( \matrix{0&\sin k'_0+i\sin k' &0&0\cr
-\sin k'_0-i\sin k'&0&-im_\psi(\kk')&0&\cr
0&im_\psi(\kk')&0&-\sin k'_0+i\sin k'\cr 
0&0&\sin k'_0-i\sin k'&0\cr}\right)\Eq(bn7)$$

Hence if we perform the change of variables 
$$\psi^-_{1,\kk'}=
{1\over\sqrt{2}}(\psi^{'(1)}_{\kk'}+i\psi^{'(2)}_{\kk'})\quad
\psi^+_{1,\kk'}=
{1\over\sqrt{2}}(\psi^{'(1)}_{\kk'}-i\psi^{'(2)}_{\kk'})\Eq(bn9)$$
$$\psi^-_{-1,\kk'}=
{1\over\sqrt{2}}(\bar\psi^{'(1)}_{\kk'}+i\bar\psi^{'(2)}_{\kk'})\quad
\psi^+_{-1,\kk'}=
{1\over\sqrt{2}}(\bar\psi^{'(1)}_{\kk'}-i\bar\psi^{'(2)}_{\kk'})$$
we find by \equ(fon)
$$P^{(1)}(d\psi')P^{(2)}(d\psi')=\Eq(bn10)$$
$$\prod_{\kk'\in D_{-,-}^+} \prod_{\o=\pm 1}
d\psi^{+}_{\kk',\o}
d\psi^{-}_{\kk',\o}d\psi^{+}_{-\kk',\o}
d\psi^{-}_{-\kk',\o}
\exp[{t\over 4 M^2}\sum_{\kk'\in D_{-,-}^+}
{\bf\tilde\x^T_{\kk'} A(\kk'){\bf\tilde\x^{(+)}}_{\kk'}}]$$
where
$${\bf\tilde\x^{T}}_\kk=(\psi^-_{\kk,1},\psi^-_{-\kk,1},
\psi^-_{\kk,-1},\psi^-_{-\kk,-1})\Eq(bn11)$$
$${\bf\tilde\x^{+,T}}_\kk=(\psi^+_{\kk,1},
\psi^+_{-\kk,1},\psi^+_{\kk,-1},\psi^+_{-\kk,-1})$$
We can make another change of variables $\psi^+_{\kk,\o}\to
\psi^+_{-\kk,\o}$ so that at the end, if
$P^{(1)}(d\psi)$ $P^{(2)}(d\psi)$ $=P(d\psi)$
$$P(d\psi)=\Eq(bn12)$$
$$\prod_{\kk\in D_{-,-}^+} \prod_{\o=\pm 1}
d\psi^{+}_{-\kk,\o}
d\psi^{-}_{\kk,\o}d\psi^{+}_{\kk,\o}d\psi^{-}_{-\kk,\o}
\exp[{t\over 2 M^2}\sum_{\kk\in D_{-,-}^+}
{\bf\hat\x^T_\kk A(\kk){\bf\hat\x^{(+)}}_\kk}]$$
where
$${\bf\hat\x^{T}}_\kk=(\psi^-_{\kk,1},\psi^-_{-\kk,1},
\psi^-_{\kk,-1},\psi^-_{-\kk,-1})$$
$${\bf\hat\x^{+,T}}_\kk=(\psi^+_{-\kk,1},
\psi^+_{\kk,1},
\psi^+_{-\kk,-1},\psi^+_{\kk,-1})\Eq(bn13)$$
and in terms of the original variables
$$\psi^-_{1,\kk'}=
{1\over\sqrt{2}}(\psi^{'(1)}_{\kk'}+i\psi^{'(2)}_{\kk'})\quad
\psi^{+}_{1,-\kk'}=
{1\over\sqrt{2}}(\psi^{'(1)}_{\kk'}-i\psi^{'(2)}_{\kk'})\Eq(bn14)$$
$$\psi^-_{-1,\kk}=
{1\over\sqrt{2}}(\bar\psi^{'(1)}_{\kk'}+
i\bar\psi^{'(2)}_{\kk'})\quad
\psi^+_{-1,-\kk}=
{1\over\sqrt{2}}(\bar\psi^{'(1)}_{\kk'}-i\bar\psi^{'(2)}_{\kk'})$$
In the following we will call $\kk'$ simply $\kk$.
If we write
$$\psi^\pm_\xx={1\over M^2}\sum_{\kk\in D_{-,-}} e^{\pm i\kk\xx}\psi^\pm_\kk\Eq(bn15)$$
we can rewrite \equ(bn14) in the coordinate space as
$$\psi^-_{1,\xx}=
{1\over\sqrt{2}}(\psi^{'(1)}_\xx+i\psi^{'(2)}_\xx)\quad
\psi^+_{1,\xx}=
{1\over\sqrt{2}}(\psi^{'(1)}_\xx-i\psi^{'(2)}_\xx)\Eq(c65)$$
$$\psi^-_{-1,\xx}=
{1\over\sqrt{2}}(\bar\psi^{'(1)}_\xx+i\bar\psi^{'(2)}_\xx)\quad
\psi^+_{-1,\xx}=
{1\over\sqrt{2}}(\bar\psi^{'(1)}_\xx-i\bar\psi^{'(2)}_\xx)$$
or in terms of the original variables 
$$\psi^-_{1,\xx}=
{1\over\sqrt{2}}
(e^{i\pp_{\e^{(1)},\e^{'(1)}}\xx}\psi^{(1)}_\xx+
ie^{i\pp_{\e^{(2)},\e^{'(2)}}\xx}
\psi^{(2)}_\xx)$$
$$\psi^+_{1,\xx}=
{1\over\sqrt{2}}(e^{i\pp_{\e^{(1)},\e^{'(1)}}\xx}
\psi^{(1)}_\xx-ie^{i\pp_{\e^{(2)},\e^{'(2)}}\xx}\psi^{(2)}_\xx)\Eq(c66)$$
$$\psi^-_{-1,\xx}=
{1\over\sqrt{2}}(e^{i\pp_{\e^{(1)},\e^{'(1)}}\xx}
\bar\psi^{(1)}_\xx+ie^{i\pp_{\e^{(2)},\e^{'(2)}}\xx}\bar\psi^{(2)}_\xx)$$
$$\psi^+_{-1,\xx}=
{1\over\sqrt{2}}(e^{i\pp_{\e^{(1)},\e^{'(1)}}\xx}
\bar\psi^{(1)}_\xx-ie^{i\pp_{\e^{(2)},\e^{'(2)}}\xx}\bar\psi^{(2)}_\xx)$$

If $\kk\in D_{-,-}^+$ then $-\kk\in D^-_{-,-}$ with $D_{-,-}^+\bigcup
D^-_{-,-}=D_{-,-}$
so that
$$P(d\psi)=\Eq(mis)$$
$$\prod_{\kk\in D_{-,-}} \prod_{\o=\pm 1}
d\psi^{+}_{\kk,\o}
d\psi^{-}_{\kk,\o}
\exp[{t\over 2M^2}\sum_{\kk\in D_{-,-}}
{\bf\bar\x^T_\kk \tilde A(\kk){\bf\bar\x^{(+)}}_\kk}]$$
$$\tilde A(\kk)= 
\left( \matrix{i\sin k+\sin k_0 & i m_\psi(\kk)\cr
-i m_\psi(\kk) & i\sin k-\sin k_0 \cr}\right)$$
$${\bf\tilde\x^{T}}_\kk=(\psi^-_{\kk,1},\psi^-_{\kk,-1})
\quad{\bf\tilde\x^{+,T}}_\kk=(\psi^+_{\kk,1},
\psi^+_{\kk,-1})\Eq(bn18)$$
\*
{\it Remark.} In the physical language, the 
change of variables \equ(c66) means that one is describing
the system in terms of {\it Dirac fermions} instead in terms of 
{\it Majorana fermions}.
\*
A similar computation can be done for $P(d\chi)$; proceeding exactly
as above we find
$$P^{(1)}(d\chi^{(1)})P^{(2)}(d\chi^{(2)})
=P(d\chi)\Eq(hbn)$$
where
$$P(d\chi)=\prod_{\kk\in D_{-,-}} \prod_{\o=\pm 1}
d\chi^{+}_{\kk,\o}
d\chi^{-}_{\kk,\o}
\exp[{t\over 2 M^2}\sum_{\kk\in D_{-,-}}
{\bf\bar\h^T_\kk \tilde B(\kk){\bf\bar\h^{(+)}}_\kk}]\Eq(misb)$$
$$\tilde B(\kk)= 
\left( \matrix{i\sin k+\sin k_0 & i m_\chi(k)\cr
-i m_\chi(k) & i\sin k-\sin k_0 \cr}\right)$$
$${\bf\bar\h^{T}}_\kk=(\chi^-_{\kk,1},\chi^-_{\kk,-1})
\quad{\bf\bar\h^{+,T}}_\kk=(\chi^+_{\kk,1},
\chi^+_{\kk,-1})\Eq(bn18b)$$
Note that
$t m_\chi(\kk)=2(\sqrt{2}+1+t)+2t(2-\cos k_0-\cos k_1)$, and
the mass of the $\chi$ fields never vanishes.
It holds that
$$\int P(d\chi)\,
\chi^{-}_{\xx,\o}\chi^{+}_{\yy,\o'} =
g^{(\chi)}_{\o,\o'}(\xx-\yy)\;,\Eq(2.9c1)$$
where
$$g^{(\chi)}_{\o,\o'}(\xx-\yy)={1\over 2 t M^2}
\sum_{\kk}e^{-i\kk(\xx-\yy)}
[\tilde B^{-1}(\kk)]_{\o,\o'}\;,\Eq(2.92kk)$$
and $\tilde B^{-1}(\kk)$ is 
the inverse of the $\tilde B(\kk)$ defined in
\equ(misb).
 
If we set
$$A(\kk) = \det \tilde B(\kk) =
-\sin^2 k_0 - \sin^2 k- [m^\chi(\kk)]^2 \; ,\Eq(2.93kk) $$
then
$$ B^{-1}(\kk)= {1\over A(\kk)}
\left( \matrix{-\sin k_0+i \sin k & -i m^\chi(\kk)\cr
im^\chi(\kk) & \sin k_0+i \sin k \cr}\right) \;.\Eq(2.94kk)$$
Similar formulas hold for $g^{(\psi)}(\xx-\yy)$ and,
if $i=\psi,\chi$, the following bounds holds,
for any $N>1$
$$|g^{(i)}_{\o,\o}(\xx-\yy)|\le {1\over 1+|{\bf d}_M(\xx-\yy)|}
{C_N\over 1+|m_i{\bf d}_M(\xx-\yy)|^N}\Eq(klm2)$$
$$|g^{(i)}_{\o,-\o}(\xx-\yy)|\le 
{ |m_i| C_N\over 1+|m_i{\bf d}_M(\xx-\yy)|^N}\Eq(klm1)$$
where 
$${\bf d}_M(\xx-\yy)=({M\over\pi}\sin({\pi(x-y)\over M}),
{M\over\pi}\sin({\pi(x_0-y_0)\over M})\Eq(qwe)$$

The same transformations are done for the term $e^{S^\nu_{\e,\e'}}$
which can be written as the product of three terms similar to 
\equ(20), \equ(21) and \equ(c69) with $t$ replaced by $\nu$
and $\sqrt{2}-1$ in \equ(20) and $\sqrt{2}+1$
in \equ(21) replaced by $0$; then \equ(20),\equ(21)
can be written in terms of Dirac fermions 
as \equ(bn18) and \equ(bn18b) with $t$ 
replaced by $\nu$ and $\sqrt{2}-1$ or $\sqrt{2}+1$
replaced by $0$.

We have then written the partition function as Grassmann
integral over the $\chi$-fields, which are massive
with a mass $O(1)$, and
the $\psi$-fields with a small mass $O(t-t_c)$ close to the 
critical temperature; we integrate out first the {\it
heavy fermions} $\chi$
to get an effective theory in terms of the 
{\it light fermions}
$\psi$ only.

\*
\vskip1cm
\section(2a,Integration of heavy fermions)
\vskip.5cm
\sub(1.4b) {\it Local interactions.}
By the change of variables in the preceding section we can write
\equ(nm1) as
$$Z_{2I}^{\e^{(1)},\e^{'(1)},\e^{(2)},\e^{'(2)}}=e^{M^2\NN}
\int P(d\psi)
\int P(d\chi)
e^{\tilde Q_{\e}(\chi,\psi)}
e^{Q(\chi,\psi)}
e^{\VV(\psi,\chi)}\Eq(c70)$$
where $\NN$ is a normalization constant,
$P(\psi)$ and $P(d\chi)$ are given by \equ(mis), \equ(misb),
$\VV$, $Q$, $\tilde Q_\e$ are obtained respectively from
\equ(c69) and \equ(2.36xx)
by the change of variables 
\equ(c66).
We write 
$$\VV(\psi,\chi)=\VV_2(\psi)+\VV_4(\psi)+\VV_\chi(\psi,\chi)\Eq(81)$$
where $\VV_2(\psi)$ is a sum of monomials
bilinear in the $\psi,\psi^+$ variables, 
$\VV_4(\psi)$ is a sum of monomials
quartic in the $\psi,\psi^+$ and
$\VV_\chi(\psi,\chi)$ is sum of monomials
bilinear or quartic each one containing at least
a $\chi_\o$ field. 
It holds that, from \equ(c63),
$$\bar H^{(\a)}_\xx H^{(\a)}_\xx={i\over 2}\psi^{(\a)}_\xx
\bar\psi^{(\a)}_\xx+R_1^{(\a)}\quad
\bar V^{(\a)}_\xx V^{(\a)}_\xx={i\over 2}\psi^{(\a)}_\xx
\bar\psi^{(\a)}_\xx+R_2^{(\a)}\Eq(c71)$$
where $R_i^{(\a)}$ is sum of monomials bilinear in the fields
$\chi,\bar\chi,\psi,\bar\psi$ and containing
at least one field $\bar\chi$ or $\chi$. 
Moreover, from \equ(c65)
$$\psi_\xx^{(1)}\bar\psi_\xx^{(1)}={1\over2}
[\psi_{1,\xx}\psi_{-1,\xx}+\psi_{1,\xx}\psi^+_{-1,\xx}
+\psi^+_{1,\xx}\psi_{-1,\xx}+
\psi^+_{1,\xx}\psi^+_{-1,\xx}]\Eq(c80)$$
$$\psi_\xx^{(2)}\bar\psi_\xx^{(2)}={1\over2}
[-\psi_{1,\xx}\psi_{-1,\xx}-\psi^+_{1,\xx}\psi^+_{-1,\xx}
+\psi^+_{1,\xx}\psi_{-1,\xx}+
\psi_{1,\xx}\psi^+_{-1,\xx}]$$
hence
$$\psi_\xx^{(\a)}\bar\psi_\xx^{(\a)}
\psi_\xx^{(\a)}\bar\psi_\xx^{(\a)}=0$$
$$\psi_\xx^{(1)}\bar\psi_\xx^{(1)}
\psi_\xx^{(2)}\bar\psi_\xx^{(2)}=e^{-2i(\pp_{\e^{(1)},
\e^{(1)}}+\pp_{\e^{(1)},
\e^{(1)}})\xx}
\psi^+_{1,\xx}
\psi^+_{-1,\xx}\psi^-_{-1,\xx}\psi^-_{1,\xx}\Eq(p1)$$
and from \equ(c62)
$$\VV_4(\psi)=-2e^{-2i(\pp_{\e^{(1)},
\e^{(1)}}+\pp_{\e^{(1)},
\e^{(1)}})\xx}
(\tilde\l_a+\tilde\l_b)\sum_\xx\psi^+_{1,\xx}
\psi^+_{-1,\xx}\psi^-_{-1,\xx}\psi^-_{1,\xx}+\VV_4^R(\psi)\Eq(v4)$$
and $\VV_4^R$ is a sum of quartic monomials
with coupling $O(\l)$
in which at least a $\partial\psi$ field. 
In the same way it is easy to check that 
$$\VV_2={1\over 2M^2}\sum_{\kk,\o}\{
[\nu+f_1(\l)][i\o+
\cos k_0+\cos k-2]\psi_{\kk,\o}\psi^+_{\kk,-\o}+$$
$$[\nu+f_2(\l)][\o\sin k_0+i\sin k]
\psi_{\kk,\o}\psi^+_{\kk,\o}\}\Eq(91)$$
where $|f_1|,|f_2|\le C|\l|$.
\vskip.5cm
\sub(1.4bi) {\it Integration of the heavy fields}
We now integrate the heavy $\chi$ fields
$$\int \bar P(d\psi)
e^{M^2\NN^{(1)}+\VV^{(1)}(\psi)}=
\int P(d\psi) \int P(d\chi)
e^{Q(\chi,\psi)}
e^{\tilde Q_\e(\chi,\psi)}
e^{\VV(\psi,\chi)}\Eq(mas)$$
where $\NN^{(1)}$ is a constant.
Let us consider a set of coordinates
$\xx_1,...,\xx_{2n}$ which are not all different
one to the other and we will denote by $\sum^*_{\xx_1,...,\xx_{2n}}$
the sum over all the distinct $\le 2n$ variables. 
We prove the following result.
\*
{\bf Theorem 2.} {\it 
There exists an $\e$ such that,
for $|\l|,|\nu|\le\e$
$$\VV^{(1)}=
\sum_{n\ge 1}\sum_{\{\e,\a,\o\}}\sum^*_{\xx_1,..,\xx_{2n}}
W_{n}(\xx_1,..,\xx_{2n})
\partial^{\a_1}\psi^{\e_1}_{\xx_1,\o_1}...
\partial^{\a_{2n}}\psi_{\xx_{2n},\o_{2n}}^{\e_{2n}}\Eq(bnv)$$
and, for $n> 2$
$$
\sum^*_{\xx_1,...\xx_{2n}}|
W_{n}(\xx_1,...\xx_n)|
\le M^2 C^n|\l|^{n/2}\Eq(bnv1)$$
For $n=2$
$$\sum^*_{\xx_1,..,\xx_{4}}
W_{2}(\xx_1,..,\xx_{2n})
\partial^{\a_1}\psi^{\e_1}_{\xx_1,\o_1}...
\partial^{\a_{2n}}\psi_{\xx_{2n},\o_{2n}}^{\e_{2n}}=$$
$$\VV_4(\psi)+\sum^*_{\xx_1,..,\xx_{4}}\sum_{\{\e,\a,\o\}}
W_{2}(\xx_1,..,\xx_{4})
\partial^{\a_1}\psi^{\e_1}_{\xx_1,\o_1}...
\partial^{\a_{2n}}\psi_{\xx_{2n},\o_{2n}}^{\e_{2n}}\Eq(v4444)$$
with
$
\sum^*_{\xx_1,..,\xx_{4}}|W_{2}(\xx_1,..,\xx_{4})|\le M^2
C |\l|$ and  
$\hat W_{2}({\bf 0},...,{\bf 0})=O(\l^2)$
; for $n=1$
$$\sum^*_{\xx_1,\xx_{2}}\sum_{\{\e,\a,\o\}}
W_{1}(\xx_1,\xx_{2})
\partial^{\a_1}\psi^{\e_1,\o_1}_{\xx_1}
\partial^{\a_2}\psi_{\xx_{2},\o_2}^{\e_{2}}=$$
$$\VV_2(\psi)+\sum^*_{\xx_1,\xx_{2}}\sum_{\{\e,\a,\o\}}
W_{1}(\xx_1,\xx_2)
\partial^{\a_1}\psi^{\e_1}_{\xx_1,\o_1}
\partial^{\a_2}\psi_{\xx_{2n},\o_{2n}}^{\e_{2}}\Eq(ve4444)$$
with
$
\sum^*_{\xx_1,\xx_{2}}| 
W_{1}(\xx_1,\xx_{2})|\le M^2
C |\l|$; $\VV_2$ and $\VV_4$ are given by \equ(v4) and \equ(91).
Finally
$$\bar P(d\psi)=
\prod_{\kk\in D_{-,-}} \prod_{\o=\pm 1}
d\psi^{+}_{\kk,\o}
d\psi^{-}_{\kk,\o}\exp[-{t\over M^2}\sum_{\kk\in D_{-,-}}
\psi^+_{\kk,\o}\tilde A_{\o,\o'}(\kk)\psi^-_{\kk,\o'}]\Eq(int11)$$
with 
$$\tilde A(\kk)={1\over C_0+f_{0,0}(\kk)} 
\left( \matrix{\tilde Z(i\sin k+\sin k_0)+f_{1,1}(\kk) 
& -i(t-t_c)C_0+f_{1,2}(\kk)\cr
i(t-t_c)C_0+f_{2,1}(\kk) & \tilde Z(i\sin k-\sin k_0)+f_{2,2}(\kk)\cr}\right)$$
with $C_0=(t+1+\sqrt{2})^2$ and $\tilde Z=
{t\over 2}[(2 t+2\sqrt{2} t)+(2\sqrt{2}+3+t^2)]$ and
$f_{0,0}(\kk)$, $f_{i,j}(\kk)$, $i,j=1,2$
analytic $O(\kk^2)$ functions.
}
\*
{\it Remark 1.} The fact that
$\hat W_{2}({\bf 0},...,{\bf 0})=O(\l^2)$
can be checked by an explicit computations
of all the contributions with coupling $O(\l)$
to $W_2$, noting that they can be only
obtained contracting a terms quartic in the $\chi$ fields
with one of the addend of \equ(c69); 
each of such terms carries a derivative
in the coordinate space, hence the Fourier transform of
such terms is vanishing at zero 
momentum.
\*
{\it Remark 2.} 
The fermionic
integration $\bar P(d\psi)$ reduces, neglecting
the functions $f_{0,0}(\kk)$, $f_{i,j}(\kk)$
which are $O(\kk^2)$, to the integration
of a system of Dirac fermions on a lattice
with mass $O(t-t_c)$. Hence, apart from the functions
$f_{0,0}(\kk)$, $f_{i,j}(\kk)$, it coincides with
the {\it free} action of the Thirring model. 
Note however that the interaction
in the Thirring model is quartic and invariant 
under the transformation $\psi^{\e}_{\xx,\o}\to e^{i\e\a_\o}
\psi^{\e}_{\xx,\o}$ in the massless case;
both this properties are not true for \equ(bnv).
\*
{\it Proof.} 
We start from the definition
of truncated expectation:
$$\EE^T_\chi(X;n)={\partial^n\over\partial\l^n}\log\int P(d\chi)
e^{\l X(\chi)}|_{\l=0}\Eq(hn1)$$
so that,
calling
$$\bar \VV(\chi,\psi)=Q(\chi,\psi)
+\tilde Q_\e(\chi,\psi)+
\VV(\chi,\psi)\Eq(hn2)$$
we obtain
$$M^2\NN^{(1)}+\VV^{(1)}(\psi)=\log \int P(d\chi)e^{-\bar \VV(\chi,\psi)}
=\sum_{n=0}^\io{(-1)^n\over n!}\EE^T_\chi(V(.,\psi;n))\Eq(3.30)$$
We label each one of the monomials (whose
number will be called $C_0$)
in $\bar\VV$ by an index $v_i$, so that
each monomial can be written as
$$\sum_{\xx_{v_i}} v(\xx_{v_i})
\prod_{f\in \tilde P_{v_i}}\partial^{\a(f)}
\psi^{\e(f)}_{\o(f),\xx(f)}
\prod_{f\in P_{v_i}}\partial^{\a(f)}\chi^{\e(f)}_{\o(f),\xx(f)}\Eq(hn3)$$
where $\xx_{v_i}$ is the total set of coordinates
associated to $v_i$ and $P_{v_i}$ and $\tilde P_{v_i}$
are set of indices labeling the $\chi$ or $\psi$-fields.
We can write 
$$\VV^{(1)}(\psi)=\sum_{\tilde P_{v_0}\not=0}
\VV^{(1)}(\tilde P_{v_0})\;,\Eq(3.31aa)$$
$$\VV^{(1)}(\tilde P_{v_0})= \sum_{\xx_{v_0}}
[\prod_{f\in\tilde P_{v_0}}
\partial^{\a(f)}\psi_{\o(f),\xx(f)}^{\e(f)}] 
K_{\tilde P_{v_0}}^{(0)}(\xx_{v_0})\;,
\Eq(3.31b)$$
$$K_{\tilde P_{v_0}}^{(0)}(\xx_{v_0})=\sum_{n=1}^\io
{1\over n!} \sum_{v_1,..,v_n}
\EE^T_{\chi}[
\tilde\chi(P_{v_1}),\ldots,
\tilde\chi(P_{v_n})]
\prod_{i=1}^n v_i(\xx_{v_i})\;,\Eq(3.31c)$$
where 
$\sum_{v_1,...v_n}\le C_0^n$, 
$\tilde P_{v_0}=\bigcup_i \tilde P_{v_i}$ and $\xx_{v_0}=\bigcup_i
\xx_{v_i}$.
We use now the well known expression for $\EE^T$ (see for instance [Le])
$$\EE^T_{\chi}(\tilde\chi(P_1),...,\tilde\chi(P_s))=
\sum_{T}\prod_{l\in T}  
g^\chi_{\o^-,\o^+}(\xx_l-\yy_l)
\int dP_{T}(\tt) \det G^{T}(\tt)\Eq(3.38)$$
where:

-)$P$ is a set of indices, and
$$\tilde\chi(P)=\prod_{f\in P}
\partial^{\a(f)}\chi^{\e(f)}_{\xx(f),\o(f)}\Eq(hn4)$$

-)$T$ is a set of lines forming an {\it anchored tree} between
the cluster of poins $P_1,..,P_s$ \ie $T$ is a set
of lines which becomes a tree if one identifies all the points
in the same clusters.

-)$\tt=\{t_{i,i'}\in [0,1], 1\le i,i' \le s\}$, $dP_{T}(\tt)$
is a probability measure with support on a set of $\tt$ such that
$t_{i,i'}=\uu_i\cdot\uu_{i'}$ for some family of vectors $\uu_i\in \RRR^s$ of
unit norm.

-)$G^{T}(\tt)$ is a $(n-s+1)\times (n-s+1)$ matrix, whose
elements are given by $G^{T}_{ij,i'j'}=t_{i,i'}
\tilde g_{\o^-,\o^+}(\xx_{ij}-\yy_{i'j'})$ with
$(f^-_{ij}, f^+_{i'j'})$ not belonging to $T$.

If $s=1$ the sum over $T$ is empty, but we can still
use the above equation by iterpreting the r.h.s.
as $1$ if $P_1$ is empty, and $det G(P_1)$ otherwise.

We bound the determinant using the well known
{\it Gram-Hadamard inequality}, stating
that, if $M$ is a square matrix with elements $M_{ij}$ of the form
$M_{ij}=<A_i,B_j>$, where $A_i$, $B_j$ are vectors in a Hilbert space
with
scalar product $<\cdot,\cdot>$, then
$$|\det M|\le \prod_i ||A_i||\cdot ||B_i||\;.\Eq(3.49)$$
where $||\cdot||$ is the norm induced by the scalar product.

Let $\HH=\RRR^s\otimes \HH_0$, where $\HH_0$ is the Hilbert space of complex
four dimensional vectors $F(\kk)=(F_1(\kk),\ldots,F_4(\kk)$), $F_i(\kk)$
being a function on the set $\DD_{-,-}$, with scalar product
$$<F,G>=\sum_{i=1}^4 {1\over M^2}\sum_{\kk} F^*_i(\kk) G_i(\kk)\;.
\Eq(3.96)$$
and it is easy to veriy that
$$G^{T}_{ij,i'j'}=t_{i,i'} 
g^{(\chi)}_{\o^-_l,\o^+_l}(\xx_{ij}-\yy_{i'j'})
=<\uu_i\otimes A_{\xx(f^-_{ij}),\o(f^-_{ij})},
\uu_{i'}\otimes B_{\xx(f^+_{i'j'}),\o(f^+_{i'j'})}>\;,\Eq(3.97)$$
where $\uu_i\in \RRR^s$, $i=1,\ldots,s$, are the vectors such that
$t_{i,i'}=\uu_i\cdot\uu_{i'}$, and
$$\eqalign{
A_{\xx,\o}(\kk)&=e^{i\kk'\xx}{1\over
\sqrt{-A(\kk)}} \cdot \cases{
(-\sin k_0+i\sin k,0,-i m_\chi(\kk),0),& if $\o=+1$,\cr
(0,im_\chi(\kk),0,m_\chi(\kk)),& if $\o=-1$,\cr}\cr
B_{\xx,\o}&=e^{i\kk'\yy}{1\over
\sqrt{-A(\kk)}} \cdot \cases{
(1,1,0,0),& if $\o=+1$,\cr
(0,0,1,(\sin k_0+i\sin k)/m_\chi(\kk)),& if $\o=-1$.\cr}\cr}\Eq(3.98)$$

Hence from \equ(3.49) we immediately find
$$|G^{T}_{ij,i'j'}|\le C_1^n\Eq(bond)$$
Note that $C_1$ is an $O(1)$ constant; this follows from the fact
that the Grassmann variable $\chi$ has an $O(1)$ mass.
Finally we get
$$
\sum_{\xx_{v_0}}|K_{\tilde P_{v_0}}^{(1)}(\xx_{v_0})|
\le \sum_{n=1}^\io{1\over n!}
\sum_{v_1,..,v_n}
\sum_{\xx_{v_1},...,\xx_{v_n}} C_1^n
\sum_{T}[\prod_{l\in T}  
|\tilde g_\chi(\xx_l-\yy_l)|]
\prod_{i=1}^n |v_i(\xx_{v_i})|\Eq(**)$$
where 
we have used that 
$\int dP_{T}(\tt)=1$. 
Regarding the sum over $T$, 
it is empty if $n=1$. If $n>1$ 
the number of anchored trees 
with $d_i$ lines branching from the vertex
$v_i$ can be bounded, 
by using Caley's formula, by
$${(s_v-2)!\over (d_1-1)!...(d_{s_v}-1)!} |P_{v_1}|^{d_1}...
|P_{v_{n}}|^{d_{n}}\;;\Eq(klxxz)$$
hence the number of addenda in 
$\sum_{T}$ is bounded by
$n!C_2^n$. Finally $T$ and the $\bigcup_i \xx_{v_i}$
form a tree connecting all points, so that
using that the propagator is massive
and that the interactions are short ranged
$\sum_{\xx_{v_1},... \xx_{v_n}} 
\sum_{T}[\prod_{l\in T}  
|\tilde g_\chi(\xx_l-\yy_l)|]
\prod_{i=1}^n |v_i(\xx_{v_i})|\le C_3^n|\l|^{\tilde n}M^2$,
where $\tilde n$ is the number of coupling $O(\l)$.

Let us consider the case $|P_{v_0}|\ge 4$.
Note that
if to $v_i$ are associated only terms
from $\VV(\psi,\chi)$,
then $\tilde n=n$. The same bound holds
for $M$ large enough,
if there are $v_i$ associated with
$\tilde Q_\e$; 
in fact such terms are vanishing as $M\to\io$
(one has in the bounds an extra $M^{-1}$ 
for any of such $v_i$,
for dimensional reasons).
Let us consider now 
the case in which 
there are end-points associated to $Q(\psi,\chi)$,
which have $O(1)$ coupling;
there are
at most $|\tilde P_{v_0}|$
end-points 
associated with $Q(\psi,\chi)$. In fact in $Q(\psi,\chi)$ 
there are only terms of the form
$\psi\chi$,
so at most the number of them is equal to the number 
of $\psi$ fields.
If we call $n_\l\le\tilde n$ 
the number of vertices quartic in the fields
it is clear that
$n_\l\ge \max\{1,|\tilde P_{v_0}|/2-1\}$; hence
$$
\sum_{\xx_{v_0}}|K_{\tilde P_{v_0}}^{(1)}(\xx_{v_0})|
\le M^2 \sum_{\tilde n=1}^\io 
C^{\tilde n+|\tilde P_{v_0}|}|\l|^{\tilde n\over 2}
|\l|^{\max\{{1\over2},|\tilde P_{v_0}|/4-1/2\}}\Eq(nm10o)$$
and \equ(bnv1) holds for
$|\tilde P_{v_0}|\ge 4$ (in the r.h.s. of \equ(nm10o)
the sum over the number
of $v_i$ to which are associated quadratic monomials
with coupling $O(\l)$ is already done). 

Consider now the case $|\tilde P_{v_0}|=2$; in this case
there are terms $\l$ independent, obtained
when to all the $v_i$ are associated with elements of
$Q(\psi,\chi)$.
It is convenient to include all such terms in the free
measure, as they cannot 
be considered as perturbations (they are not $O(\l)$).
Instead of computing all such terms, we can proceed
in a more rapid way by noting that
$$\int\bar P(d\psi)=\int P(d\psi)\int P(d\chi)e^{Q(\psi,\chi)}\Eq(mis12)$$
and if, if $<X>_0=\int \bar P^{0}(d\psi)X$,
it holds
$$<\psi^-_{\xx,1}\psi^+_{\yy,1}>_0
={1\over 2}<(\psi^{(1)}_\xx+i\psi^{(2)}_\xx)
(\psi^{(1)}_\yy-i\psi^{(2)}_\yy)>_0
=<\psi^{(1)}_\xx\psi^{(1)}_\yy>_0\Eq(vbd)$$
as $<\psi^{(1)}\psi^{(2)}>_0=0$ if $\l=0$ (the two systems
are decoupled).
In the same way
$$<\psi^-_{\xx,-1}\psi^+_{\yy,-1}>_0
={1\over 2}<(\bar\psi^{(1)}_\xx+i\bar\psi^{(2)}_\xx)
(\bar\psi^{(1)}_\yy-i\bar\psi^{(2)}_\yy)>_0
=<\bar\psi^{(1)}_\xx\bar\psi^{(1)}_\yy>_0
\Eq(hjn4)$$
and finally
$$<\psi^-_{\xx,1}\psi^+_{\yy,-1}>_0
={1\over 2}<(\psi^{(1)}_\xx+i\psi^{(2)}_\xx)
(\bar\psi^{(1)}_\yy-i\bar\psi^{(2)}_\yy)>_0
=<\psi^{(1)}_\xx\bar\psi^{(1)}_\yy>_0\Eq(hjn5)$$
Note that $<\psi^\a_\xx\psi^\a_\yy>_0$
are the Ising correlation computed in [MPW], [SP]
where is found that
$$<\psi^{(\a)}_\xx\psi^{(\a)}_\yy>_0=
{1\over M^2}
\sum_\kk {e^{-i\kk(\xx-\yy)}\over\D}c_{1,1}(\kk)\Eq(sall1) $$
$$<\bar\psi^{(\a)}_\xx\bar\psi^{(\a)}_\yy>_0={1\over M^2}
\sum_\kk {e^{-i\kk(\xx-\yy)}\over\D}c_{2,2}(\kk)\Eq(sall2)$$
$$<\psi^{(\a)}_\xx\bar\psi^{(\a)}_\yy>_0=
{1\over M^2}
\sum_\kk {e^{-i\kk(\xx-\yy)}\over\D}c_{1,2}(\kk)\Eq(hjn6)$$
where
$$\D=2t(1-t^2)(2-\cos k_0-\cos k)+(t+1+\sqrt{2})^2(t-\sqrt{2}+1)^2\Eq(hjn8)$$
and 
$$c_{1,1}(\kk)={t\over 2}
[(2t+2\sqrt{2}t)(\sin k_0\cos k-i\sin k\cos k_0)
+(2\sqrt{2}+3+t^2)(\sin k_0-i\sin k)]\Eq(11111)$$
$$c_{2,2}(\kk)=
{t\over 2}
[(2t+2\sqrt{2}t)(\sin k_0\cos k-i\sin k\cos k_0)
+(2\sqrt{2}+3+t^2)(-\sin k_0-i\sin k)]\Eq(bveee)$$
$$c_{1,2}(\kk)=-c_{1,2}(\kk)=
{-i\over 2}
[(2\sqrt{2}t^2+4 t^2)\cos k\cos k_0+$$
$$
(t^3+2t\sqrt{2}+t)(\cos k+\cos k_0)-2-2\sqrt{2}+2t^2]\Eq(vvvaq)$$
\*
Note that
$${c_{1,1}(\kk)c_{2,2}(\kk)-c_{1,2}(\kk)c_{2,1}(\kk)\over\D(\kk)}
-(t+1+\sqrt{2})^2=f(\kk)$$
where $f(\kk)=O(\kk^2)$ and with bounded derivatives.

\vskip.5cm
\section(2z,Renormalization Group for light fermions)
\vskip.5cm
\sub(1.4bz) {\it Multiscale analysis}.
We start considering $Z^{-,-,-,-}_{2I}$
in the sum \equ(nm), written as in \equ(mas).
We introduce a {\sl scaling parameter} 
$\g>1$ and a positive function $\c(\kk) \in C^{\io}$ such that
$$ \c(\kk) = \c(-\kk) = \cases{
1 & if $|\kk| <t_0 \= a_0  /\g \;,$ \cr
0 & if $|\kk| >a_0 \; ,$\cr}\Eq(2.30)$$
where
$$|\kk|=\sqrt{\sin k_0^2+\sin k^2}\;.\Eq(2.31)$$
We define also, for any integer $h\le 0$,
$$f_h(\kk)= \c(\g^{-h}\kk)-\c(\g^{-h+1}\kk)\; ;\Eq(2.36)$$
we have, for any $h_M<0$,
$$\c(\kk) = \sum_{h=h_M+1}^0 f_h(\kk) +\c(\g^{-h_M}\kk)\; .
\Eq(2.37)$$
Note that, if $h\le 0$, $f_h(\kk) = 0$ for $|\kk|
<t_0\g^{h-1}$ or $|\kk| >t_0 \g^{h+1}$, and $f_h(\kk)=
1$, if $|\kk| =t_0\g^h$.
Therefore
$$f_h(\kk)=0\quad \forall h< h_{M} =\min \{h:t_0\g^{h+1} >
\sqrt{2}(\p M^{-1})^2\}\;,\Eq(2.40)$$
and
$$1=\sum_{h=h_{M} }^1 f_h(\kk)\qquad f_1=1-\chi(\kk)\; .\Eq(2.41)$$

We define a sequence of {\it effective potentials}
$\VV^{(h)}(\psi)$ defined iteratively in the following way; assuming
that we have integrated the scales $h=1,0,-1,-2,...,h+1$
$$e^{-M^2 E_{M}} = \int P_{Z_h,m_h,C_h}(d\psi^{(\le h)}) \, e^{-\VV^{(h)}
(\sqrt{Z_h}\psi^{(\le h)})-M^2 E_h}\;,\quad \VV^{(h)}(0)=0\;,\Eq(2.65)$$
where
$$\eqalign{
&P_{Z_h,m_h,C_h}(d\psi^{(\le h)}) =
\prod_{\kk:C^{-1}_h(\kk)>0}\prod_{\o=\pm1}
{d\hat\psi^{(\le h)+}_{\kk,\o}
d\hat\psi^{(\le h)-}_{\kk',\o}\over \NN(\kk)}
\cdot\cr
&\exp \left\{-{1\over M^2} \sum_{\kk:C^{-1}_h(\kk)>0} \,C_h(\kk) Z_h
\sum_{\o,\o'=\pm1} \hat\psi^{(\le h)+}_{\kk,\o} T^{(h+1)}_{\o,\o'}
\hat\psi^{(\le h)-}_{\kk,\o'}\right\}\;,\cr} \Eq(2.66)$$
$$C_h(\kk)^{-1}=\sum_{j=h_{M}}^h f_j(\kk')\;,\Eq(2.68)$$
and the $2\times2$ matrix $T_{h}(\kk')$ is given by
$${1\over C_0+f_{0,0}(\kk)} 
\left( \matrix{\tilde Z(i\sin k+\sin k_0)+f_{1,1}(\kk)Z_h^{-1} 
& -im_{h-1}(\kk)+f_{1,2}(\kk)Z_h^{-1}\cr
im_{h-1}(\kk)+f_{2,1}(\kk)Z_h^{-1} & 
\tilde Z(i\sin k-\sin k_0)+f_{2,2}(\kk)Z_h^{-1}\cr}\right)\Eq(2.69)$$
with $m_1=C_0(t-t_c)$ and $C_0, \tilde Z$ defined in Theorem 2.

Finally $\VV^{(h)}$ is given by
$$
\VV^{(h)}(\psi^{(\le h)}) = \sum_{n=1}^\io 
\sum^*_{\xx_1,...,\xx_{2n},\atop \ss,\oo,\a}
\prod_{i=1}^{2n}\partial^{\a_i}\psi^{(\le h)\s_i}_{\xx_i,\o_i}
W_{2n,\ss,\oo}^{(h)}(\xx_1,...,\xx_{2n})
\Eq(2.70a)$$

\sub(1.4bzxx) {\it The localization operator}.
We define an $\LL$ operation, for $h\le 0$, in the following way:

\*
1) If $2n=4$, then
$$\LL \hat W_{4,\ss,\oo}^{(h)}(\kk_1,\kk_2,\kk_3)=
\hat W_{4,\ss,\oo}^{(h)}(\bk++,\bk++,\bk++)\;,\Eq(2.72)$$
where
$$\bk\h{\h'} = \left(\h{\p\over M},\h'{\p\over M}\right)\;.\Eq(2.73)$$
\*
2) If $2n=2$ then
$$\LL 
\hat W_{2,\ss,\oo}^{(h)}(\kk)= 
\fra14 \sum_{\h,\h'=\pm 1}
[\hat W_{2,\ss,\oo}^{(h)}(\bk\h{\h'})\;\cdot$$
$$\cdot\;+\hat W_{2,\ss,\oo}^{(h)}(\bk\h{\h'})
(\h {L\over \p}\sin k  +
\h'{\b\over \p} \sin k_0)]\Eq(2.74)$$
\*
3) In all the other cases
$$\LL \hat W_{2n,\ss,\oo}^{h}(\kk_1,\ldots,\kk_{2n-1})=0\;.\Eq(2.77)$$
\*
By \equ(2.72) the operator $\LL$
satisfies the relation $\RR \LL =0$.
\*
{\it Remark.} First note that in the limit $M\to\io$
\equ(2.74) becomes simply
$$\LL 
\hat W_{2,\ss,\oo}^{(h)}(\kk)=
[\hat W_{2,\ss,\oo}^{(h)}({\bf 0})
+k_0\partial_{k_0}\hat W_{2,\ss,\oo}^{(h)}({\bf 0})+
k\partial_{k}\hat W_{2,\ss,\oo}^{a(h)}({\bf 0})],\Eq(2.74xx)$$
hence $\LL 
\hat W_{2,\ss,\oo}^{(h)}(\kk)$ has to be understood
as a discrete version of the Taylor expansion up to order $1$.
Moreover the localization operator acting 
on the effective potential in the $x$-space representation
can be seen as an operator on monomials of Grassmann
variables; for instance \equ(2.72) implies in the
$x$-space representation, in the $M\to\io$ limit
$$\LL
\sum^*_{\xx_1,...,\xx_{4}}
\prod_{i=1}^{4}\psi^{(\le h)\e_i}_{\xx_i,\o_i}
W_{4}^{(h)}(\xx_1,...,\xx_{4})=
\hat W_4({\bf 0},{\bf 0},{\bf 0})\prod_{i=1}^{4}\psi^{(\le 
h)\e_i}_{\xx_1,\o_i}\Eq(2.74mm)$$
where the r.h.s. of the above equation is always vanishing
unless $\prod_{i=1}^{4}\psi^{(\le h)\e_i}_{\xx_1,\o_i}$
is a permutation of 
$\psi^{(\le h)+}_{\xx_1,+}\psi^{(\le h)-}_{\xx_1,+}
\psi^{(\le h)+}_{\xx_1,-}\psi^{(\le h)-}_{\xx_1,-}$;
hence $\LL$ acts on a quartic monomial
producing a {\it local} expression. Analogous
considerations can be done for $n=1$.
\vskip.5cm
\sub(1.4bza) We have, before continuing, 
to exploit the consequences of some symmetries. 
There are no local terms of the form
$\psi^+_{\xx,1}\psi^-_{\xx,1}$;
such local terms can be written as
$\psi^{(1)}_\xx\psi^{(2)}_\xx$,
but the model is invariant under the transformation
$$\psi^{(1)}, 
\bar\psi^{(1)},\chi^{(1)},\bar\chi^{(1)}
\to-\psi^{(1)}, 
-\bar\psi^{(1)},-\chi^{(1)},-\bar\chi^{(1)}$$
$$\psi^{(2)}, 
\bar\psi^{(2)},\chi^{(2)},\bar\chi^{(2)}
\to\psi^{(2)}, 
\bar\psi^{(2)},\chi^{(2)},\bar\chi^{(2)},\Eq(sim1)$$
hence such terms cannot be present 
as they violate such symmetry.

There are no local terms of the form 
$\psi_{1,\xx}\psi_{-1,\xx}$
or $\psi^+_{1,\xx}\psi^+_{-1,\xx}$
(or $\psi_{1,\xx}\partial\psi_{-1,\xx}$, $\psi^+_{1,\xx}\partial\psi^+_{-1,\xx}$)
; in fact,
$$\psi_{1,\xx}\psi_{-1,\xx}={1\over 2}[\psi^{(1)}_\xx\bar\psi^{(1)}_{\xx}-
\psi^{(2)}_\xx\bar\psi^{(2)}_{\xx}+i\psi^{(1)}_\xx\bar\psi^{(2)}_{\xx}+i
\psi^{(2)}_\xx\bar\psi^{(1)}_{\xx}]
\Eq(casss)$$
and the last two terms violates the symmetry
\equ(sim1);
moreover the first two terms
are {\it odd} in the exchange
$(1),(2)
\to (2),(1)$ and
the model
is invariant in the exchange $(1),(2)
\to (2),(1)$.
Moreover the model
is invariant under complex conjugation and
the exchange
$$\psi^{(\a)}_\xx,\bar\psi^{(\a)}_{\xx}
\to\bar\psi^{(\a)}_{\xx},\psi^{(\a)}_{\xx}
\quad\chi^{(\a)}_\xx,\bar\chi^{(\a)}_{\xx}\to
\bar\chi^{(\a)}_{\xx},\chi^{(\a)}_{\xx};\Eq(sim2)$$
this follows from the fact that,
from \equ(c63), $\bar H^{(\a)}, H^{(\a)}, \bar V^{(\a)},
V^{(\a)}$, written in terms of $\bar\psi^{(\a)},\psi^{(\a)},
\bar\chi^{(\a)},\chi^{(\a)}$, are invariant
under such transformation.
Hence the coefficient of the local
part of the quartic (non vanishing) terms is real; in fact
$\hat w(0,0,0)\psi^+_{1,\xx}\psi_{1,\xx}
\psi^+_{-1,\xx}\psi_{-1,\xx}=\hat w(0,0,0)\psi^{(1)}_{\xx}
\bar\psi^{(1)}_{\xx}\psi^{(2)}_{\xx}\bar\psi^{(1)}_{\xx}$
must be equal, by the above invariance,
to $\hat w^*(0,0,0)\bar\psi^{(1)}_{\xx}   
\psi^{(1)}_{\xx}\bar\psi^{(2)}_{\xx}\psi^{(1)}_{\xx}$,
hence $\hat w(0,0,0)=\hat w^*(0,0,0)$.
Finally the combination of local terms
$\psi^+_{\xx,1}\psi^-_{\xx,-1}+
\psi^+_{\xx,-1}\psi^-_{\xx,1}$
is equal to ${1\over 2}[\psi^{(1)}_\xx\bar\psi^{(2)}_\xx-
\psi^{(2)}_\xx\bar\psi^{(1)}_\xx]$
so it cannot be present as it violates
the symmetry
\equ(sim1).
On the other hand
$\psi^+_{\xx,1}\psi^-_{\xx,-1}-
\psi^+_{\xx,-1}\psi^-_{\xx,1}$
is equal to ${1\over 2}[\psi^{(1)}_\xx\bar\psi^{(1)}_\xx+
\psi^{(2)}_\xx
\bar\psi^{(2)}_\xx]$;
hence 
the coefficient of the local
part is imaginary; in fact
$\hat w(0)[\psi^{(1)}_\xx
\bar\psi^{(1)}_\xx+
\psi^{(2)}_\xx\bar\psi^{(2)}_\xx]$
must be equal
to $\hat w^*(0)[\bar\psi^{(1)}_\xx\psi^{(1)}_\xx+
\bar\psi^{(2)}_\xx\bar
\psi^{(2)}_\xx]$,
by the invariance under complex conjugation and \equ(sim2), 
hence $\hat w(0)=-\hat w^*(0)$.
\*
\sub(1.4bzz)
By \equ(2.72),\equ(2.74),\equ(2.77)
and the symmetry relations in \S 4.3,
we can write $\LL \VV^{(h)}$ in the following way:
$$\eqalign{
&\LL\VV^{(h)}(\psi^{(\le h)})=
(s_h+\g^h n_h) F_\s^{(\le h)}-l_h F_\l^{(\le h)}
+z_{h} F_\z^{(\le h)}+\a_h F_\z^{(\le h)}\cr
&+\sum_{i=1}^2\sum_\o z_{i,\o,h} F_i^{(\le h)}
+\sum_{i=1}^2\sum_\o \tilde z_{i,\o,h} \tilde F_i^{(\le h)} 
\;\cr}\Eq(2.79) $$
where, if $|\l|,|\nu|\le\e$,  
$\tilde z_{i,\o,1},z_{i,\o,1}=O(\e)$, $s_1=O(\e)$,
$l_1=2\l{\rm sech}^2 J(a+b)+O(\e^2)$, $\nu_1=\nu+O(\e)$ 
and
$$\eqalign{
F_m^{(\le h)}&=\sum_{\o=\pm 1}
{i\o\over M^2}\sum_{\kk\in {\cal D}_{M}}
\hat\psi^{(\le h)+}_{\kk,\o}
\hat\psi^{(\le h)-}_{\kk,-\o}\;,\cr
F_\l^{(\le h)}&={1\over (M)^4}\sum_{\kk_1,...,\kk_4
\in \DD_{M}}
\hat\psi^{(\le h)+}_{\kk_1,+1}
\hat\psi^{(\le h)-}_{\kk_2,-1} \hat\psi^{(\le h)+}_{\kk_3,-1}
\hat\psi^{(\le h)-}_{\kk_4,+1}\d(\kk_1-\kk_2+\kk_3-\kk_4)\;,\cr
F_{\a}^{(\le h)}&={1\over M}
\sum_{\kk\in {\cal D}_{M}}
i\sin k\hat\psi^{(\le h)+}_{\kk,\o}
\hat\psi^{(\le h)-}_{\kk,\o},\quad
F_{\z}^{(\le h)}={1\over M}
\sum_{\kk\in {\cal D}_{M}}
\o\sin k_0\hat\psi^{(\le h)+}_{\kk,\o}
\hat\psi^{(\le h)-}_{\kk,\o}\;,\cr
\tilde F_i^{\le h}&={1\over M}\sum_{\kk\in {\cal D}_{M}}
\tilde f_i(\kk)\hat\psi^{(\le h)+}_{\kk,\o}
\hat\psi^{(\le h)-}_{\kk,-\o},\quad 
F_i^{\le h}=\sum_{\kk\in {\cal D}_{M}}
f_i\hat\psi^{(\le h)+}_{\kk,\o}
\hat\psi^{(\le h)-}_{\kk,\o}
\; ,\cr
}\Eq(mnnq)$$
with $f_1=\sin k$ and $f_2=\sin k_0$. The constants $n_h, s_h,l_h,z_h,\a_h$ 
are real.
At the end of our iterative
construction it will appear that
it is possible to write (see Remark 1 in \S 4.9)
$$\hat W_{2,\ss,\oo}^{(h)}= 
\hat W_{2,\ss,\oo}^{a(h)}+\hat W_{2,\ss,\oo}^{b(h)}\Eq(num)$$
with $\hat W_{2,\ss,\oo}^{a(h)}$  
vanishing if at least one $m_k=0$, 
for $1\ge k\ge h+1$, and
$\hat W_{2,\ss,\oo}^{b(h)}$ is the rest; we define
$$s_h=\d_{\o,-\o}[
\fra14 \sum_{\h,\h'=\pm 1}
\hat W_{2,\ss,\oo}^{a(h)}(\bk\h{\h'})]\qquad \g^h n_h=\d_{\o,-\o}
[\fra14 \sum_{\h,\h'=\pm 1}
\hat W_{2,\ss,\oo}^{b(h)}(\bk\h{\h'})].\Eq(mmm)$$
In the same way we include in 
$\sum_{i=1}^2\sum_\o z_{i,\o,h} F_i^{(\le h)}$
all the terms not in 
$z_{h} F_\z^{(\le h)}+\a_h F_\z^{(\le h)}$
with $z_h, \a_h$ real.

We {\it renormalize} the free integration
$P_{Z_h,m_h,C_h}(d\psi^{(\le h)})$ 
by adding to it part of the r.h.s. of \equ(2.79). We get
$$\eqalign{
\int P_{Z_h,m_h,C_h}(d\psi^{(\le h)}) &\, e^{-\VV^{(h)}(\sqrt{Z_h}
\psi^{(\le
h)})}=\cr
&=e^{-L\b t_h}\int P_{\tilde Z_{h-1},m_{h-1},C_h}(d\psi^{(\le h)}) \,
e^{-\tilde\VV^{(h)}(\sqrt{Z_h}\psi^{(\le h)})}\;,\cr}\Eq(2.82)$$
where $P_{\tilde Z_{h-1}, m_{h-1},C_h}(d\psi^{(\le h)})$ is obtained from
$P_{Z_h,m_h,C_h}(d\psi^{(\le h)})$ by substituting $Z_h$ with
$$\tilde Z_{h-1}(\kk)=Z_h[1+C_h^{-1}(\kk)\tilde Z^{-1}(C_0+f_{0,0}(\kk))
z_h]\;\Eq(2.84a)$$
and $m_h(\kk)$ with
$$m_{h-1}(\kk)={Z_h\over\tilde Z_{h-1}(\kk)}
[m_h(\kk)+C_h^{-1}(\kk) (C_0+f_{0,0}(\kk))
s_h]\;;\Eq(2.84)$$
moreover $\tilde\VV^{(h)}=\VV^{(h)}-Z_h s_h F_\s^{(\le h)}-Z_h z_h
(F_\z^{(\le h)}+F_\a^{(\le h)})$. 
We will call $m_h(0)\equiv m_h$.
The r.h.s of \equ(2.82) can be written as
$$e^{- M^2t_h} \int P_{Z_{h-1},m_{h-1},C_{h-1}}(d\psi^{(\le h-1)}) \int
P_{Z_{h-1},m_{h-1},f_h^{-1}}(d\psi^{(h)}) \, e^{-\tilde\VV^{(h)}
(\sqrt{Z_h}\psi^{(\le h)})}\; , \Eq(2.88) $$
where the factor $\exp(-M^2 t_h)$ in \equ(2.82) takes into account the different
normalization of the two integrations and
$$Z_{h-1}=Z_h(1+C_0 \tilde Z^{-1} z_h)\;, \qquad \tilde f_h(\kk)=Z_{h-1}[{C_h^{-1}(\kk)
\over \tilde Z_{h-1}(\kk)}-
{C_{h-1}^{-1}(\kk)\over Z_{h-1}}]\;.\Eq(2.89xx)$$
Note that $\tilde f_h(\kk)$ has the same support of $f_h(\kk)$.
The {\it single scale} propagator is 
$$\int P_{Z_{h-1},m_{h-1}, \tilde f_h^{-1}}(d\psi^{(h)})\,
\psi^{(h)-}_{\xx,\o}\psi^{(h)+}_{\yy,\o'} =
{g^{(h)}_{\o,\o'}(\xx-\yy)\over Z_{h-1}}\;,\Eq(2.91)$$
where
$$g^{(h)}_{\o,\o'}(\xx-\yy)={1\over M^2}
\sum_{\kk}e^{-i\kk(\xx-\yy)}
\tilde f_h(\kk)[T_{h}^{-1}(\kk)]_{\o,\o'}\;,\Eq(2.92)$$
and $T_{h}^{-1}(\kk)$ is the inverse of the $T_{h}(\kk)$ defined in
\equ(2.69).
The large distance behaviour of $g^{(h)}_{\o,\o'}(\xx-\yy)$
is given by, if $|z_h|\le {1\over 2}$ and $\sup_{k\ge h}|{Z_{k}\over
Z_{k-1}}|\le e^{|\l|}$,
given the positive integers $N, n_0, n_1$ and putting
$n=n_0+n_1$, and the constant $C_{N,n}$ 
$$
|\dpr_{x_0}^{n_0} \dpr_x^{n_1} g^{(h)}_{\o,\o}(\xx-\yy)|\le 
C_{N,n}
{\g^{h+n} \over 1+(\g^h|\dd(\xx-\yy))|^N}\Eq(2.102a)$$
$$|\dpr_{x_0}^{n_0} \dpr_x^{n_1} g^{(h)}_{\o,-\o}(\xx-\yy)|
\le C_{N,n}
|{m_h\over \g^h}|
{\g^{h+n}\over 1+(\g^h|\dd(\xx-\yy)|)^N}\;.\Eq(2.102)$$
where $\dpr_x$ denotes the discrete derivative.
It will be useful to write
$$g^{(h)}_{\o,\o}(\xx-\yy)=g^{(h)}_{L;\o,\o}(\xx-\yy)
+\tilde g^{(h)}_{\o,\o}(\xx-\yy)+\hat g^{(h)}_{\o,\o}(\xx-\yy)\Eq(dec)$$
with 
$$g^{(h)}_{L;\o,\o}(\xx-\yy)=
{1\over M^2}\sum_{\kk}e^{-i\kk(\xx-\yy)}
{C_0 f_h(\kk)\over -\tilde Z\o\sin k_0+i \tilde Z\sin k}\;,\Eq(2.92aa)$$
obeying to the bound
\equ(2.102a) while
$$
|\dpr_{x_0}^{n_0} \dpr_x^{n_1} \tilde g^{(h)}_{\o,\o}(\xx-\yy)|
\le C_{N,n}
{\g^{{3\over 2}h+n} \over 1+(\g^h|\dd(\xx-\yy))|^N}\Eq(2.102b).$$
$$|\dpr_{x_0}^{n_0} \dpr_x^{n_1} \hat 
g^{(h)}_{\o,\o}(\xx-\yy)|\le C_{N,n}
|{m_h\over \g^h}|^2
{\g^{h+n}\over 1+(\g^h|\dd(\xx-\yy)|)^N}\;.\Eq(2.102z)$$
Moreover
$$g^{(h)}_{\o,-\o}(\xx-\yy)=\hat g^{(h)}_{\o,-\o}(\xx-\yy)
+\tilde g^{(h)}_{\o,-\o}(\xx-\yy)\Eq(dec1)$$
with 
$$\hat g^{(h)}_{\o,-\o}(\xx-\yy)=
{1\over M^2}\sum_{\kk}e^{-i\kk(\xx-\yy)}f_h(\kk)
{-i C_0 m_h(\kk)\over \tilde Z^2\sin^2 k_0+ \tilde Z^2\sin k^2+m_h^2(\kk)}\Eq(ndd)$$
verifying \equ(2.102) and $\tilde 
g^{(h)}_{\o,-\o}(\xx-\yy)$ verifying \equ(2.102b).

We now {\it rescale} the field so that
$$\tilde\VV^{(h)}(\sqrt{Z_h}\psi^{(\le h)})=
\hat\VV^{(h)}(\sqrt{Z_{h-1}}\psi^{(\le h)})\;;\Eq(2.107xxx)$$
it follows that
$$\LL\hat\VV^{(h)}(\psi)= \g^h\nu_h F_\s^{(\le h)}+
\d_h F_\a^{(\le h)}+\l_h F_\l^{(\le h)}+
\sum_{i=1}^2\sum_\o \z_{i,\o,h} F_i^{(\le h)}+
\sum_{i=1}^2\sum_\o \tilde \z_{i,\o,h} \tilde F_i^{(\le h)} \;,\Eq(2.108xxx)$$
where
$$\nu_h ={Z_h\over Z_{h-1}}n_h\;,\quad 
\d_h={Z_h\over Z_{h-1}}(a_h-z_h)
\;,\quad \l_h=({Z_h\over Z_{h-1}})^2 l_h
\;,\Eq(2.109xxxx)$$
and $\tilde \z_{i,\o,h}={Z_h\over Z_{h-1}}\tilde z_{i,\o,h}$,
$\z_{i,\o,h}={Z_h\over Z_{h-1}}z_{i,\o,h}$.
We call 
$v_h=(\l_h,\d_h,\bigcup_{i,\o }\tilde \z_{i,\o,h},\bigcup_{i,\o
}\z_{i,\o,h})$ and the set of $(v_h,m_h,n_h)$
are called {\it running coupling constants}.

If we now define
$$e^{-\VV^{(h-1)}(\sqrt{Z_{h-1}}\psi^{(\le h-1)}) -L\b \tilde E_h}
= \int P_{Z_{h-1},m_{h-1},\tilde f_h^{-1}}(d\psi^{(h)}) \, e^{-\hat\VV^{(h)}
(\sqrt{Z_{h-1}}\psi^{(\le h)})}\;,\Eq(2.110xxx)$$
it is easy to see that $\VV^{(h-1)}(\sqrt{Z_{h-1}}\psi^{(\le h-1)})$ is of
the form \equ(2.70a) and that
$$E_{h-1} = E_h + t_h +\tilde E_h\;.\Eq(2.111)$$
It is sufficient to use the well known identity
$$M^2\tilde E_h + \VV^{(h-1)}(\sqrt{Z_{h-1}}\psi^{(\le h-1)})=
\sum_{n=1}^{\io}{1\over n!}
(-1)^{n+1}\EE^{T,n}_h(\hat\VV^{(h)}(\sqrt{Z_{h-1}}\psi^{(\le h)}))\;,
\Eq(2.112)$$
where $\EE^{T,n}_h$ denotes the {\it truncated expectation of order $n$} with
propagator $Z_{h-1}^{-1}g^{(h)}_{\o,\o'}$, see \equ(2.91), and observe that
$\psi^{(\le h)}= \psi^{(\le h-1)}+\psi^{(h)}$.
\*
Let us define
$$h^*={\rm inf}\{h: 0\ge h\ge h_{M}, 
a_0 \g^{\bh-1}\ge 4| m_{\bh}|,
\forall \bh:0\ge\bh \ge h\}\;.\Eq(2.116)$$
Of course this definition is meaningful only 
for $m_0$ such that
$|m_0|\le {a_0\over 4\g}$.

The integration of the scales from $h^*$ to $h_{M}$ will be performed ``in a
single step''.  This follows from thw following result
\*
\sub(2.12) {\cs Lemma.} {\it Assume that $h^*$ is finite 
uniformly in $M$,
so that $|{m_{h^*-1}\g^{-h^*}}|\ge \bar\k$, for a suitable constant $\bar\k$
and define
$${\bar g^{(\le h^*)}_{\o,\o'}(\xx-\yy)\over Z_{h^*-1}}\equiv \int
P_{Z_{h^*-1},m_{h^*-1},C_{h^*}}(d\psi^{(\le h^*)})
\psi_{\xx,\o}^{(\le h^*)-}\psi_{\yy,\o'}^{(\le h^*)+}\;.\Eq(2.119)$$
Then, given the positive integers $N, n_0, n_1$ and putting
$n=n_0+n_1$, there exist a constant $C_{N,n}$ such that
$$|\dpr_{x_0}^{n_0} \dpr_x^{n_1} g^{(\le h^*)}_{\o,\o'}(\xx;\yy)|
\le C_{N,n} {\g^{h^*+n}\over 1+(\g^{h^*}|\dd(\xx-\yy)|)^N}\;.\Eq(2.120)$$}
\*
\*
\sub(2.12a){\it Remark.} 
Let us now explain the main motivations of the integration
procedure discussed above. In a Renormalization Group
framework one has to identify the relevant, marginal and irrelevant
effective
interactions. By a power counting argument one sees
that the terms bilinear in the fields are {\it relevant}, 
and the terms quartic in the fields or quadratic
with a derivated (in the $x$ space) field 
are marginal. As it was shown in \S 4.3, there
is only one kind of relevant term, and this is a rather
crucial point.
The Renormalization Group flow of the relevant terms
can be controlled by the introduction of a {\it
counterterm}; as we have at our disposal only
one counterterm, it is important to have only one kind
of relevant effective interactions.
The unique relevant effective interaction has the form
$\psi^+_{\xx,1}\psi^-_{\xx,-1}-
\psi^+_{\xx,-1}\psi^-_{\xx,1}$, which can be interpreted as a
{\it mass term} in a fermionic quantum field theory like the
Thirring model. There
is however an important difference:
in fermionic models coming from QFT
if there is no mass term
in the formal free action then no mass terms are 
generated by Renormalization Group iterations, in absence
of spontaneous symmetry breaking; 
the reason is that such models
are invariant under the local gauge transformation 
$\psi^\e_{\xx,\o}\to e^{i\e\a_{\o}}
\psi^\e_{\xx,\o}$ if there is no mass term in the formal free
action.
In our model 
this is not true, as the interaction
is not invariant under this symmetry;
hence even if $t=t_c$ (or $m_1=0$)
a mass term 
can be generated in the RG iterations. Hence
we collect all the relevant terms which are
vanishing if $m_k=0$, $1\ge k\ge h+1$, in 
$s_h$, which we include in the fermionic free integration; 
the "mass"  has a non trivial
flow producing at 
the end the critical index of the correlation length.
The remaining terms are left in the effective interaction;
they are constituting the running coupling constant $\nu_h$
(in a gauge invariant theory $\nu_h=0$)
whose flow 
is controlled by a suitable choice of the counterterm $\nu$;
the running coupling constant $\nu_h$ takes into account
the renormalization of the critical temperature.
The running coupling constant $v_h$ are dimensionally marginal;
again the $\tilde \z_{i,\o,h}, \z_{i,\o,h}$ correspond to effective
interactions
which are absent from the free action, but which
could be possibly generated by the RG iterations, and they
would be absent in a QFT fermionic model.
Finally, due to the mass gap, the propagator of
the integration of
all the scales between $h^*$ and $h_{M}$ has
the same bound as the
propagator
of the integration of a single scale greater than $h^*$; this property is
used to perform the integration of all the scales
$\le h^*$ in a single step.
\*
\sub(2.2cvz) {\cs Theorem 3.} {\it Let $h> h^*\ge 0$
and, for some constants
$c_1$, if
$$\max_{k\ge h}[|v_k|+|\nu_k|]\le\e_h,\quad 
\sup_{h'>h}\Big|{m_{h'}\over m_{h'-1}}\Big|\le e^{c_1\e_h},\quad
\sup_{h'>h}\Big|{Z_{h'}\over Z_{h'-1}}\Big|\le e^{c_1\e^2_h}
\Eq(3.88oo)$$
there exists a constant $\bar\e$ (depending on $c_1$) such that,
if $\e_h\le \bar\e$,
then, for a suitable constant $c_0$, independent of $c_1$, 
as well as of $M$, the kernels in \equ(2.70a) verify
$$\sum^*_{\xx_1,...,\xx_{2n}}|W^{(h)}_n(\xx_1,...,\xx_{2n})|
\le M^2 \g^{-hD_k(P_{v_0})} (c_0\e_h)^{max(1,n-1)}\Eq(3.89oo)$$
where
$$D_k(P_{v_0})=-2+n+k\;.\Eq(3.90)$$
and $k=\sum_{i=1}^{2n}\a_i$.}
\*
\*
\sub(2.2cvz) {\it Proof.}
We write $\VV^{(h)}$ 
in terms of a {\it tree expansion}, similar to that described, for
example, in [BM].
 
\insertplot{300pt}{150pt}%
{\ins{30pt}{85pt}{$r$}\ins{50pt}{85pt}{$v_0$}\ins{130pt}{100pt}{$v$}%
\ins{35pt}{-2pt}{$h$}\ins{55pt}{-2pt}{$h+1$}\ins{135pt}{-2pt}{$h_v$}%
\ins{215pt}{-2pt}{$0$}\ins{235pt}{-2pt}{$+1$}\ins{255pt}{-2pt}{$+2$}}%
{fig51}{\eqg(1)}

\vglue.5truecm


We need some definitions and notations.
 
\0 1) Let us consider the family of all trees which can be constructed
by joining a point $r$, the {\it root}, with an ordered set of $n\ge 1$
points, the {\it endpoints} of the {\it unlabeled tree}, 
so that $r$ is not a branching point. $n$ will be called the
{\it order} of the unlabeled tree and the branching points will be called
the {\it non trivial vertices}.
The unlabeled trees are partially ordered from the root to the endpoints in
the natural way; we shall use the symbol $<$ to denote the partial order. 
Two unlabeled trees are identified if they can be superposed by a suitable
continuous deformation, so that the endpoints with the same index coincide.
It is then easy to see that the number of unlabeled trees with $n$ end-points
is bounded by $4^n$. 
We shall consider also the {\it labeled trees} (to be called simply trees in
the following); they are defined by associating some labels with the unlabeled
trees, as explained in the following items.
 
\0 2) We associate a label $h\le 0$ with the root and we denote $\TT_{h,n}$ the
corresponding set of labeled trees with $n$ endpoints. Moreover, we introduce
a family of vertical lines, labeled 
by an an integer taking values in
$[h,2]$, and we represent any 
tree $\t\in\TT_{h,n}$ so that, if $v$ is an
endpoint or a non trivial vertex, it is contained in a vertical line with
index $h_v>h$, to be called the {\it scale} of $v$, while the root is on the
line with index $h$. There is the constraint that, if $v$ is an endpoint,
$h_v>h+1$; if there is only one end-point its scale must
be equal to $h+2$,
for $h\le 0$.
 
The tree will intersect in general the vertical lines in set of
points different from the root, the endpoints and the non trivial vertices;
these points will be called {\it trivial vertices}. The set of the {\it
vertices} of $\t$ will be the union of the endpoints, the trivial vertices
and the non trivial vertices.
Note that, if $v_1$ and $v_2$ are two vertices and $v_1<v_2$, then
$h_{v_1}<h_{v_2}$.
 
Moreover, there is only one vertex immediately following
the root, which will be denoted $v_0$ and can not be an endpoint;
its scale is $h+1$.
 
\0 3) With each endpoint $v$ of scale $h_v=+2$ we associate one of the 
contributions to $\VV^{(1)}$ given by \equ(bnv);
with each endpoint $v$ of
scale $h_v\le 1$ one of the terms in 
$\LL V^{(h_v-1)}$ defined in \equ(2.108xxx).
Moreover, we impose the constraint that, if $v$ is an endpoint and 
$h_v\le 1$,
$h_v=h_{v'}+1$, if $v'$ is the non trivial vertex immediately preceding $v$.
 
\0 4) If $v$ is not an endpoint, the {\it cluster } $L_v$ with frequency $h_v$
is the set of endpoints following the vertex $v$; if $v$ is an endpoint, it is
itself a ({\it trivial}) cluster. The tree provides an organization of
endpoints into a hierarchy of clusters.
 
\0 5) We introduce a {\it field label} $f$ to distinguish the field variables
appearing in the terms associated with the endpoints as in item 3);
the set of field labels associated with the endpoint $v$ will be called $I_v$.
Analogously, if $v$ is not an endpoint, we shall
call $I_v$ the set of field labels associated with the endpoints following
the vertex $v$; $\xx(f)$, $\e(f)$ and $\o(f)$ will denote the space-time
point, the $\e$ index and the $\o$ index, respectively, of the
field variable with label $f$.

If $h\le 0$, the effective potential can be written in the
following way:
$$\VV^{(h)}(\sqrt{Z_h}\psi^{(\le h)}) + M^2 \tilde E_{h+1}=
\sum_{n=1}^\io\sum_{\t\in\TT_{h,n}} \VV^{(h)}(\t,\sqrt{Z_h}\psi^{(\le
h)})\;,\Eq(3.29oo)$$
where, if $v_0$ is the first vertex of $\t$ and $\t_1,..,\t_s$
($s=s_{v_0}$) are the subtrees of $\t$ with root $v_0$,\\
$\VV^{(h)}(\t,\sqrt{Z_h}\psi^{(\le h)})$ is defined inductively by the
relation
$$\eqalign{
&\qquad \VV^{(h)}(\t,\sqrt{Z_h}\psi^{(\le h)})=\cr &{(-1)^{s+1}\over s!}
\EE^T_{h+1}[ \bar\VV^{(h+1)}(\t_1,\sqrt{Z_h}\psi^{(\le h+1)});\ldots;
\bar\VV^{(h+1)}(\t_{s},\sqrt{Z_h}\psi^{(\le h+1)})]\;,\cr}\Eq(3.30oo)$$
and $\bar\VV^{(h+1)}(\t_i,\sqrt{Z_h}\psi^{(\le h+1)})$

\0 a) is equal to $\RR\VV^{(h+1)}(\t_i,\sqrt{Z_h}\psi^{(\le h+1)})$ if the
subtree $\t_i$ is not trivial;

\0 b) if $\t_i$ is trivial and $h\le -1$, it is equal to one of
the terms in $\LL\VV^{(h+1)}(\sqrt{Z_h}\psi^{(\le h+1)})$ or, if
$h=0$, to one of the terms contributing to
$\VV^{(1)}(\psi^{\le 1})$ \equ(bnv).

In \equ(3.30oo) $\EE^T_{h+1}$ denotes the truncated expectation with
respect to the measure $\prod_{\o} P(d\psi_\o^{(h+1)})$.
The r.h.s. of \equ(3.30oo) can be written more
explicitly in the following way. Given $\t\in\TT_{0,n}$, there are
$n$ endpoints of scale $2$ 
and only another one vertex, $v_0$, of
scale $1$; let us call $v_1,\ldots, v_n$ the endpoints. We choose,
in any set $I_{v_i}$, a subset $Q_{v_i}$ and we define
$P_{v_0}=\cup_i Q_{v_i}$. We have
$$\VV^{(0)}(\t, \psi^{(\le 0)})=\sum_{P_{v_0}}
\VV^{(0)}(\t,P_{v_0})\;,\Eq(3.31aoo)$$
$$\VV^{(0)}(\t,P_{v_0})= \int d\xx_{v_0}
\tilde\psi^{\le 0}(P_{v_0})
K_{\t,P_{v_0}}^{(1)}(\xx_{v_0})\;, \Eq(3.31boo)$$
$$K_{\t,P_{v_0}}^{(1)}(\xx_{v_0})={1\over n!} \EE^T_{1}[
\tilde\psi^{(1)}(P_{v_1}\bs Q_{v_1}),\ldots,
\tilde\psi^{(1)}(P_{v_n}\bs Q_{v_n})] \prod_{i=1}^n
K^{(2)}_{v_i}(\xx_{v_i})\;,\Eq(3.31coo)$$
where we use the definitions 
$\tilde\psi^{(\le h)}(P_v) =\prod_{f\in P_v} \partial^{m(f)}\psi^{(\le
h)\e(f)}_{\xx(f),\o(f)}
$, and
$K^{(2)}_{v_i}(\xx_{v_i})$ are the kernels in \equ(bnv).
We now write $\VV^{(0)}$ as $\LL \VV^{(0)}+\RR \VV^{(0)}$, with
$\LL \VV^{(0)}$ defined as in \equ(2.79), 
and we write for $\RR\VV^{(0)}$ a
decomposition similar to the previous one, with
$\RR\VV^{(0)}(\t,P_{v_0})$ in place of
$\VV^{(0)}(\t,P_{v_0})$.
By iterating the previous procedure,
one gets for $\VV^{(h)}(\t,\psi^{(\le h)})$, for any
$\t\in\TT_{h,n}$, the representation described below (see [GM]
and [BM]
for details).

We associate with any vertex $v$ of the tree a subset $P_v$ of
$I_v$, the {\it external fields} of $v$. These subsets must
satisfy various constraints. First of all, if $v$ is not an
endpoint and $v_1,\ldots,v_{s_v}$ are the vertices immediately
following it, then $P_v \subset \cup_i P_{v_i}$; if $v$ is an
endpoint, $P_v=I_v$. We shall denote $Q_{v_i}$ the intersection of
$P_v$ and $P_{v_i}$; this definition implies that $P_v=\cup_i
Q_{v_i}$. The subsets $P_{v_i}\bs Q_{v_i}$, whose union ${\cal
I}_v$ will be made, by definition, of the {\it internal fields} of
$v$, have to be non empty, if $s_v>1$.
Given $\t\in\TT_{h,n}$, there are many possible choices of the
subsets $P_v$, $v\in\t$, compatible with all the constraints; we
shall denote $\PP_\t$ the family of all these choices and $\bP$
the elements of $\PP_\t$.
We can write
$$\VV^{(h)}(\t,\psi^{(\le h)})=\sum_{\bP\in\PP_\t}
\VV^{(h)}(\t,\bP)\;.\Eq(3.32)$$
$\VV^{(h)}(\t,\bP)$ can be represented as
$$\VV^{(h)}(\t,\bP)=\sqrt{Z_h}^{|P_{v_0}|}\int d\xx_{v_0} \tilde\psi^{(\le h)}
(P_{v_0}) K_{\t,\bP}^{(h+1)}(\xx_{v_0})\;,\Eq(3.33)$$
with $K_{\t,\bP}^{(h+1)}(\xx_{v_0})$ defined inductively
(recall that $h_{v_0}=h+1$) by the equation, valid for any
$v\in\t$ which is not an endpoint,
$$K_{\t,\bP}^{(h_v)}(\xx_v)={1\over s_v !}({Z_{h_v}\over Z_{h_v-1}})^{|P_v|\over 2}
\prod_{i=1}^{s_v} [K^{(h_v+1)}_{v_i}(\xx_{v_i})]\; \;\EE^T_{h_v}[
\tilde\psi^{(h_v)}(P_{v_1}\bs Q_{v_1}),\ldots,
\tilde\psi^{(h_v)}(P_{v_{s_v}}\bs
Q_{v_{s_v}})]\;,\Eq(3.35)$$
Moreover, if $v$ is an endpoint, $K^{(2)}_{v}(\xx_{v})$ is defined
as in \equ(bnv) if $h_v=2$, otherwise
$$K^{(h_v)}_{v}(\xx_{v})= \cases{
\l_{h_v-1}\d(\xx_1-\xx_2)\d(\xx_2-\xx_3)\d(\xx_3-\xx_4)
& if $v$ is of type $\l$,\cr
\g^{h_v-1}\n_{h_v-1}\d(\xx-\yy) & if $v$ is of
type $\n$,\cr 
\d_{h_v-1}\d(\xx-\yy) & if $v$ is of
type $\d$,\cr 
\z_{h_v-1,\o,i}
\d(\xx-\yy),\quad 
\tilde \z_{h_v-1,\o,i}
\d(\xx-\yy) & if 
$v$ is of
type $z$,$\tilde z$}\Eq(3.37)$$
If $v_i$ is not an
endpoint,
$$K^{(h_v+1)}_{v_i}(\xx_{v_i}) = 
\RR K_{\t_i,\bP^{(i)}}^{(h_v+1)}
(\xx_{v_i})\;,\Eq(3.37a)$$
where $\t_i$ is the subtree of $\t$ starting from $v$ and passing
through $v_i$ (hence with root the vertex immediately preceding
$v$), $\bP^{(i)}$ are the restrictions to $\t_i$ of
$\bP$. 

\equ(3.32) is not the final form of our expansion, since we
further decompose $\VV^{(h)}(\t,\bP)$, by using the 
representation of the truncated expectation \equ(3.38).
If we apply the expansion \equ(3.38) in each non trivial vertex of
$\t$, we get an expression of the form
$$\VV^{(h)}(\t,\bP) = \sqrt{Z_h}^{|P_{v_0}|}\sum_{T\in {\bf T}} \int d\xx_{v_0}
\tilde\psi^{(\le h)}(P_{v_0}) W_{\t,\bP,T}^{(h)}(\xx_{v_0})\;,\Eq(3.38a)$$
where ${\bf T}$ is a special family of graphs on the set of points
$\xx_{v_0}$, obtained by putting together an anchored tree graph
$T_v$ for each non trivial vertex $v$. Note that any graph $T\in
{\bf T}$ becomes a tree graph on $\xx_{v_0}$, if one identifies
all the points in the sets $x_v$, for any vertex $v$ which is also
an endpoint.

Taking into account the effect of the $\RR$ operation
we obtain (see [BM], \S 3 for a detailed 
proof of a similar formula)
$$\eqalign{&
W_{\t,\bP, T}(\xx_{v_0})=\sum_{\a\in A_T}  \left[\prod_{i=1}^n
K_{v_i^*}^{h_i}\right]
\Bigg\{\prod_{v\,\atop\hbox{\ottorm not e.p.}}{1\over s_v!} \int
dP_{T_v}(\tt_v)\det G_\a^{h_v,T_v}(\tt_v)\cdot\cr
&[\prod_{v\,\atop\hbox{\ottorm not e.p.}}({Z_{h_v}\over
Z_{h_v-1}})^{|P_v|\over2}]\cdot\Big[\prod_{l\in T_v}
\partial^{q_\a(f^-_l)}
\partial^{q_\a(f^+_l)} [(\xx_{l}- \yy_{l})^{b_\a(l)}
\dpr^{m_l} \tilde
g^{(h_v)}_{\o^-_l,\o^+_l}(\xx_l-\yy_l)]\Big]\Bigg\}\;,\cr}\Eq(3.40)$$
where ``e.p.'' is an abbreviation of ``endpoint'' and, together
with the definitions used before, we are using the following ones:

\item{1)} $A_T$ is a set of indices which allows to distinguish the different
terms produced by the non trivial $\RR$ operations and the
iterative decomposition of the zeros;

\item{2)} $v^*_1,\ldots,v^*_n$ are the endpoints of $\t$ and
$h_i=h_{v_i^*}$;

\item{3)} $b_\a(l)$, $q_\a(f^-_l)$ and $q_\a(f^+_l)$ are positive
integers $\le 3$; $\hat\dpr^0=I$;

\item{4)} if $v$ is a non trivial vertex (so that $s_v>1$),the elements
$G^{h_v,T_v}_{\a, ij,i'j'}$ of $G_\a^{h_v,T_v}(\tt_v)$ are of the
form
$$G^{h_v,T_v}_{\a, ij,i'j'}=t_{i,i'}
\hat\partial^{q_\a(f^-_{ij})}
\hat\partial^{q_\a(f^+_{i'j'})}\dpr^{m(f_l^-)}
\dpr^{m(f_l^-)} \tilde
g^{(h_v)}_{\o^-,\o^+}(\xx_{ij}-\yy_{i'j'})\;;\Eq(3.40a)$$
if $v$ is trivial, $T_v$ is empty and $\int dP_{T_v}(\tt_v)\det
G_\a^{h_v,T_v}(\tt_v)$ has to be interpreted as $1$, if $|{\cal
I}_v|=0$ (${\cal I}_v$ is the set of internal fields of $v$),
otherwise it is the determinant of a matrix of the form
\equ(3.40a) with $t_{i,i'}=1$.

It would be very difficult to give a precise description of the various
contributions to the sum over $A_T$, but fortunately we only need to know
some very general properties, in particular that $|A_T|\le C^n$
for some constant $C$ and that
for any $\a\in A_T$, the following inequality is satisfied
$$\Big[\prod_{f\in I_{v_0}} \g^{h_\a(f) q_\a(f)} \Big]
\Big[\prod_{l\in T} \g^{-h_\a(l) b_\a(l)} \Big] \le
\prod_{v\,\hbox{\ottorm not e.p.}} \g^{-z(P_v)}\;,\Eq(3.83ooo)$$
where $h_\a(f)=h_{v_0}-1$ if $f\in P_{v_0}$, otherwise it is the scale of
the vertex where the field with label $f$ is contracted;
$h_\a(l)=h_v$, if
$l\in T_v$ and
$$z(P_v)=\cases{
1 & if $|P_v|=4$,\cr
2 & if $|P_v|=2$,\cr
0 & otherwise.\cr
}\Eq(3.84)$$

By a standard computation (see for instance [BM], \S 3)
by bounding the determinant by
the Gram-Hadamard inequality (see \equ(3.49)) we obtain
$$\eqalign{
&\int d\xx_{v_0} |W_{\t,\bP,{\bf T}}(\xx_{v_0})|\le
C^n M^2\e_h^n \g^{-h D_k(P_{v_0})}\;\cdot\cr
&\cdot\; \prod_{v\,\hbox{\ottorm not e.p.}} \left\{ {1\over s_v!}
C^{\sum_{i=1}^{s_v}|P_{v_i}|-|P_v|}({Z_{h_v}\over
Z_{h_v-1}})^{|P_v|\over2}
\g^{-[-2+{|P_v|\over 2}+z(P_v)]}\right\},\cr}\Eq(3.105)$$
with $-2+{|P_v|\over 2}+z(P_v)>0$.
%
In order to perform the sums note that
the number of unlabeled trees is $\le
4^n$; fixed an unlabeled tree, the number of terms in the sum over the
various labels of the tree is bounded by $C^n$, except the sums over the scale
labels. 
In order to bound the sums over the scale labels and $\bP$ we first use
the inequality
$$\eqalign{
&\prod_{v\,\hbox{\ottorm not e.p.}}
\g^{-[-2+{|P_v|\over 2}+z(P_v)]}\le\cr
&\le [\prod_{\tilde v} \g^{-2\a(h_{\tilde v}-h_{\tilde v'})}]
[\prod_{v\,\hbox{\ottorm not e.p.}}\g^{-2\a|P_v|}],\cr}\Eq(3.111)$$
where $\tilde v$ are the non trivial vertices, and $\tilde v'$ is the
non trivial vertex immediately preceding $\tilde v$ or the root. The
factors $\g^{-2\a(h_{\tilde v}-h_{\tilde v'})}$ in the r.h.s.
of \equ(3.111) allow to bound the sums over the scale labels by $C^n$;
$\a$ is a suitable constant (one finds $\a={1\over 40}$).
 
Finally the sum over $\bP$ can be bounded by using the following combinatorial
inequality, trivial for $\g$ large enough. 
Let $\{p_v, v\in \t\}$ a set of integers such that
$p_v\le \sum_{i=1}^{s_v} p_{v_i}$ for all $v\in\t$ which are not endpoints;
then
$$\prod_{v\,\hbox{\ottorm not e.p.}} \sum_{p_v} \g^{-{p_v\over 40}}
\le C^n\;.\Eq(3.112)$$
It follows that
$$\sum_{\bP\atop |P_{v_0}|=2m}\prod_{v\,\hbox{\ottorm not e.p.}}
\g^{-{|P_v|\over 40}}\le \prod_{v\,\hbox{\ottorm not e.p.}} \sum_{p_v}
\g^{-{p_v\over 40}} \le C^n\;.\Eq(3.113)$$
\*
\sub(2.2cvz2){\it Remark 1.} The decomposition in \equ(num)
respects the determinant structure of the truncated
expectations, as
we can decompose the propagators as in \equ(dec),\equ(dec1),
obtaining, for any tree with $n$ end-points,
a family of $C^n$ different contributions to $\hat W_2^{(h)}$,
which can be bounded as before. 
Hence we can include in 
$\hat W^{(a)}$ the terms with at least a propagator
$\hat g^{(k)}_{\o,-\o}$, defined in
\equ(ndd).

{\it Remark 2.} If the tree $\t$ has an end-point on scale
$k$, the bound \equ(3.89oo) can be improved by a factor $\g^{\a(h-k)}$,
as it follows immediately from \equ(3.111). In particular
if to an end-point is associated a term $\RR V^{(1)}$
there is an extra factor $\g^{\a h}$; this is a {\it short memory}
property.
\*
\sub(2.2cvz1) The above results 
were proved for $Z^{-,-,-,-}_{2I}$;
a similar analysis can be repeated 
for $Z_{2I}^{\e^{(1)},\e^{'(1)},
\e^{(2)},\e^{'(2)}}$, for any value of
$\e^{(1)},\e^{'(1)}, \e^{(2)},\e^{'(2)}$.
The only difference
is that one has in addition the function $\tilde Q_\e$ 
in the interaction and the oscillating functions
$e^{i\pp_M\xx}$. 
One can split $\VV^1$ in a part identical
to the one for $Z_{-,-,-,-}$, called $\bar \VV^{(1)}$, 
and the rest; hence we repeat the multiscale 
analysis by writing $\VV^{(h)}=
\bar\VV^{(h)}+\VV^{'(h)}$,
with $\VV^{'(h)}$ 
given by a sum of trees with at least an
end-point associated to $\VV^{(1)}-\bar \VV^{(1)}$;
we
define the localization operators
acting non trivially
only on $\bar\VV^{(h)}$ (and defined as above).
It is easy to see that the terms from trees with at least
one end-point 
associated to
$\VV^{(1)}-\bar\VV^{(1)}$ are vanishing in the limit $M\to\io$;
in fact the bounds for such terms
is improved by the factor 
${\g^{-h^*}\over M}$, by simply dimensional considerations, and 
we will see in the following section
that $\g^{-h^*}$ is a finite number independent
from $M$, hence such terms are vanishing in the limit $M\to\io$.
\vskip.5cm
\section(2z,The flow of the running coupling constants)
\vskip.5cm
\*
\sub(5.1)By the analysis of the preceding section
it follows that the running coupling constants $v_k,\nu_k,m_k$, $1\ge k\ge h^*$,
verify a set of recursive equations
called {\it Beta function equations}, of the form

$$\nu_{h-1}=\g\nu_h+\b^h_\nu(v_h,\nu_h;...;v_1,\nu_1)$$
$$v_{h-1}=v_h+\b^h_v(v_h,\nu_h;...;v_1,\nu_1)\Eq(gg1)$$
$${m_{h-1}\over m_h}=1+\b_m^h(v_h,\nu_h;...;v_1,\nu_1)$$
$${Z_{h-1}\over Z_h}=1+\b_z^h(v_h,\nu_h;...;v_1,\nu_1)$$
By repeating the analysis 
for proving Theorem 3 to the functions
$\b^h_i$ we have that, if \equ(3.88oo) holds,
then the $\b^h_i$ are expressed by convergent series.
We want to show that there exists a positive constant $a_3$ 
such that, if $|t-t_c|\ge e^{-{1\over a_3\l^2}}$, it is possible to find a
function $\n_1$ (hence a function $\nu$)
so that \equ(3.88oo) holds
(if $\l$ is small enough).
Iterating the first of \equ(gg1) we find
$$ \n_h = \g^{-h+1} \left[ \n_1 + \sum_{k=h+1}^1 \g^{k-2}
\b^k_\n(\n_k,\ldots,\n_0)\right] \;,\Eq(5.27)$$
where now the functions $\b_\nu^{k}$ are thought as functions of $\n_k,\ldots,
\n_1$ only.

If we put $h=h^*$ in \equ(5.27), we get the following identity:
$$ \n_{h^*} = \g^{-h^*+1} \left[ \n_1 + \sum_{k=h^*+1}^1 \g^{k-2}
\b^k_\n(\n_k,\ldots,\n_0)\right] \;,\Eq(5.27a)$$

and we look for a $\nu_1$ verifying
$$\n_1 = - \sum_{k=h^*+1}^1\g^{k-2}\b_\n^k(\n_k,\ldots,\n_1)\;.\Eq(5.28)$$
so that
$$\n_h  = -\g^{-h}\sum_{k=h^*+1}^h \g^{k-1} 
\b_\n^k(\n_k,\ldots,\n_1)
\;,\qquad h^*< h\le 1\;.\Eq(5.29)$$

Let be $\tilde\nu\equiv\{ \tilde\nu_k, h^*\le h\le 1\}$ 
and $||\tilde\nu||=\sup_{h^*\le h\le 1}|\tilde\nu_h|$;
we call $M$ the set of $\tilde\nu$ with bounded norm.
We decouple the beta function equations
\equ(gg1) imagining that in the
last three
equations of \equ(gg1) $\nu_k$ is replaced by $\tilde\nu_k$
acting as a parameter; we call $m_h(\tilde\nu), v_k(\tilde \n)$
the solution of the second and third of \equ(gg1) as functions of
the parameter $\tilde\nu$. We shall prove the following lemma.

\*
\sub(5.5) 
{\cs Lemma.} 
{\it There exists $\e$ and $a_3$ 
such that, for $|t-t_c|\ge e^{-1\over a_3\l_1^2}$, $|\l|\le\e$
and any $\tilde\nu$ such that $||\tilde\n||, ||\tilde\nu'||
\le C_\nu|\l|$, it holds, if $m_0=|t-t_c|$ and $C,C_\n$
are constants
$$|\l_h(\tilde\nu)-\l_1|\le {|\l_1|\over 4}
\qquad |\l_h(\tilde\nu)-\l_h(\tilde\n')|\le C |\l| \max_{k> h}|\tilde\n_k-\tilde\nu'_k|
\Eq(gg3)$$
$$|\d_h(\tilde\nu)|\le {|\l_1|\over 4}
\qquad |\d_h(\tilde\nu)-\d_h(\tilde\n')|\le C |\l|
\max_{k> h}|\tilde\n_k-\tilde\nu'_k|
$$
$$|\tilde \z_{i,\o,h}(\tilde\nu)|\le |\tilde \z_{0,\o,h}|+|\l|
\quad|\tilde \z_{i,\o,h}(\tilde\nu)
-\tilde \z_{i,\o,h}(\tilde\nu')|\le C |\l|  
\max_{k> h}|\tilde\n_k-\tilde\nu'_k|$$
$$|\z_{i,\o,h}(\tilde\nu)|\le |\z_{0,\o,h}|+|\l|
\quad|\z_{i,\o,h}(\tilde\nu)
-\z_{i,\o,h}(\tilde\nu')|\le C |\l| 
\max_{k> h}|\tilde\n_k-\tilde\nu'_k|$$
$$|m_0|\g^{c_1 a_2 \l_1h}\le |m_h(\tilde\nu)|\le|m_0|\g^{c_2 a_2 \l_1h}
\qquad|{m_h(\tilde\nu)\over m_h(\tilde\nu')}-1|\le  
\max_{k> h}|\tilde\n_k-\tilde\nu'_k||\l|^{-1}
$$
$$\g^{-c_3 a_4 \l_1^2 h}\le |Z_h(\tilde\nu)|\le \g^{-c_4 a_4 \l^2_1h}
\qquad|{Z_{h-1}(\tilde\nu)\over Z_{h}(\tilde\nu)}-{Z_{h-1}(\tilde\nu')\over Z_{h}(\tilde\nu')}|\le 
\max_{j> h}|\tilde\n_k-\tilde\nu'_k|
$$
with $a_2,a_4, c_1,c_2, c_3,c_4$ positive constants.

%
}
\*
{\it Remark.} An obvious corollary of the above statement
is that there exits a finite (uniformly in $M$) $h^*$ such that
$m_{h^*-1}\g^{h^*-1}\ge\k$
for a suitable constant $\k$.
\*
\sub(5.11) {\it Proof.} 
The proof is done by induction. 
By iterating
the second equation of \equ(gg1) we find
$$\l_{h-1}(\tilde\n)-\l_1=\sum_{k=h}^1\b_\l^{k}(v_h(\tilde\n),
\tilde\nu_h;...;
v_1(\tilde\n),\tilde\nu_1)\Eq(gg5)$$
and by induction 
$$|\b_\l^{k}(v_h(\tilde\n),\tilde\nu_h;...;
v_1(\tilde\nu),\tilde\nu_1)|\le C_1
[|\l_1|^3+\l_1^2{|m_h|\over\g^h}+\l_1^2\g^{\a h}].\Eq(gg6)$$
The first addend of the r.h.s. of \equ(gg6)
is a bound on the terms containing only
propagators $g^{(h)}_{L,\o}(\xx-\yy)$,
see \equ(2.92aa), and trees with
end-points $v$ with $h_v\le 1$, and the bound follows
from the fact that the second order terms cancels out.
The second addend is a bound on the terms containing at least a
propagator $\hat g^{(h)}_{\o_1,\o_2}(\xx-\yy)$
\equ(ndd),\equ(2.102z) (the bound follows from the short
memory property and \equ(2.102), \equ(2.102z)).
The third addend is a bound on the sum of terms with
at least a propagator $\tilde g^{(h)}_{\o_1,\o_2}(\xx-\yy)$
or from trees 
with at least an
end-point $v$ with $h_v=2$, and we have used \equ(2.102b).

Inserting \equ(gg6)
in \equ(gg5) we get
$$|\l_{h-1}(\tilde\n)-\l_1|\le C_1 
[|h||\l_1|^3+\l_1^2\sum_{k=h}^0
[{|m_k|\over\g^k}+\g^{\a k}]|\le {|\l_1|\over 4}\Eq(gg7)$$
for $a_3$ large enough and $\l$ small enough, 
where we have used that 
$\sum_{k=h}^0{|m_k|\over\g^k}$ is bounded (by induction and the 
definition of $h^*$). 

Moreover the expansion for
$\l_{h-1}(\tilde\nu)-\l_{h-1}(\tilde\nu')$
is given by a sum of terms similar to the ones for
$\l_{h-1}(\tilde\nu)$ in which a $v_k$
is replaced by $v_k(\tilde\nu)-v_k(\tilde\nu')$
bounded by $C\l^2\max_{j>k}|\tilde\nu_j-\tilde\nu'_j|$
or $m_k$ or $Z_k$ replaced by 
the relative difference;
one finds, for $a_3$ large and $\l$ small enough 
$$|\l_{h-1}(\tilde\nu)-\l_{h-1}(\tilde\nu')|\le C_2 
[|h| |\l|^3$$
$$+|\l|\sum_{k=h}^0
[{|m_k|\over\g^k}|+\g^{\a k}
]]\max_{k> h}|\tilde\n_k-\tilde\nu'_k|\le 
C|\l|\max_{k> h}|\tilde\n_k-\tilde\nu'_k|,\Eq(papll)$$
where the first term 
is a bound on the terms containing only
propagators $g^{(h)}_{L,\o}(\xx-\yy)$, and trees with
end-points $v$ with $h_v\le 1$,
containing at least a 
$v_k(\tilde\nu)-v_k(\tilde\nu')$
or a $\tilde\nu_k-\tilde\nu'_k$ (note also
that the first term with end-point $\nu$ is $\l_h^2\nu_h^2$).
A similar analysis can be repeated for $\d_h$.

The recursive equation for $\tilde \z_{i,\o,h}$ is
$$\tilde \z_{i,\o,h-1}(\tilde\nu)
=\tilde \z_{i,\o,h}(\tilde\nu)+
\tilde \z_{i,\o,h}(\l^2_h a_{\tilde z}+\b_z)+
R^{(h)}_{\tilde z}\Eq(lam3)$$
with $|\b_z^h|\le C_3|\l|^3$ and 
$|R_{\tilde z}^h|\le C_4[\g^{\a h}+{|m_h|\over\g^{h}}]\l^2$.
In $\l^2_h(a_2+\b_z)$ in \equ(lam3) are the contributions
from terms with only diagonal propagators $g_{L,\o}^{(h)}(\xx)$
and from trees with endpoints $v$ with $h_v\le 1$;
there is
necessarily at least a vertex $\tilde \z$.
In fact assume that a contribution has $n_\l$ end-points of type $\l$
and $n$ end-points of type $\nu$; there are then
$n_\l-{1\over 2}+{n\over 2}$ propagators
$\o=1$ and $n_\l-{1\over 2}+{n\over 2}$
propagators $\o=-1$, so that $n$ must be {\it odd};
the total number of propagators
$2n_\l-1+n$ is an {\it even} number;
the derivated integrand is odd under the exchange $\kk\to-\kk$ 
and so it is vanishing.
Moreover there is no contribution of order $\l_h \tilde z_h$
(they corresponds to {\it tadpole} graphs
, whose derivative is zero). In $R^{(h)}_{\tilde z}$ are
the trees with at least an end 
point $v$ with $h_v=2$, or with a propagator $\hat g_{\o_1,\o_2}^{(k)}$,
or with a propagator $\tilde g_{\o_1,\o_2}^{(k)}$.
It holds by iterating, for $a_3$ large and $\l$ small enough,  
$|\tilde \z_{i,\o, h}|\le |\tilde z_{i,\o,0}|+|\l|$. Moreover
$$|\tilde \z_{i,\o,h-1}(\tilde\nu)-
\tilde \z_{i,\o,h-1}(\tilde\nu')|\le C_3\max_{j> h}
|\tilde\n_k-\tilde\nu'_k|
[|\l|^3 |h|+|\l|\sum_{k=h}^0
[{|m_k|\over\g^k}+\g^{\a k}
]]\le$$
$$C|\l|\max_{j> h}|\tilde\n_k-\tilde\nu'_k|\Eq(gg24)$$
Similar computations can be repeated for 
$\z_{i,\o,h}$.

By definition (see \equ(num))
$m_{h-1}(\tilde\nu)$ is given by a sum over 
terms with at least a non diagonal propagator $\hat
g^{(h)}_{\o,-\o}(\xx-\yy)$,
hence
$${m_{h-1}(\tilde\nu)\over m_h(\tilde\nu)}=1+\l_h(-a_2+h_m)\Eq(aaasz)$$
where $a_2>0$ is a constant and $|h_m|\le C |\l_1|$; then 
(assuming $\l_1>0$; similar computations can be repeated for
$\l_1<0$), for $\l$ small enough
$$(1-{5\over 4}a_2 
\l_1)\le{|m_{h-1}(\tilde\nu)|\over |m_h(\tilde\nu)|}
\le (1-{3\over 4} a_2
\l_1)\Eq(gg23)$$
Then
$$|m_{h-1}(\tilde\nu)|\ge |m_0| \g^{2 a_2
\l_1(h-1)}
{(1-{5\over 4} a_2 \l_1)
\over \g^{
-2 a_2\l_1}}\ge
|m_0| \g^{2 a_2\l_1(h-1)}\Eq(gg22)$$
and
$$|m_{h-1}(\tilde\nu)|\le |m_0|
\g^{{a_2\over 2}\l_1(h-1)}{(1-{3\over 4}
a_2 \l_1)
\over \g^{-\l_1 {a_2\over 2}}}\le
|m_0| \g^{{a_2\over 2}\l_1(h-1)}\Eq(gg21)$$
Finally, for $a_3$ large enough
$$|{m_{h-1}(\tilde\n)\over m_{h-1}(\tilde\n')}-1|=
|{\prod_{k=h}^1(1+\b_m^k(\tilde \n))-
\prod_{k=h}^1(1+\b_m^k(\tilde \n'))
\over\prod_{k=h}^1(1+\b_m^k(\tilde \n
'))}|$$
$$\le C_4\log\b\max_{0\le k\le h}
|v_k(\tilde v)-v_k(\tilde v')|\le 
\max_{j> h}|\tilde\n_k-\tilde\nu'_k||\l|^{-1}\Eq(gg50)$$
Finally the last of \equ(gg3) are found in a similar way, noting that
${Z_{h-1}(\tilde\nu)\over Z_h(\tilde\nu)}=1+\l^2_h(a_4+h_s)$,
with $|h_s|\le C|\l_1|$.
\*
\sub(1.4bzzz)
{\cs Lemma.} 
{\it There exists $\e$, $a_3$ and $C_\nu$ such that, 
for $|t-t_c|\ge e^{-1\over a_3\l^2}$ and $|\l|\le\e$,
there exists a $\n_1(\l)$ 
so that $\max_{k\ge h^*}|\nu_k|\le C_\nu|\l|$.}
\*
{\it Proof.} It is sufficient to look for a fixed
point for the operator ${\bf T}:M\to M$, 
and ${\bf T}$ is defined in the following way,
if $\tilde\nu'={\bf T}(\tilde\n)$, see \equ(5.29):
$$\tilde\n'_h  = -\g^{-h}\sum_{k=h^*+1}^h \g^{k-1} \b^\n_k(v_k(\tilde\nu),\tilde \n_k,\ldots,\n_0(\tilde\nu),\tilde\nu_0)
\;,\qquad h^*< h\le 1\;.\Eq(5.29a)$$

We want to prove that it is possible to choose the constant
$C_\n\ge 1$ in Lemma (5.2)
so that, if $|\l|$ is small enough, the set
$\FF=\{\tilde\n\in M: ||\tilde\n||\le C_\nu |\l|\}$ is
invariant under ${\bf T}$ 
and ${\bf T}$ is a contraction on it. This
is sufficient to prove the lemma, since $M$ is a Banach
space, as one can easily show.

By \equ(5.29a) and Theorem 3
$$|\tilde\n'_h|\le
\sum_{j=h^*+1}^h\g^{-h+j-1}
\left[C_{1,\n} |\l|+ \sum_{n=2}^{\io}c^n|\l|^n\right]
\;,\Eq(5.11a)$$
where $C_{1,\n}$ is a constant depending
on the
first order contribution (\ie the {\it tadpole}). So, 
for a proper $C_\nu$,
$||\tilde\n'||\le C_\nu|\l|$.

We then show that ${\bf T}$ 
is a contraction
on $\FF$. In fact, given
$\tilde\n_1,\tilde\n_2\in\FF$, by using Theorem 3
and Lemma (5.2), we see that, for $\l$ small enough
$$\tilde\n'_{1,h}-\tilde\n'_{2,h}
= -\g^{-h}\sum_{k=h^*+1}^h \g^{k-1}\times$$
$$[\b^\n_k(v_k(\tilde\nu_{1}),
\tilde \n_{k,1},\ldots,v_1(\tilde\nu_{1}),\tilde\nu_{1,1})-
\b^\n_k(v_k(\tilde\nu_{2}),
\tilde \n_{k,2},\ldots,v_1(\tilde\nu_{2}),
\tilde\nu_{1,2})]
\Eq(la22)$$
and 
$$|\tilde\n'_{1,h}-\tilde\n'_{2,h}|\le C_1 |\l|
[\g^{-h}\sum_{k=h^*+1}^h \g^{k-1}]
[\sum_{j\ge h}\g^{\a(h-j)}\max_{i\ge j}
|\tilde\n_{1,i}-\tilde\n_{2,i}|]\le 
C_2|\l|||\tilde\n_1-\tilde\n_2||
$$
with $C_1,C_2$ constants, 
so that ${\bf T}$ is a contraction.
\*
\vskip.5cm
\sub(1.bzz) If $a=b=0$ one has the case treated in [SP],
corresponding to two independent Ising
models with a quartic or a next to nearest neighbor
interaction. 
In such a case 
the local part of the quartic terms is vanishing by Pauli 
principle,
so that if $|P_v|=4$ one can apply "freely" a first order
renormalization obtaining an additional
$\g^{-(h_v-h_{v'})}$, for any $v$ such that $|P_v|=4$,
in the bounds.
At each step one can include all the quadratic running coupling constants
in the free integrations; their beta function is $\tilde \z_{i,h-1}=
\tilde \z_{i,h}
+O(\g^{\a h}\l)$ and $\z_{i,h-1}=\z_{i,h}
+O(\g^{\a h}\l)$,
as the beta function is sum over all the trees
with end-points necessarily at scale $h_v=2$ (
contrary to our case, as the local
part of the quartic terms is vanishing, 
and there are
no end-points associated to the quadratic terms
as they are included in the free integration). Hence all 
the quadratic couplings are $O(\l)$ and they do not change the scaling
properties of the propagator on a single scale; the qualitative 
behaviour
of $C_v$ for $\l=0$ or $\l\not=0$ are then the same
for temperatures up to $t_c$.
\vskip1cm
\section(2z,Correlation functions)
\vskip.5cm 
\sub(1.4c) {\it Flow of observables.}
We consider the following functional integral
$$e^{\SS(\phi)}=\int P(d\psi)
e^{\VV^{(1)}(\psi)+\BB(\phi,\psi)}
\;\Eq(6.2)$$
where $\VV^{(1)}$ given by
\equ(mas) and $\BB=\BB^1+\BB^2$ with
$$\BB^{(1)}(\psi,\phi)=
\int d\xx \phi^{(1)}(\xx)
[\psi^{+}_{\xx,1} \psi^{-}_{\xx,-1}+
\psi^{-}_{\xx,1} \psi^{+}_{\xx,-1}]
\;\Eq(6.3)$$
$$\BB^{(2)}(\psi,\phi)=
\int d\xx \phi^{(2)}(\xx)
[\psi^{+}_{\xx,1} \psi^{+}_{\xx,-1}
+\psi^{-}_{\xx,1} \psi^{-}_{\xx,-1}]
\;.\Eq(6.3b)$$

After integrating the fields $\psi^{(1)},...\psi^{(h+1)}$, $0\ge h\ge h^*$,
we find
$$e^{\SS(\phi)}=e^{-L\b E_h+S^{(h+1)}(\phi)}\int P_{Z_h, m_h,C_h}(d\psi^{\le
h})e^{-\VV^{(h)}(\sqrt{Z_h}\psi^{(\le h)})+\BB^{(h)}
(\sqrt{Z_h}\psi^{(\le h)},\phi)}\;,\Eq(6.4)$$
where $P_{Z_h,m_h,C_h}(d\psi^{(\le h)})$ 
and $\VV^{h}$ are given by \equ(2.66) and \equ(2.70a), 
respectively, while $S^{(h+1)}$ $(\phi)$, which
denotes the sum over all the terms dependent on $\phi$ but independent of
the $\psi$ field, and $\BB^{(h)}(\psi^{(\le h)}, \phi)$, which denotes the
sum over all the terms containing at least one $\phi$ field and two $\psi$
fields, can be represented in the form
$$S^{(h+1)}(\phi)=\sum_{m=1}^\io\int d\xx_1\cdots d\xx_m
S^{(h+1)}_m(\xx_1,\ldots,\xx_m)
\Big[\prod_{i=1}^m\phi^{(\a_i)}(\xx_i)\Big]\Eq(6.5)$$
$$\eqalign{
&\BB^{(h)}(\psi^{(\le h)},\phi)=\sum_{m=1}^\io\sum_{n=1}^{\io} \sum_{\ss,\oo}
\int d\xx_1\cdots d\xx_m d\yy_1 \cdots d\yy_{2n} \;\cdot\cr
&\qquad\cdot\; B^{(h)}_{m,2n,\ss,\oo}(\xx_1,\ldots,\xx_m;\yy_1,\ldots,\yy_{2n})
\Big[\prod_{i=1}^m\phi^{(\a_i)}(\xx_i)\Big] \Big[\prod_{i=1}^{2n}
\psi^{(\le h)\s_i}_{\yy_i,\o_i}\Big]\;.\cr}\Eq(6.6)$$
 
Since the field $\phi$ is equivalent, from the point of view of dimensional
considerations, to two $\psi$ fields, 
the only terms in the r.h.s. of
\equ(6.6) which are not irrelevant are those with $m=1$ and $n=1$, which are
marginal. The localization $\LL$ is defined equal to zero
except in the following cases

$$\LL\int d\xx d\yy d\zz W^h_{1,2}(\xx;\yy,\zz)\phi^{(1)}(\xx)
\psi^{\s}_{\yy,1} \psi^{\s}_{\zz,-1}
=\int d\xx d\yy d\zz W^h_{1,2}(\xx;\yy,\zz)\phi^{(1)}(\xx)
\psi^{\s}_{\xx,1} \psi^{\s}_{\xx,-1}\Eq(locdfc)$$
$$\LL\int d\xx d\yy d\zz W^h_{1,2}(\xx;\yy,\zz)\phi^{(2)}(\xx)
\psi^{\s}_{\yy,1} \psi^{-\s}_{\zz,-1}
=\int d\xx d\yy d\zz W^h_{1,2}(\xx;\yy,\zz)\phi^{(2)}(\xx)
\psi^{\s}_{\xx,1} \psi^{-\s}_{\xx,-1}$$

Hence
$$\LL \BB^{(h)}(\psi^{(\le h)},\phi)={Z^{(1)}_h\over Z_h} F_1^{(\le h)}
+{Z^{(2)}_h\over Z_h} F_2^{(\le h)}\;,\Eq(6.12)$$
where $Z^{(1)}_h$ and $Z^{(2)}_h$ are real numbers, such that
$Z^{(1)}_1=Z^{(2)}_1=1$ and
$$F_1^{(\le h)}=
\int d\xx \phi^{(1)}(\xx)
[\psi^{(\le h)+}_{\xx,1}\psi^{(\le h)-}_{\xx,-1}
+\psi^{(\le h)-}_{\xx,1}\psi^{(\le h)+}_{\xx,-1}]
\;,\Eq(6.13)$$
$$F_2^{(\le h)}=\int d\xx \phi^{(2)}(\xx)
[\psi^{(\le h)+}_{\xx,1}\psi^{(\le h)+}_{\xx,-1}
+\psi^{(\le h)-}_{\xx,1}\psi^{(\le h)-}_{\xx,-1}]
\;.\Eq(6.14)$$
 
By using the notation of the preceding section
$$\eqalign{
&e^{-M^2 t_h} \int P_{\tilde Z_{h-1},m_{h-1},C_h}(d\psi^{(\le h)})
e^{-\tilde\VV^{(h)}(\sqrt{Z_h}\psi^{(\le h)})+\BB^{(h)}
(\sqrt{Z_h}\psi^{(\le h)},\phi)}\;=\cr
&= e^{-M^2 t_h} \int P_{Z_{h-1},m_{h-1},C_{h-1}}(d\psi^{(\le h-1)})\;\cdot\cr
&\cdot\; \int P_{Z_{h-1},m_{h-1},\tilde f_h^{-1}}(d\psi^{(h)})
e^{-\hat\VV^{(h)}(\sqrt{Z_{h-1}}\psi^{(\le h)})+\BB^{(h)}
(\sqrt{Z_{h-1}}\psi^{(\le h)},\phi)}\;,\cr}\Eq(6.15)$$
where 
$\BB^{(h-1)}(\psi^{(\le h-1)},\phi)$ and $S^{(h)}(\phi)$
are then defined by
$$\eqalign{
&e^{-\VV^{(h-1)}(\sqrt{Z_{h-1}}\psi^{(\le h-1)})+\BB^{(h-1)}
(\sqrt{Z_{h-1}}\psi^{(\le h-1)},\phi)-L\b\tilde E_h+\tilde S^{(h)}(\phi)}=\cr
&=\int P_{Z_{h-1},m_{h-1},\tilde f_h^{-1}}(d\psi^{(h)})
e^{-\hat\VV^{(h)}(\sqrt{Z_{h-1}}\psi^{(\le h)})+\hat\BB^{(h)}
(\sqrt{Z_{h-1}}\psi^{(\le h)},\phi)}\;.\cr}\Eq(6.17)$$
 
The definitions \equ(6.12) easily imply that
$${Z^{(1)}_{h-1}\over Z_{h-1}} = {Z^{(1)}_{h}\over Z_{h}}
[1 + a_1\l_h+h_1]\qquad
{Z^{(2)}_{h-1}\over Z_{h-1}}  = {Z^{(2)}_{h}\over Z_{h}}
[1 - a_1\l_h+h_2]
\Eq(6.18)$$
with $|h_i|\le C\l^2$, $i=1,2$;  proceeding as in the
proof of Lemma 5.2 
$$\g^{-\l_1 c_4 h}< {Z^{(1)}_h\over Z_h}< \g^{-\l_1 c_3 h}\quad
\g^{\l_1 c_1 h}< {Z^{(2)}_h\over Z_h}< \g^{\l_1 c_2 h}\;.\Eq(5.54xx)$$

The fields of scale between $h^*$ and $h_{L,\b}$ are integrated
in a single step, 
and
it follows that
$$S(\phi)= -L\b E_{L,\b}+ S^{(h)}(\phi)=
-L\b E_{L,\b}+ \sum_{h=h^*}^1 \tilde S^{(h)}(\phi)\;;\Eq(6.20)$$
As a $\phi$ fields is dimensionally analogous
to two external fields, we get, proceeding as in the proof
of Theorem 3 (see also [BM] \S 5 for a detailed proof
of similar bounds in a related case)
$$|{\partial^2\over\partial \phi^{(1)}(\xx)
\partial\phi^{(1)}(\yy)}\tilde S^{(h)}(\phi)||_{\phi=0}
\le \g^{2h} [{Z^{(1)}_h\over Z_h}]^2 {C_N\over 1+(\g^h|d(\xx-\yy)|)^N}\Eq(spal1)$$
$$|{\partial^2\over\partial \phi^{(2)}(\xx)
\partial\phi^{(2)}(\yy)}\tilde S^{(h)}(\phi)||_{\phi=0}
\le \g^{2h} [{Z^{(2)}_h\over Z_h}]^2 {C_N\over 1+(\g^h|d(\xx-\yy)|)^N}\Eq(spal2)$$
and if $\xx_{\a}=\xx,\yy$ and $\phi_\a=\phi_1,\phi_2$
$$|{\partial^3\tilde S^{(h)}(\phi)
\over\partial \phi^{(\a_1)}(\xx_1)
\partial\phi^{(\a_2)}(\xx_2) \partial \phi^{(\a_3)}(\xx_3)
}||_{\phi=0}
\le \g^{-h}\g^{4h}[\prod_{i=1}^3
{Z^{(\a_i)}_h\over Z_h} ]{C_N\over 1+(\g^h|d(\xx-\yy)|)^N}\Eq(spal3)$$
$$|{\partial^4\tilde S^{(h)}(\phi)\over\partial \phi^{(\a_1)}(\xx_1)
\partial\phi^{(\a_2)}(\xx_2) 
\partial \phi^{(\a_3)}(\xx_3)\partial \phi_{\a_4}(\xx_4)
}||_{\phi=0}\le$$
$$\g^{-2h}\g^{6h}[\prod_{i=1}^4
{Z^{(\a_i)}_h\over Z_h}]  
{C_N\over 1+(\g^h|d(\xx-\yy)|)^N}\Eq(Spla4)$$

\vskip.5cm 
\sub(1.4cz) {\it Correlation functions.}
The correlation function $<\s^{(\a)}_\xx\s^{(\a)}_{\xx'}
\s^{(\a)}_\yy\s^{(\a)}_{\yy'}>_T$,
where $\xx'=(x+1,x_0)$ or $(x,x_0+1)$ is given by
$$<\s^{(\a)}_\xx\s^{(\a)}_{\xx'}
\s^{(\a)}_\yy\s^{(\a)}_{\yy'}>_T={\partial\over
\partial J^{(\a)}_{\xx,\xx'}}
{\partial\over\partial J^{(\a)}_{\yy,\yy'}}\log Z_{2I}(\{
J^{(\a)}_{\xx,\xx'}\})|_{\{J_{\xx,\xx'}\}=\{J\}} \Eq(fgc)$$
If $O^{(\a)}(\xx)=\s^{(\a)}_{x,x_0}
\s^{(\a)}_{x+1,x_0}$ each derivative
produces a factor
$${\rm sech}^2 J \bar H^{\a}_\xx H^{(\a)}_{x+1,x_0}+\tanh J +
{\partial
\over \partial J_{\xx;x+1,x_0}}\VV\Eq(sklvv)$$
with $\VV$ given by \equ(c61).
We define, using \equ(l9),\equ(c61)
$$
<O^{(\a)}(\xx)
O^{(\a)}(\yy)>_{\L;\e^{(1)},\e^{'(1)},\e^{(2)},\e^{'(2)}}=$$
$${\int [\prod_{\a=1}^2
P^{(\a)}_{\e^{(\a)},\e^{'(\a)}}
(dH^{(\a)},dV^{(\a)})]
e^{\VV}
O^{(\a)}(\xx)O^{(\a)}(\yy)\over 
\int [\prod_{\a=1}^2
P^{(\a)}_{\e^{(\a)},\e^{'(\a)}}
(dH^{(\a)}, dV^{(\a)})]
e^{\VV}}\Eq(fg2)$$
Moreover we call
$$
<O^{(\a)}(\xx)
O^{(\a)}(\yy)>_{\L,T;\e^{(1)},\e^{'(1)},\e^{(2)},\e^{'(2)}}=
<O^{(\a)}(\xx)O^{(\a)}(\yy)
>_{\L;\e^{(1)},\e^{'(1)},\e^{(2)},\e^{'(2)}}$$
$$-<O^{(\a)}(\xx)
>_{\L;\e^{(1)},\e^{'(1)},\e^{(2)},\e^{'(2)}}
<O^{(\a)}(\xx)
>_{\L;\e^{(1)},\e^{'(1)},\e^{(2)},\e^{'(2)}}\Eq(bhdd)$$
so that
$$<O^{(\a)}(\xx)O^{(\a)}(\xx)>_{\L,T}=
\sum_{\e^{(1)},\e^{'(1)}}
(-1)^{\d_{\e^{(1)},\e^{'(1)}}}
\sum_{\e^{(2)},\e^{'(2)}}(-1)^{\d_{\e^{(2)},\e^{'(2)}}}$$
$${
Z_{2I}^{\e^{(1)},\e^{'(1)},\e^{(2)},\e^{'(2)}}
<O^{(\a)}(\xx)O^{(\a)}(\yy)>_{\L,T;\e^{(1)},\e^{'(1)},\e^{(2)},\e^{'(2)}}\over
{\sum_{\e^{(1)},\e^{'(1)}}
(-1)^{\d_{\e^{(1)},\e^{'(1)}}}
\sum_{\e^{(2)},\e^{'(2)}}(-1)^{\d_{\e^{(2)},\e^{'(2)}}} 
Z_{2I}^{\e^{(1)},\e^{'(1)},\e^{(2)},\e^{'(2)}}}}\Eq(sg3)$$

Suppose that $\xx$ and $\yy$
are fixed to an $M$ independent value; then 
$$\lim_{M\to\io}
<O^{(\a)}(\xx)O^{(\a)}(\yy)>_{\L,T;\e^1,\e'^1,\e^2,\e'^2}
-<O^{(\a)}(\xx)O^{(\a)}(\yy)>_{\L,T;-,-,-,-}=0.
\Eq(como)$$
In fact, as explained in \S 4.10,
the l.h.s. of \equ(como)
can be written as a sum of trees, and in each of them
there is a factor $e^{i\pp_{\e,\e'}\xx}-1$ or
an end-point associated to $\tilde Q_{\e^1,\e'^1,\e^2,\e'^2}$.
With respect to the previous bounds,  
simply dimensional analysis says
there is now an extra factor
${\g^{-h^*}\over M}$ in the bounds 
so it is vanishing in the limit $M\to\io$,
if $\xx,\yy$ and $t-t_c$ are fixed to an $M$-independent value.
We can then simply study 
$<O^{(\a)}(\xx)O^{(\a)}(\yy)>_{\L,T;-,-,-,-}$
which is given by the Grassmann integral \equ(fg2)
with $O^{(\a)}$ given by \equ(sklvv). 
By performing the change of variables
\equ(c63) and \equ(c66), we get a sum
of averages of monomials in the $\psi$ and $\chi$
fields; we integrate the $\chi$-fields, as discussed
in \S 3,
and we obtain a sum of Grassmann integrals of monomials
in the $\psi$ fields; remembering that
$$\bar H^{\a}_{\xx} H^{(\a)}_{\xx} 
\bar H^{\a}_{\yy} H^{(\a)}_{\yy}=
{1\over 16}[\psi^-_{1,\xx}\psi^-_{-1,\xx}
\psi^+_{1,\yy}\psi^+_{-1,\yy}+\psi^-_{1,\yy}\psi^-_{-1,\yy}
\psi^+_{1,\xx}\psi^+_{-1,\xx}+\Eq(ama8)$$
$$\psi^-_{1,\xx}\psi^+_{-1,\xx}
\psi^+_{1,\yy}
\psi^-_{-1,\yy}+\psi^-_{1,\xx}\psi^+_{-1,\xx}
\psi^-_{1,\yy}
\psi^+_{-1,\yy}+
\psi^+_{1,\xx}\psi_{-1,\xx}^-\psi^-_{1,\yy}
\psi^+_{-1,\yy}+
\psi^+_{1,\xx}\psi_{-1,\xx}^-\psi^+_{1,\yy}
\psi^-_{-1,\yy}]$$
we find the first two of \equ(jkazz) and \equ(bvqii)
by \equ(6.2), \equ(6.18), \equ(6.20)
and \equ(spal1),\equ(spal2).
In $\O^{(\a),c}$ we include to contributions
to the correlation function corresponding
to monomials with six or more $\psi$ fields, and
from \equ(spal3), \equ(Spla4) the last of \equ(jkazz)
follows.

Finally the specific heat is obtained in a similar way,
noting that we have to
sum over $\xx-\yy$, and this produces
an extra $\g^{2h}$ in 
the r.h.s. of \equ(spal1),\equ(spal2),\equ(spal3),\equ(Spla4).
\vskip1cm
{\bf Acknowledgments.} This paper was partly written
in the stimulating atmosphere of the
Institute for Advanced Studies, in Princeton. I thank
Prof. Spencer for his invitation
and for many clarifying discussions about his work [PS]. 
{\baselineskip=12pt
\vskip1cm
\centerline{\titolo References}
\*
\halign{\hbox to 1.2truecm {[#]\hss} &
        \vtop{\advance\hsize by -1.25 truecm \0#}\cr
B&{R. Baxter, Exactly solved models in statistical mechanics,
Academic Press (1982)}\cr
BG& {G. Benfatto, G. Gallavotti:
Renormalization group. Physics notes 1, Princeton University
Press (1995). }\cr
BGPS& {G. Benfatto, G. Gallavotti, A. Procacci, B. Scoppola:
Beta Functions and Schwinger Functions for a Many Fermions System in One
Dimension.
{\it  Comm. Math. Phys.} {\bf  160}, 93--171 (1994). }\cr
BM& {G. Benfatto, V.Mastropietro. 
Renormalization group, hidden symmetries 
and approximate Ward identities in the $XYZ$ model. 
{\it Rev. Math. Phys.} 13 (2001), no. 11, 1323--143;
and to appear in {\it Comm. Math. Phys.}}\cr
GM& {Gentile,.G, Mastropietro.V. 
Renormalization group for one-dimensional fermions. 
A review on mathematical results. {\it Phys. Rep.} 352 (2001), no. 4-6,
273--43}\cr
K&{P.W.Kasteleyn, Dimer Statistics and phase transitions, {\it J. Math.Phys.}
4,287 (1963)}\cr
F&{C. Fan, On critical properties of the Ashkin-Teller model,
{\it Phis. Rev. B}, 6, 136-136 (1972)}\cr
H&{C.Hurst, New approach to the Ising problem,
, {\it J.Math. Phys.} 7,2, 305-310 (1966)}\cr
ID&{ C. Itzykson, J. Drouffe, "Statistical field theory: 1," Cambridge Univ.
Press, 1989.}\cr
Le& {A. Lesniewski:
Effective action for the Yukawa 2 quantum field Theory.
{\it Comm. Math. Phys.} {\bf 108}, 437-467 (1987). }\cr
Li& {H. Lieb, Exact solution of the problem of entropy of two-dimensional
ice,
{\it Phys. Rev. Lett.}, 18, 692-694, (1967)}\cr
LP&{A.Luther, I.Peschel. Calculations
of critical exponents in two dimension from quantum
field theory in one dimension.
{\it Phys. Rev. B} 12, 3908-3917 (1975)}\cr
MW&{B. McCoy, T. Wu, "The two-dimensional Ising model," 
Harvard Univ. Press, 1973.}\cr
MPW&{E.Montroll, R.Potts, J.Ward. Correlation
and spontaneous magnetization of the two dimensional 
Ising model. {\it J. Math. Phys.} 4,308 (1963)}\cr
N& {M.P.M. den Nijs. Derivation of extended scaling relations between
critical exponents in two dimensional models from the one dimensional
Luttinger model, {\it Phys. Rev. B}, 23, 11 (1981)
6111-6125}\cr
PB& {A.M.M. Pruisken, A.C. Brown.
Universality fot the critical lines of the eight vertex,
Ashkin-Teller and Gaussian models, {\it Phys. Rev.} B, 23, 3 (1981)
1459-1468}\cr  
PS& {H. Pinson, T.Spencer. Universality in 2D critical Ising model.
To appear in Comm. Math. Phys.}\cr
S& { S. Samuel, 
The use of anticommuting variable integrals in statistical mechanics'',
{\it J. Math. Phys.} 21 (1980) 2806}\cr
Spe& {T. Spencer: A mathematical approach to universality in
two dimensions. {\it Physica A} {\bf 279}, 250-259 (2000). }\cr
SML& { T. Schultz, D. Mattis, E. Lieb, Two-dimensional
Ising model as a soluble problem
of many Fermions, {\it Rev. Mod. Phys.} 36 (1964) 856.}\cr
W& {F.W.Wu The Ising model with four spin interaction
{\it Phys. Rev.} B 4, 2312-2314 (1971)}\cr}
\bye

\*
\section(2, Ashkin Teller model)
\*
In the case $J_1\not=J_2$.
there is an extra massive term,
if $\D=m_1-m_2$
$$\D\sum_{\o=\pm 1}\int d\kk [\psi^+_{\kk,1}
\psi^+_{-\kk,-1}+\psi^-_{\kk,1}
\psi^-_{-\kk,-1}]$$
This running coupling constant has an its own flow,
$\D\g^{\h' h}$. the propagator has the following form,
if $m=m_1+m_2$
$$\exp[{1\over M^2}\sum_{\kk\in D_{\e,\e'}^+}
{\bf\tilde\x^T_\kk \tilde A(\kk){\bf\tilde\x^{(+)}}_\kk}]$$
$$\tilde A T (\kk)= 
\left( \matrix{\sin k-i \sin k_0 & i m(k)&0&\D\cr
-i m(k) & \sin k+i \sin k_0&\D & 0 \cr
\D & 0& \sin k-i \sin k_0 & i m(k)\cr
0 & \D &-i m(k) & \sin k+i \sin k_0
\cr}\right)$$
$${\bf\tilde\x^{T}}_\kk=(\tilde\psi_{\kk,1},\tilde\psi_{\kk,-1},
\tilde\psi^+_{-\kk,1},\tilde\psi^+_{-\kk,-1}
)$$
one can repat the analysis as above; one has a flow $m_h=m\g^{\h h}$
and $\D_h=\D\g^{\h' h}$, and everything is as before if $m_h>>\D_h$.

\*
\section(6,The bibliography)
\*
\vskip.cm
[PB] A.M.M. Pruisken, A.C. Brown.
{\it Universality fot the critical lines of the eight vertex, 
Ashkin-Teller and Gaussian models}, Phys. Rev. B, 23, 3 (1981)
1459-1468
\vskip.5cm
[N] M.P.M. den Nijs. {\it Derivation of extended scaling relations between
critical exponents in two dimensional models from the one dimensional
Luttinger model}, Phys. Rev. B, 23, 11 (1981)
6111-6125
\vskip1cm
F.W.Wu The Ising model with four spin interaction
 Phys. Rev. B 4, 2312-2314 (1971)
\bye